\documentclass[]{aa}

\usepackage[fleqn]{amsmath}
\usepackage{amssymb}
\usepackage{textcomp}
\usepackage{graphicx}
\usepackage{txfonts}
\usepackage{natbib}
\usepackage{url}
\usepackage{multirow}
\usepackage{subfigure}
\usepackage[dvipsnames]{xcolor}
\usepackage{placeins}
\usepackage[version=3]{mhchem}
\usepackage{multicol}
\usepackage{lscape}
\usepackage{chemarr}

\begin{document}

\title{The molecular composition of the planet-forming regions of protoplanetary disks 
across the luminosity regime}
\titlerunning{The molecular composition of protoplanetary disks}

\author{Catherine Walsh\inst{\ref{inst1}} 
\and Hideko Nomura\inst{\ref{inst2}} 
\and E.~F.~van~Dishoeck\inst{\ref{inst1},\ref{inst3}}}
\authorrunning{Walsh, C. et al.}

\institute{Leiden Observatory, Leiden University, P.~O.~Box 9513, 2300 RA Leiden, The Netherlands \\
\email{cwalsh@strw.leidenuniv.nl} \label {inst1} 
\and Department of Earth and Planetary Sciences, Tokyo Institute of Technology, 2-12-1 Ookayama, 
Meguro-ku, Tokyo 152-8551, Japan \label{inst2}
\and Max-Planck-Institut f\"{u}r extraterretrische Physik, Giessenbachstrasse 1, 
85748 Garching, Germany \label{inst3}}

\date{Received 15 June 2015 / Accepted 30 July 2015}

\abstract
{Near- to mid-infrared observations of molecular emission from protoplanetary disks 
show that the inner regions are rich in small organic volatiles 
(e.g., \ce{C2H2} and \ce{HCN}).  
Trends in the data suggest that disks around cooler stars ($T_\mathrm{eff}$~$\approx$~3000~K) 
are potentially (i) more carbon-rich; and (ii) more molecule-rich than their hotter 
counterparts ($T_\mathrm{eff}$~$\gtrsim$~4000~K).}
{To explore the chemical composition of the planet-forming region ($<$~10AU) 
of protoplanetary disks around stars over a range of spectral types 
(from M~dwarf to Herbig~Ae) and compare with the observed trends.}
{Self-consistent models of the physical structure of a protoplanetary disk around 
stars of different spectral types are coupled with a comprehensive gas-grain chemical 
network to map the molecular abundances in the planet-forming zone.  
The effects of (i) \ce{N2} self shielding; (ii) X-ray-induced chemistry; and (iii) initial 
abundances, are investigated. 
The chemical composition in the `observable' atmosphere is compared 
with that in the disk midplane where the bulk of the planet-building reservoir resides.}
{M~dwarf disk atmospheres are 
relatively more molecule rich than those for T~Tauri or Herbig~Ae disks.    
The weak far-UV flux helps retain this complexity which is enhanced by 
X-ray-induced ion-molecule chemistry.  
\ce{N2} self shielding has only a small effect in the disk molecular 
layer and does not explain the higher \ce{C2H2}/\ce{HCN} ratios 
observed towards cooler stars.  
The models underproduce the \ce{OH}/\ce{H2O} column density ratios constrained 
in Herbig~Ae disks, despite reproducing (within an order of magnitude) 
the absolute value for \ce{OH}: the inclusion of self shielding for
\ce{H2O} photodissociation only increases this discrepancy.   
One possible explanation is the adopted disk structure. 
Alternatively, the `hot' \ce{H2O} ($T$~$\gtrsim$~300~K) chemistry may be more 
complex than assumed.
The results for the atmosphere are independent of the assumed initial abundances; 
however, the composition of the disk midplane is sensitive to the initial main 
elemental reservoirs. 
The models show that the gas in the inner disk is generally more carbon rich 
than the midplane ices. This effect is most significant for disks around cooler stars.  
Furthermore, the atmospheric C/O ratio appears larger than it actually is when calculated 
using observable tracers only.  
This is because gas-phase \ce{O2} is predicted to be a significant reservoir of 
atmospheric oxygen.}
{The models suggest that the gas in the inner regions of 
disks around cooler stars is more carbon rich; however, 
calculations of the molecular emission are necessary to definitively confirm 
whether the chemical trends reproduce the observed trends.}


\maketitle

\section{Introduction}
\label{introduction}

Protoplanetary disks provide the ingredients - dust, gas, and ice -  for 
planets and planetesimals such as comets 
\cite[for a review, see, e.g.,][]{williams11}.  
In disks around low-mass stars ($\lesssim$~2~$M_\odot$), 
planetary systems are thought to form relatively close to the 
parent star ($\lesssim$~10~AU); hence, the chemical composition of the inner disk region 
sets the initial conditions and available elemental components for planetary systems.  
The molecular material within $\approx$~10~AU is generally dense 
($\gtrsim$~10$^{8}$~cm$^{-3}$) and can reach high temperatures 
($\gtrsim$~100~K) allowing (ro)vibrational excitation of molecules which 
emit radiation at near- to mid-infrared (IR) wavelengths.  
The physical conditions within this region drive the chemistry 
towards the formation of small, simple, stable molecules 
most of which, fortunately, also have strong (ro)vibrational transitions.  

Near- to mid-IR spectroscopy of nearby primordial protoplanetary disks 
has demonstrated that the inner planet-forming regions are rich in organic volatiles.  
The {\em Spitzer Space Telescope} allowed the first detection 
of simple organic molecules in protoplanetary disks at IR wavelengths.  
\citet{lahuis06} reported absorption bands of \ce{C2H2}, \ce{HCN}, and \ce{CO2} in 
the spectrum of the low-mass young stellar object, IRS~46, attributed to 
absorption from hot molecules in a disk within a few AU of the embedded star.  
The following year, \citet{gibb07} detected absorption bands from 
\ce{CO}, \ce{C2H2}, and \ce{HCN} in the disk around a member of the binary system, GV~Tau, 
using NIRSPEC on Keck.  
Ongoing efforts have detected additional molecules in either emission 
or absorption, including, \ce{OH}, \ce{H2O}, and \ce{CH4}, in several nearby disks 
using both {\em Spitzer} and ground-based facilities
\citep[see, e.g.,][]{carr08,salyk08,pascucci09,pontoppidan10,carr11,fedele11,
salyk11,mandell12,bast13,najita13,gibb13,pascucci13}.  
Because of dust opacity, such observations probe the composition of 
the disk atmosphere only.  
It remains unclear whether the atmospheric composition 
is representative of that of the disk midplane  
within which planetesimals sweep up the bulk of their material.  

Several interesting trends have been noticed in the IR observations.  
\citet{pascucci08,pascucci09} presented results from 
a low-resolution ($R$~$\approx$~64~-~128) {\em Spitzer}/IRS survey of more than 
60 sources ranging from brown dwarfs ($T_\mathrm{eff}$~$\approx$~3000~K) 
to Sun-like stars ($T_\mathrm{eff}$~$\approx$~5000~K).  
The observations demonstrated an underabundance 
of HCN relative to \ce{C2H2} in disks around 
M~dwarfs compared with those around T~Tauri stars.  
The authors postulate this could be due to the difference in FUV 
(far-ultraviolet) luminosity: 
M~dwarf stars have insufficient FUV photons to dissociate \ce{N2}, 
the main nitrogen reservoir, thereby trapping elemental nitrogen which would 
otherwise be available to form other nitrogen-containing species, e.g., HCN.  
The authors have since published additional observations
at higher spectral resolution ($R$~$\approx$~600) and find the same result: 
the ratio of \ce{C2H2}/\ce{HCN} line flux and relative column density decreases 
with increasing spectral type \citep{pascucci13}.   
They also find that the \ce{HCN}/\ce{H2O} line flux ratios are higher for M~dwarf stars 
than for T~Tauri stars leading the authors to conclude that the C/O 
ratio in the inner regions of M~dwarf and brown dwarf disks is higher ($\approx$~1) 
than that for disks around T~Tauri stars ($<$~1).  
Within the subset of T~Tauri disks, \citet{najita13} postulated 
that a second trend was present.
They found a correlation between 
the \ce{HCN}/\ce{H2O} line flux ratio and disk mass.   
Their hypothesis is that planetesimal formation is more efficient in higher 
mass disks and is able to lock up a significant fraction of oxygen 
(in the form of water ice) thereby increasing the C/O ratio in the inner 
region of the disk.  
In summary, the C/O ratio in the inner regions of protoplanetary disks appears 
to increase with decreasing spectral type, and within a particular sub 
class of star-disk systems, to increase with increasing disk mass.

\citet{pontoppidan10} conducted a similar survey in the 10~-~36~$\mu$m wavelength range, 
with a source list also covering Herbig Ae/Be stars ($T_\mathrm{eff}$~$\gtrsim$~10,000~K).   
They detected strong \ce{H2O} mid-IR line emission from 22 T~Tauri stars in their sample, 
with a detection rate on the order of 2/3; however, to their surprise, no disks in their sample of 
25 Herbig stars exhibited water (nor \ce{OH}) line emission.  
At near-IR wavelengths, \citet{fedele11} conducted a high-resolution spectroscopic 
survey ($L$-band) of Herbig~Ae/Be disks.  
\ce{OH} emission was detected in only four objects, mainly flared disks, and, similar to 
that found by \citet{pontoppidan10}, \ce{H2O} was not detected.  
Both sets of authors suggested that the stronger FUV flux from the Herbig~Ae/Be stars 
dissociates molecules in the unshielded inner disk region thereby lowering the line flux 
in relation to the stellar luminosity.  
Observations from {\em Herschel} support this hypothesis: 
most of the Herbig~Ae/Be disks in the GASPS and DIGIT key programs exhibit strong 
OI emission 63~$\mu$m with a sub sample also showing OH emission \citep{meeus12,fedele13}.  
Many of these sources also have strong OI emission at $6300~\AA$ \citep[see e.g.,][]{acke05}.
In contrast, only HD~163296 (which is a `flat' or settled disk) has a robust (>~3~$\sigma$) 
water detection between 50 and 220~$\mu$m \citep{fedele12,fedele13}.   

One of the first chemical models which concentrated 
{\em solely} on the inner disk ($\lesssim10$~AU) of a T~Tauri star is presented in 
\citet{markwick02}; however, the authors neglected the influence 
of the stellar and interstellar radiation fields on the disk physics and
chemistry using the reasoning that viscous heating will dominate 
the inner disk structure.   
The initial detections of volatiles in the inner regions of protoplanetary 
disks \citep{lahuis06,gibb07,carr08} prompted a flurry of astrochemical 
models mainly focussed on T~Tauri disks 
\citep[e.g,][]{agundez08,willacy09,woods09,walsh10,najita11}.  
These models differed somewhat in their level of complexity.   
\citet{agundez08} adapted a model used for photon-dominated regions (PDRs) 
but neglected heating due to UV excess emission and X-rays and also assumed 
that the dust and gas temperatures were equal.   
\citet{willacy09} and \citet{woods09} calculated the gas temperature by solving the 
equation of thermal balance; however, the dust temperature and density 
were assumed and heating by UV excess emission from the star was neglected 
\citep{dalessio06}.  
\citet{najita11} adopted a similar approach albeit using a reduced chemical network.  
\citet{walsh10} used a {\em self-consistent} model for the protoplanetary 
disk structure including the effects of heating by 
UV excess emission and X-rays \citep{nomura05,nomura07} and a chemical 
network similar in complexity to the work of \citet{willacy09}, excluding 
deuterium.  
Several works have also concentrated solely on water production in 
the inner regions of protoplanetary disks 
\citep{glassgold09,bethell09,meijerink09,woitke09,adamkovics14,du14}; however, 
to date, there has been no detailed study of chemistry in the inner regions 
of M~dwarf nor Herbig~Ae stars, especially to address the trends seen 
in the \ce{C2H2}/\ce{HCN} line ratios. 

In this work, we compute the physical and chemical structure of the planet-forming 
region ($\lesssim$~10~AU) of a protoplanetary disk around stars of different spectral types: 
(i) an M~dwarf star; (ii) a T~Tauri star; and (ii) a Herbig~Ae star.  
Our aim is to investigate whether the stellar radiation field plays a role 
in the observed trends in \ce{C2H2}/\ce{HCN} and \ce{OH}/\ce{H2O} ratios derived from the 
mid-IR observations.  
Given the proposed importance of \ce{N2} photodissociation for the formation of HCN, 
we investigate the effect of \ce{N2} self shielding 
using recently computed shielding functions \citep{li13a} on the subsequent nitrogen 
chemistry.  
We also use these models to probe the connection between the 
`observable' gas emission from the disk atmosphere 
(at near- to mid-IR wavelengths) with the 
chemical composition of the midplane within which forming planets and 
planetesimals sweep up the bulk of their elemental building blocks.  

The remainder of the paper is structured as described.  
In Sect.~\ref{diskmodels}, we outline the methods for 
computing the disk physical and chemical structure, in Sect.~\ref{results} 
we present our results, and in Sects.~\ref{discussion} and \ref{summary}, we 
discuss the implications of this work and summarise the main results, respectively.

\section{Protoplanetary disk models}
\label{diskmodels}

\subsection{Physical model}
\label{physicalmodel}

The physical structure of each disk model is calculated using the 
methods outlined in \citet{nomura05} with the addition of X-ray heating as 
described in \citet{nomura07}.  
Because the methodology is covered in detail in a series of previous publications 
\citep[see, e.g.,][]{nomura05,nomura07,walsh10,walsh12,walsh14},  
here, we highlight the important parameters only.

\begin{figure}
\includegraphics[width=0.5\textwidth]{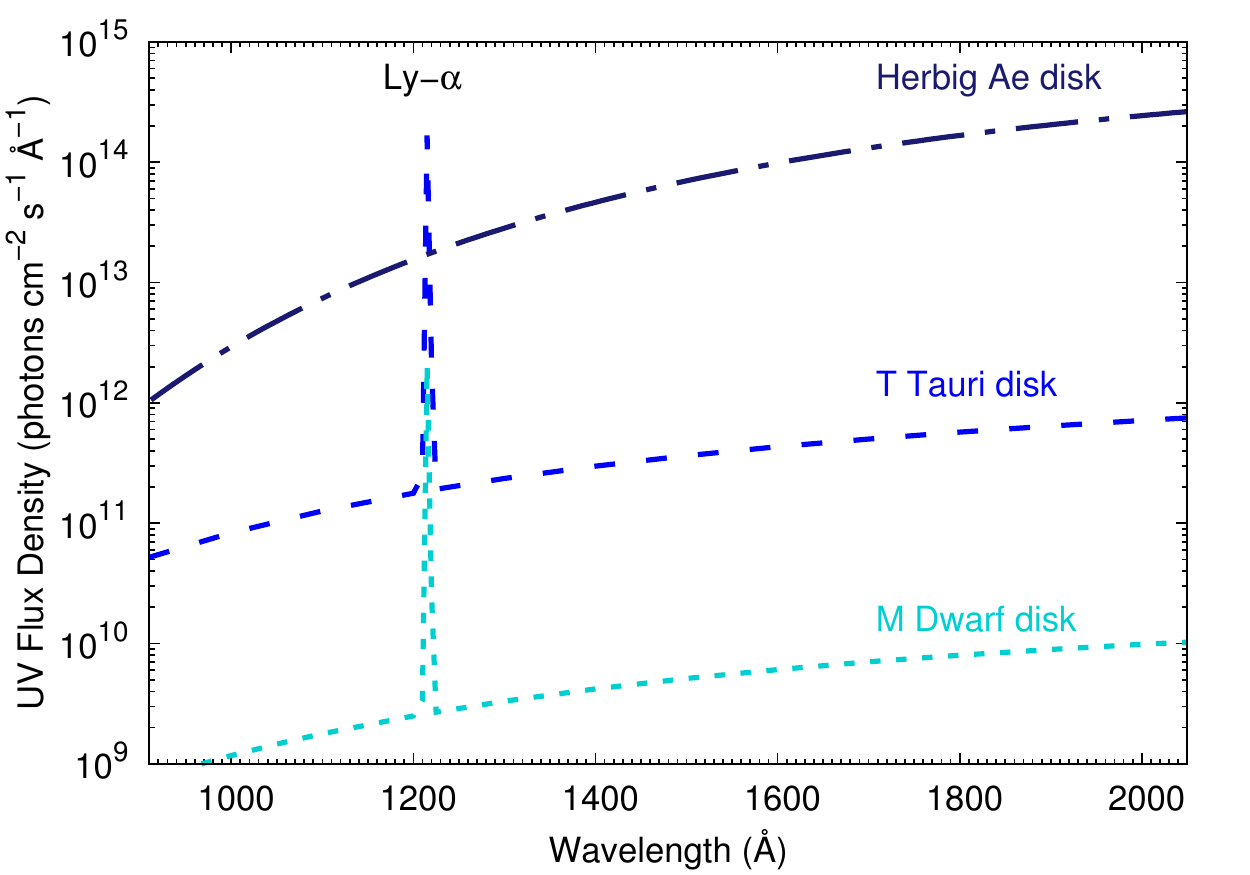}
\caption{FUV radiation fields at a distance of 1~AU from the central star (see text for details).}
\label{figure1}
\end{figure}

For each disk model, we assume the disk is steady, axisymmetric, and 
in Keplerian rotation about the central star.  
We parametrise the kinematic viscosity via the dimensionless 
$\alpha$ parameter which scales the maximum size of turbulent eddies 
by the disk scale height, $H$, and the sound speed of the gas, $c_s$, 
i.e., $\nu$~$\approx$~$\alpha\,H\,c_s$.  
For protoplanetary disks, $\alpha$~$\approx$~0.01.  
We model the structure of a disk surrounding a star of three different 
spectral types: an M~dwarf star, a T~Tauri star, and a Herbig~Ae star.  
We list the stellar mass, $M_\star$, stellar radius, $R_\star$, and effective temperature, $T_\star$, 
of each host star in Table~\ref{table1}. 
We also list the adopted mass accretion rate of each star-disk system, $\dot{M}$, 
and the gas mass surface density at 10~AU, $\Sigma_\mathrm{10AU}$.   
We assume the T~Tauri and Herbig~Ae systems have a mass accretion 
rate typical for pre-main-sequence stars, $\sim$~10$^{-8}$~$M_\odot$ yr$^{-1}$.  
Because the accretion signatures from lower-mass stars are 
not as strong as for the higher-mass systems, we assume a lower accretion 
rate for the M~dwarf system, $\sim$~10$^{-9}$~$M_\odot$ yr$^{-1}$ \citep[see, e.g.,][]{herczeg09}.
We calculate the dust temperature assuming local radiative equilibrium between 
the absorption and reemission of radiation by dust grains.  
For the calculation of the FUV extinction by dust grains, we 
adopt a dust-grain size distribution which replicates the extinction 
curve observed in dense clouds \citep{weingartner01}.
We calculate the gas temperature assuming detailed thermal balance between the 
heating and cooling of the gas. 
We include heating via photoelectric emission from dust grains induced 
by FUV photons and heating due to the X-ray ionisation of H atoms. 
The cooling mechanisms included are gas-grain collisions and line transitions.  
The radiation field of each star is simulated as a black body at the stellar 
effective temperature (as listed in Table~\ref{table1}) with UV excess emission 
scaled to the relative mass accretion rates for the M~dwarf and T~Tauri disks.  
The stellar FUV (912~--~2100~$\AA$) radiation field at 1~AU is shown in Figure~\ref{figure1} for 
all three central stars.  
The UV excess emission has two components derived from a best-fit model 
of the observed TW~Hya spectrum: a diluted black-body spectrum 
to simulate bremsstrahlung emission ($T_\mathrm{br}\sim25,000$~K) and Lyman-$\alpha$ line emission.  
The Lyman-$\alpha$ line is modelled as a Gaussian with a width 
$\approx$~2~$\AA$~and the peak flux is determined assuming a line/continuum luminosity ratio of $10^{3}$ 
\citep[see][and references therein]{nomura05}. 
Because young stars also exhibit strong X-ray emission, we use 
a TW~Hya-like X-ray spectrum (generated by fitting the observed {\em XMM-Newton} spectrum) 
for the M~dwarf and T~Tauri stars, with a total luminosity, $L_x$~$\sim$~10$^{30}$~erg~s$^{-1}$ 
\citep{preibisch05} and assume $L_x$~$\approx$~3~$\times$~10$^{29}$~erg~s$^{-1}$ and 
$T_x$~$\approx$~1.0~keV for the X-ray spectrum of the Herbig~Ae star 
\citep[see, e.g.,][]{zinnecker94,hamaguchi05}.  

\begin{table*}
\caption{Star and disk parameters. \label{table1}}
\centering
\def\arraystretch{1.5}
\begin{tabular}{lccccccc}
\hline\hline
Star      & $M_\star$    & $R_\star$    & $T_\star$ & $\dot{M}$             & $\Sigma_\mathrm{10AU}$ & $L_{x}$        & UV Excess?  \\ 
          & ($M_\odot$)  & ($R_\odot$)  & (K)       & ($M_\odot$ yr$^{-1}$) & (g cm$^{-2}$)          & (erg s$^{-1}$) &             \\ 
\hline
M Dwarf   & 0.1  & 0.7 & 3,000   & $10^{-9}$  & 1.0 & 10$^{30}$ & Y \\
T Tauri   & 0.5  & 2.0 & 4,000   & $10^{-8}$  & 12  & 10$^{30}$ & Y \\
Herbig Ae & 2.5  & 2.0 & 10,000  & $10^{-8}$  & 56  & 10$^{29}$ & N \\
\hline
\end{tabular}
\tablefoot{The total disk mass will depend on the assumed outer disk radius which in this work, is $>>$~10~AU. 
Instead, we list the gas mass surface density at a radius of 10~AU.}
\end{table*}

\subsection{Chemical model}
\label{chemicalmodel}

The network we use to calculate the disk chemical evolution includes 
gas-phase reactions, gas-grain interactions 
(freezeout and desorption), and grain-surface chemistry.  

\subsubsection{Gas-phase network}
\label{gasphasenetwork}
The basis for the gas-phase chemistry is the complete network from the recent 
release of the UMIST Database for Astrochemistry (UDfA) termed `\textsc{Rate}12' which is 
publicly available\footnotemark~\citep{mcelroy13}.  
\footnotetext{\url{http://www.udfa.net}}
\textsc{Rate}12 includes gas-phase two-body reactions, 
photodissociation and photoionisation, direct cosmic-ray ionisation, 
and cosmic-ray-induced photodissociation and ionisation.  
In this work, the photodissociation and photoionisation rates are calculated by integrating over 
the specific reaction cross section for each species and the calculated FUV spectrum at each point 
in the disk \citep{walsh12} using the cross sections from \citet{vandishoeck06}.
Similar to previous work, we have supplemented this gas-phase network with direct X-ray ionisation reactions
and X-ray-induced ionisation and dissociation processes 
\citep[as described in][and see Sect.~\ref{xraychemistry}]{walsh12}.  
We have also added a set of three-body reactions compiled for use 
in combustion chemistry models\footnotemark~\citep[see, e.g.,][]{baulch05}~which 
are necessary because three-body processes become increasingly important 
in the inner disk region where the density and temperature are sufficiently high 
($\gtrsim$~10$^{10}$~cm$^{-3}$ and $\gtrsim$~1000~K). 
We have also included at least one collisional dissociation 
reaction (AB~+~M~$\rightarrow$~A~+~B~+~M) for all neutral species 
expected to be abundant in the inner disk.  
We include a small chemical network 
involving vibrationally excited, or `hot', \ce{H2}.  
For each gas-phase neutral-neutral reaction involving \ce{H2} which also 
possesses an activation barrier, we include a duplicate reaction 
involving hot \ce{H2} with a barrier reduced by the internal energy 
of the excited \ce{H2} \citep[$\approx$~30,163~K, see, e.g.,][and references therein]{bruderer12}.

\footnotetext{\url{http://kinetics.nist.gov/kinetics/index.jsp}}

\subsubsection{Gas-grain interactions}
We allow the freezeout (adsorption) of molecules on dust grains forming 
ice mantles and the desorption (sublimation) of ices via thermal desorption  
and photodesorption \citep[][]{tielens82,hasegawa92,walsh10,walsh12}.  
We adopt the set of molecular binding energies compiled for use in conjunction 
with \textsc{Rate}12 \citep{mcelroy13}.  
We have updated the binding energies in light of recent measurements 
for \ce{HCN} \citep{noble12}.
To simplify the calculation of the gas-grain interaction rates, we 
assume compact spherical grains with a radius of 0.1~$\mu$m and a fixed 
density of $\sim$~10$^{-12}$ relative to the gas number density.  
Each grain thus has $\sim$~10$^{6}$ surface binding sites.  
We include photodesorption by both external photons and photons 
generated internally via the interaction of cosmic rays with \ce{H2} 
molecules.  
We use experimentally determined photodesorption yields where available 
\citep[see, e.g.,][]{oberg09a,oberg09b,oberg09c}. 
For all other species we use a yield of ~10$^{-3}$ molecules photon$^{-1}$. 
In the calculation of the freezeout rates, we assume a sticking 
coefficient, $S$~$\sim$~1, for all species except H, for which 
we use a temperature-dependent expression which takes into account both physisorption 
and chemisorption and describes the decreased sticking probability at 
higher temperatures \citep{sha05,cuppen10b}. 
We assume the rate of \ce{H2} formation equates to half the 
rate of arrival of H atoms on dust grain surfaces. 

\subsubsection{Grain-surface network}
For completeness, we also supplement our reaction scheme with grain-surface 
association reactions extracted from the publicly available 
Ohio State University (OSU) network\footnotemark~\citep{garrod08}.  
For those species important in grain-surface chemical reaction schemes, 
e.g., the \ce{CH3O} radical, which are not included in \textsc{Rate}12, we also 
extract the corresponding gas-phase formation and destruction reactions 
from the OSU network.  
The grain-surface network has been further updated to include 
all studied routes to water formation under 
interstellar and circumstellar conditions \citep{cuppen10a,lamberts13}.
The grain-surface reaction rates are calculated assuming the Langmuir-Hinshelwood 
mechanism only, and using the rate-equation method as described in \citet{hasegawa92}.  
We limit the chemically `active' zone to the top two monolayers of the
ice mantle.  
We assume the size of the barrier to surface diffusion is 0.3~$\times$ the 
binding energy; in this way, volatile species diffuse at a faster rate than 
strongly bound species.  
This value lies at the optimistic end of the range determined by recent 
off-lattice kinetic Monte Carlo simulations of surface diffusion of 
\ce{CO} and \ce{CO2} on crystalline water ice \citep{karssemeijer14}. 
This allows the efficient formation of complex organic molecules 
via radical-radical association reactions at $\gtrsim$~20~K 
\citep[see, e.g.,][]{vasyunin13,walsh14}.  
For the lightest reactants, H and \ce{H2}, we use either the classical diffusion rate or the 
quantum tunnelling rate depending on which is fastest \citep[see, e.g.,][]{tielens82,hasegawa92}. 
For the latter rates, we follow \citet{garrod11} and adopt a rectangular barrier 
of width 1.5~$\AA$.  
We also include reaction-diffusion competition in which the reaction 
probability is determined by the relative rates between the 
barrier-mediated reaction and thermal 
diffusion \citep[see, e.g.,][]{chang07,garrod11}.  
Although still relatively simplistic, this takes into account the increased 
probability of reaction in the limit where the thermal diffusion of 
the reactants away from a common binding site is slow compared with the 
barrier-mediated reaction rate. 

\footnotetext{\url{http://faculty.virginia.edu/ericherb/research.html}}

\subsubsection{Photodissociation}
One further important process now included is a more 
robust description of the photodissociation rate of \ce{N2} 
which includes the effects of self shielding and mutual shielding by \ce{H} 
and \ce{H2}. 
\ce{N2} is important because it is thought to be the main nitrogen-bearing 
molecule in interstellar and circumstellar media.  
Self (and mutual) shielding occurs predominantly for those species 
which dissociate via line transitions and occurs when foreground 
material removes photons necessary for dissociation deeper into the cloud.  
In this way, the photodissociation rate of molecules which can self shield 
is reduced relative to the dissociation rates for those species 
which dissociate via the absorption of continuum photons only.  
\ce{H2} and \ce{CO} are famous examples of molecules which can 
self shield \citep[see, e.g.,][]{federman79,glassgold85,vandishoeck88,lee96}.  
\ce{H2} also possesses line transitions which overlap with dissociative states 
of CO and \ce{N2} leading to shielding of CO and \ce{N2} by foreground 
\ce{H2}, a process termed `mutual' shielding.  

The inclusion of the self (and mutual) shielding of \ce{N2} is now possible 
due to the work by \citet{li13a} in which they present parametrised 
shielding functions calculated using a high-resolution model spectrum of \ce{N2}.  
These shielding functions are publicly available for download\footnotemark~to use 
in astrochemical models.
To use the computed shielding functions which are parametrised in temperature and 
\ce{H}, \ce{H2}, and \ce{N2} column density, one needs to {\em a priori} calculate 
the foreground column densities of each species.  
In protoplanetary disk models this is not trivial because the dissociating radiation 
can have multiple sources (e.g., stellar photons and interstellar photons).  
In addition, the chemistry needs to be calculated in series 
(from the inside outwards and from the surface downwards) 
rather than in parallel which can significantly increase computation time. 
For these reasons, a more pragmatic approach is adopted here, similar to that used in \citet{visser11}: 
an `effective' shielding column is calculated based on the calculated FUV integrated 
flux relative to that assuming no intervening material.
The FUV extinction, $\tau_\mathrm{UV}$, is given by
\begin{equation}
\tau_\mathrm{UV}(R,Z) = - \ln \left [ \frac{G_\mathrm{UV}(R,Z)}{G_\star(R,Z)+ G_\mathrm{ext}} \right ], \\  
\end{equation}
where $G_\mathrm{UV}(R,Z)$ and $G_\star(R,Z)$ are the 
calculated and the geometrically diluted unattenuated stellar FUV integrated fluxes at 
a grid point $(R,Z)$, and $G_\mathrm{ext}$ is the external unattenuated FUV flux.  
The effective visual extinction is related to the UV extinction via the empirical relation, 
$A'_\mathrm{v}(R,Z)$~$\approx$~$\tau_\mathrm{UV}(R,Z)/3.02$~mag, and the `effective' 
\ce{H2} column density is calculated using, 
$N'_\ce{H2}(R,Z)$~$\approx$~0.5~$\times$~(1.59~$\times$~10$^{21}$)~$\times$~$A'_\mathrm{v}(R,Z)$~cm$^{-2}$ 
\citep{bohlin78}.
For simplicity, the effective shielding column densities for \ce{N2} 
at each point in the disk are estimated by assuming \ce{N2} has a fixed 
(rather conservative) fractional abundance, $\sim$~10$^{-5}$, with respect to \ce{H2}.  
Finally, the photodissociation rate for \ce{N2} is given by,
\begin{equation}
k_\mathrm{ph}^\ce{N2}(R,Z) = 
k_0^\ce{N2}(R,Z) \times \theta_\ce{N2} \left[N'_\ce{H2}(R,Z),N'_\ce{N2}(R,Z),T(R,Z) \right] \quad \mathrm{s}^{-1},    
\end{equation}
where $k_0^\ce{N2}(R,Z) = \int_\lambda G_\mathrm{UV}(R,Z,\lambda)\,\sigma_\ce{N2}(\lambda) \, d\lambda$ is the 
unshielded photodissociation rate for \ce{N2} and $\theta_\ce{N2}$ is the 
shielding function which is a function of \ce{H2} and \ce{N2} column density and temperature, 
$T$.    

We adopt a similar method for the computation of the \ce{H2} and CO 
photodissociation rates using shielding functions calculated by 
\citet{lee96} and \citet{visser09}, respectively.  

\subsubsection{X-ray-induced reactions}
\label{xraychemistry}

We include a set of X-ray-induced reactions which we duplicate from the existing 
set of cosmic-ray-induced reactions contained in {\sc Rate}12 \citep{mcelroy13}.  
The reaction rates are estimated by scaling the cosmic-ray-induced reaction rate 
by the ratio of the local X-ray and cosmic-ray ionisation rates, 
i.e., $k_\mathrm{XR} \approx k_\mathrm{CR}\times(\zeta_\mathrm{XR}/\zeta_\mathrm{CR})$.  
This is a common assumption in chemical models of X-ray irradiated environments 
\citep[see, e.g.,][]{maloney96,stauber05}.  
The `secondary' X-ray ionisation rate, $\zeta_\mathrm{XR}$, 
is calculated at each grid point in the disk 
by taking into account the local X-ray spectrum and the explicit elemental 
composition of the gas \citep{glassgold97}. 
We also include a set of `primary' X-ray ionisation reactions 
\citep[for further details see][]{walsh12}. 

\footnotetext{\url{http://home.strw.leidenuniv.nl/~ewine/photo/}}

\subsubsection{Initial Abundances}

To generate a set of initial abundances for input into the disk model, 
we run a dark cloud model 
($T_\mathrm{gas}$~=~$T_\mathrm{dust}$~=~10~K, $n$~=~10$^{4}$~cm$^{-3}$, and $A_\mathrm{V}$~=~10~mag).   
We use the low-metal elemental abundances from \citet{graedel82} supplemented with updated values 
for O, C, and N based on diffuse cloud observations: $3.2\times10^{-4}$, $1.4\times10^{-4}$, and $7.5\times10^{-5}$
relative to total hydrogen nuclei density, respectively \citep{cardelli91,cardelli96,meyer98}.  
In this way, we begin the disk calculations with an appreciable ice 
reservoir on the grain mantle built up over the lifetime of the pre-stellar core 
prior to disk formation.  

In Table~\ref{table2} we list the abundances (with respect to 
total H nuclei density) of abundant C-, N-, and O-bearing species 
at times of 1.0, 3.2, and 10.0 $\times10^{5}$ years.  
We limit the listed species to those which have an abundance 
$\gtrsim$~1\% that of water ice, the dominant O-bearing species 
at late times ($>$~10$^5$ years).  
Over the relatively large time steps listed, the trend from atomic to 
molecular gas can be seen, as can the freezeout of 
volatile species formed in the gas phase, such as \ce{CO} and \ce{N2}. 
Ice species formed in situ via hydrogenation of atoms
on and within the grain mantle (\ce{CH4}, \ce{NH3}, 
and \ce{H2O}) show a general trend 
of increasing abundance towards late times ($\sim$~$10^{6}$ years) 
as does the abundance of \ce{CO2} ice which is formed primarily via the reaction 
between \ce{CO} and \ce{OH}.  
The behaviour of \ce{H2CO} and \ce{CH3OH} 
is more complex: although both are formed via the hydrogenation 
of CO ice, over time, processing of the ice by the cosmic-ray-induced UV field  
causes a depletion in \ce{CH3OH} at late times in favour of \ce{H2CO}.  

The median relative ice abundances in dense, quiescent clouds measured 
in absorption against background stars
is 100:31:38:4 for \ce{H2O}:\ce{CO}:\ce{CO2}:\ce{CH3OH} \citep[][]{oberg11}.  
These values are also in line with those measured in low-mass 
protostellar envelopes which also include measurements for \ce{NH3} and \ce{CH4} 
\citep[\ce{H2O}:\ce{NH3}:\ce{CH4} = 100:5:5,][]{oberg11}.  
We opt to use initial abundances at a time of 3.2~$\times10^{5}$ years
which corresponds to an ice ratio of 100:15:6:1 for 
\ce{H2O}:\ce{CO}:\ce{CO2}:\ce{CH3OH} and 100:7:22 for \ce{H2O}:\ce{NH3}:\ce{CH4}. 
This is a compromise between the set of abundances at early and late times: 
at the former, the CO abundance is low compared with observations, whereas 
at the latter, the \ce{CH3OH} abundance is low compared with observations.  
However, for the physical conditions in the inner disk, we expect the 
initial abundances to be important only in the midplane of the disk where 
the chemical timescales can be long compared with the disk lifetime.  

\begin{table}
\caption{Abundances with respect to total H nuclei density for the dark cloud model and used 
as initial abundances in the disk model. \label{table2}}
\centering
\def\arraystretch{1.5}
\begin{tabular}{lccc}
\hline\hline
Species & 1.0$\times10^{5}$~yrs & 3.2$\times10^{5}$~yrs & 1.0$\times10^{6}$~yrs \\
\hline
\multicolumn{4}{c}{Gas species} \\
\hline
\ce{H}     & 7.7(-05) & 5.2(-05) & 3.6(-05)  \\
\ce{H2}    & 5.0(-01) & 5.0(-01) & 5.0(-01)  \\
\ce{He}    & 9.8(-02) & 9.8(-02) & 9.8(-02)  \\
\ce{C}     & 4.4(-05) & 2.2(-06) & 5.1(-09)  \\
\ce{N}     & 5.1(-05) & 2.8(-05) & 5.8(-07)  \\
\ce{O}     & 1.9(-04) & 1.0(-04) & 7.5(-06)  \\
\ce{CH4}   & 1.0(-06) & 8.8(-07) & 6.5(-08)  \\
\ce{CO}    & 5.5(-05) & 7.4(-05) & 1.9(-05)  \\
\ce{N2}    & 3.2(-06) & 6.0(-06) & 4.2(-06)  \\
\ce{O2}    & 8.2(-09) & 1.8(-07) & 5.4(-06)  \\
\hline
\multicolumn{4}{c}{Ice species} \\
\hline
\ce{CH4}    & 1.7(-05) & 2.5(-05) & 2.8(-05)  \\
\ce{NH3}    & 5.8(-06) & 7.4(-06) & 8.4(-06)  \\
\ce{H2O}    & 6.1(-05) & 1.1(-04) & 1.8(-04)  \\
\ce{CO}     & 6.9(-07) & 1.6(-05) & 6.7(-05)  \\
\ce{N2}     & 2.4(-06) & 1.1(-05) & 2.8(-05)  \\
\ce{H2CO}   & 2.4(-08) & 1.5(-07) & 2.3(-06)  \\
\ce{CH3OH}  & 1.6(-06) & 1.5(-06) & 9.3(-07)  \\
\ce{CO2}    & 2.5(-06) & 6.3(-06) & 1.3(-05)  \\
\hline
\end{tabular}
\tablefoot{$a(b)$ means $a$~$\times$~10$^{b}$}
\end{table}

\section{Results}
\label{results}

\subsection{Disk physical structure}
\label{diskphysicalstructure}

\begin{figure*}
\includegraphics[width=1.0\textwidth]{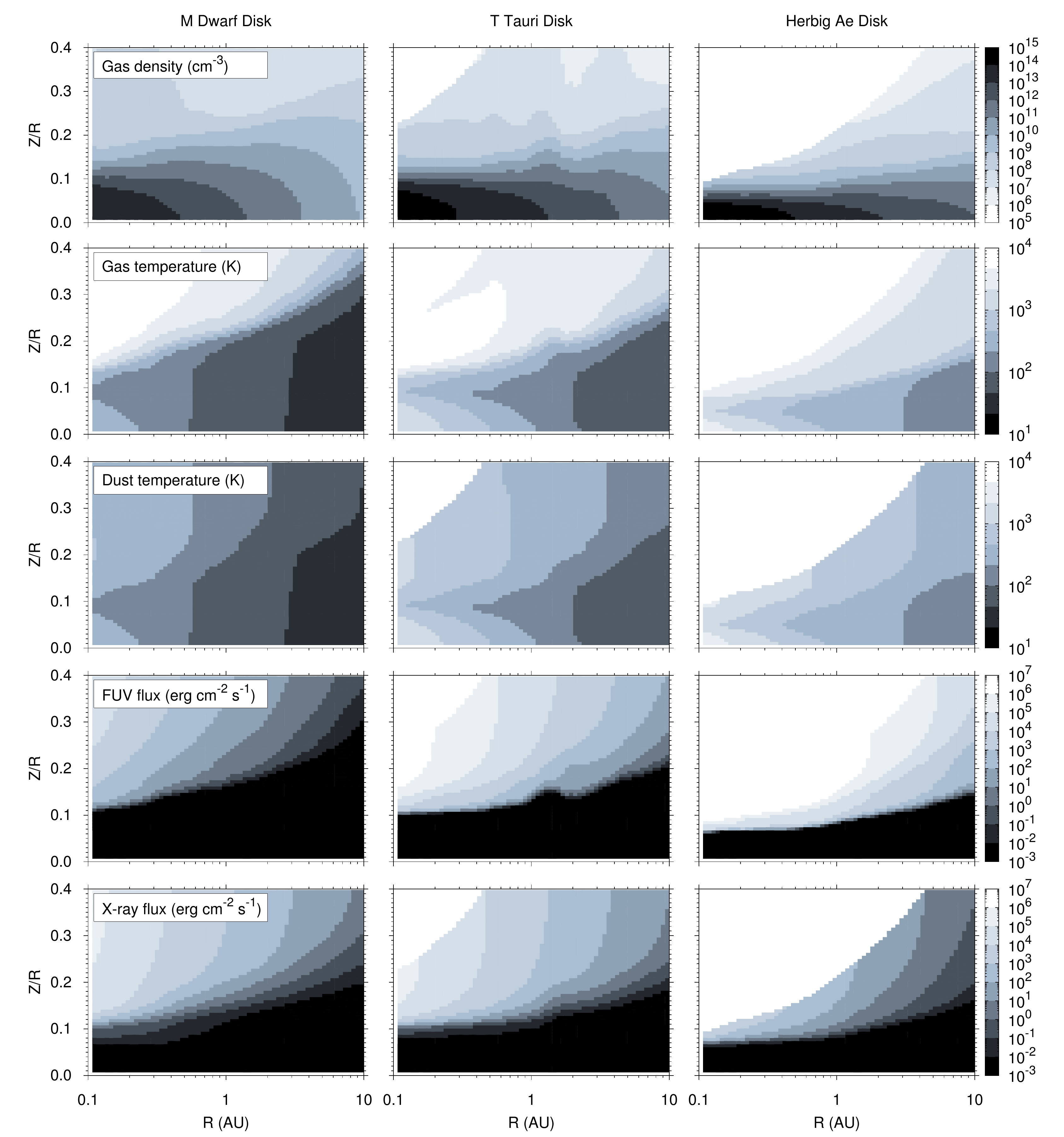}
\caption{Number density (top row), gas temperature (second row), dust temperature 
(middle row), far-UV flux (forth row) and X-ray flux (bottom row) as a function of 
disk radius and disk height (scaled by the radius) for the M dwarf disk (left column), 
the T Tauri disk (middle column) and the Herbig Ae disk (right column).}
\label{figure2}
\end{figure*}

In Figure~\ref{figure2} we display the physical structure of 
each disk as a function of radius and height (scaled by the radius) 
for the M~dwarf disk (left-hand column), 
T~Tauri disk (middle column), and Herbig~Ae disk (right-hand column).  
The lower density in the atmosphere of the Herbig~Ae disk 
is because the scale height of the Herbig disk is smaller than that 
for the disks around the lower-mass stars ($H = c_s/\Omega \propto M_\star^{-0.5}$, 
where $c_s$ and $\Omega$ are the sound speed and 
keplerian angular velocity, respectively).  
The surface density of the M~dwarf disk is around an order of magnitude lower than 
the other two objects; hence, the lower number density of gas throughout. 
In all three disks, the gas and dust temperatures decouple 
in the disk atmosphere such that the gas is significantly hotter than the  
dust.  
There are several general trends with increasing spectral type: 
(i) the gas and dust temperatures increase; 
(ii) the strength of the FUV flux in the disk surface 
increases; (iii) the strength of the X-ray flux in the disk surface 
decreases.  
For the M~dwarf and T~Tauri disks, the X-rays penetrate deeper into the disk atmosphere 
than the FUV photons.  
For all three disks, the midplane is effectively shielded 
from all sources of external radiation, including the central star and the 
interstellar radiation field.  
The increasing importance of viscous heating in the midplane 
is indicated by a temperature inversion below which the temperature 
begins to increase with depth
(see the second and third rows of Figure~\ref{figure2}). 

\subsection{\ce{C2H2} and \ce{HCN}}
\label{C2H2andHCN}

\subsubsection{Chemical structure}

\begin{figure*}
\includegraphics[width=1.0\textwidth]{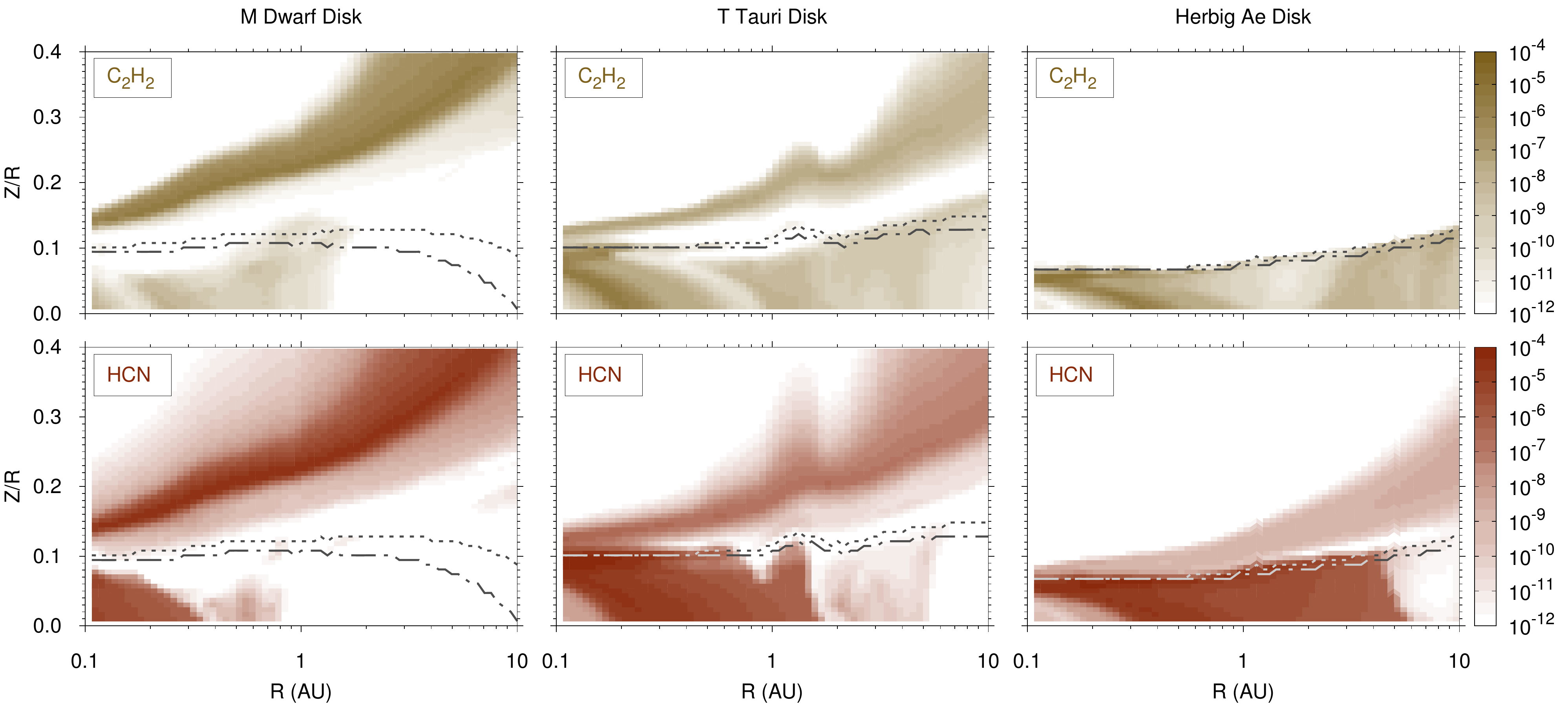}
\caption{Fractional abundance relative to total gas number density of 
\ce{C2H2} (top row) and \ce{HCN} (bottom row) for the M~dwarf disk (left-hand column), 
T~Tauri disk (middle column), and Herbig~Ae disk (right-hand column). 
The dotted and dot-dashed lines indicate the dust column density (integrated from the surface downwards) 
at which $\tau$~$\approx$~1 at 3~$\mu$m and 14~$\mu$m, respectively.}
\label{figure3}
\end{figure*}

In Figure~\ref{figure3} we display the 
fractional abundance with respect to gas number density of 
\ce{C2H2} and \ce{HCN} as a function of disk radius, $R$, and 
disk height divided by radius, $Z/R$. 
The dotted and dot-dashed lines represent the $\tau$~=~1 surface at 
3 and 14~$\mu$m, respectively.  
These are determined using the 
dust opacity table adopted in the computation of the disk physical structure.  
Throughout the remainder of the paper, we adopt the ad hoc definition of the 
disk atmosphere as the material above the $\tau$(14~$\mu$m)~=~1 surface.

As the stellar effective temperature increases, several trends are evident: 
(i) the disk molecular layer is pushed deeper into the disk atmosphere; 
(ii) the fractional abundances of \ce{C2H2} and \ce{HCN} decrease in the atmosphere; and 
(iii) the extent over which both species reach a significant abundance increases 
in the disk midplane. 
\ce{C2H2} and \ce{HCN} are relatively abundant in the molecular layer of the M~dwarf 
disk reaching maximum fractional abundances of $\approx~$~5~$\times$~10$^{-6}$ 
and $\approx$~5~$\times$~10$^{-5}$, respectively.  
The peak \ce{C2H2} fractional abundance in the molecular layer of 
the T~Tauri disk is $\approx$~~1~$\times$10$^{-7}$, whereas
that for the Herbig~Ae disk is negligible ($<$~10$^{-11}$).  
The peak \ce{HCN} fractional abundance in the molecular layer 
for the T~Tauri and Herbig~Ae disks is around two ($\sim$~10$^{-7}$) and four 
($\sim$~10$^{-9}$) orders of magnitude lower than that for the M~dwarf disk.  
Thus, the model results suggest that the {\em relative} molecular 
complexity in the disk atmosphere decreases with increasing stellar 
effective temperature in line with increased photodestruction.  

\ce{C2H2} reaches a relatively large fractional abundance ($\gtrsim$~10$^{-6}$)
only in specific regions in the midplane of the T~Tauri and Herbig~Ae disks whereas 
in the M~dwarf midplane it is much lower ($\lesssim$~10$^{-9}$). 
In contrast, \ce{HCN} reaches a relatively 
high fractional abundance ($\gtrsim$~$10^{-6}$) 
over a greater spatial extent when compared with \ce{C2H2}.  
An apparent HCN `snow line' moves outwards as the disk midplane temperature 
increases (see Figure~\ref{figure3}).  
However, emission from midplane HCN at near- to mid-IR wavelengths is likely obscured 
by dust in the disk atmosphere as shown by the locations of the 
$\tau$~=~1 surface at 3 and 14~$\mu$m in Figure~\ref{figure3}. 

\subsubsection{Chemistry of \ce{C2H2} and HCN}

\begin{figure}
\includegraphics[width=0.5\textwidth]{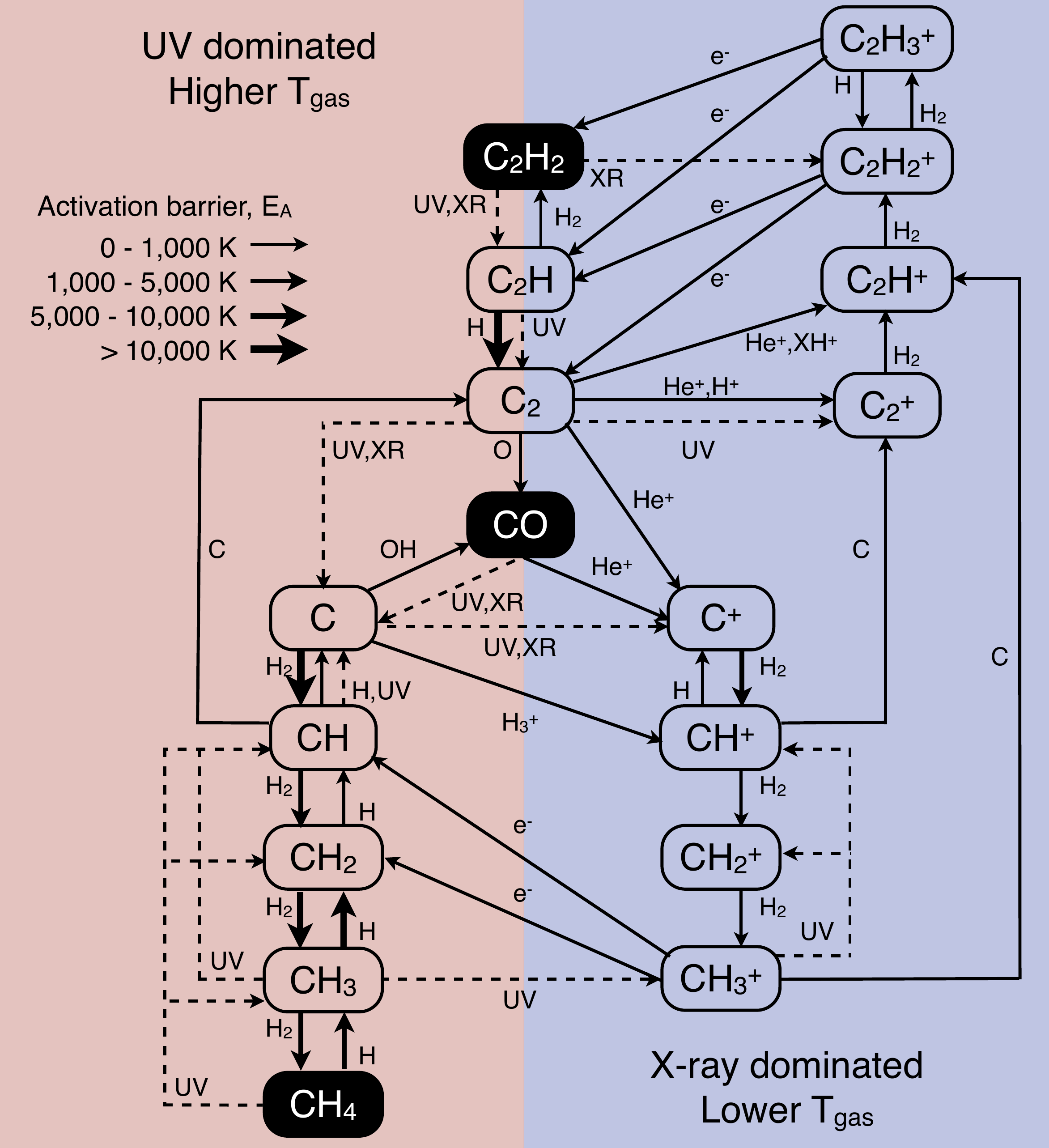}
\caption{Dominant reactions controlling the abundance of \ce{C2H2} 
and related species at the position of peak fractional abundance in the disk 
atmospheres at a radius of 1~AU.  
The solid lines indicate gas-phase two-body reactions and the dotted lines 
indicate UV or X-ray-induced reactions.  
The width of the arrows for the gas-phase reactions 
shows the magnitude of the activation barrier.  
The left-hand side, highlighted in red, is the chemistry most dominant 
in UV-dominated gas with a higher gas temperature (e.g., in Herbig~Ae disks), whereas 
that highlighted in blue on the right-hand side is that which dominates in 
X-ray irradiated gas with a lower gas temperature (e.g., in X-ray irradiated M~dwarf 
disks).}
\label{figure3a}
\end{figure}

\begin{figure}
\includegraphics[width=0.5\textwidth]{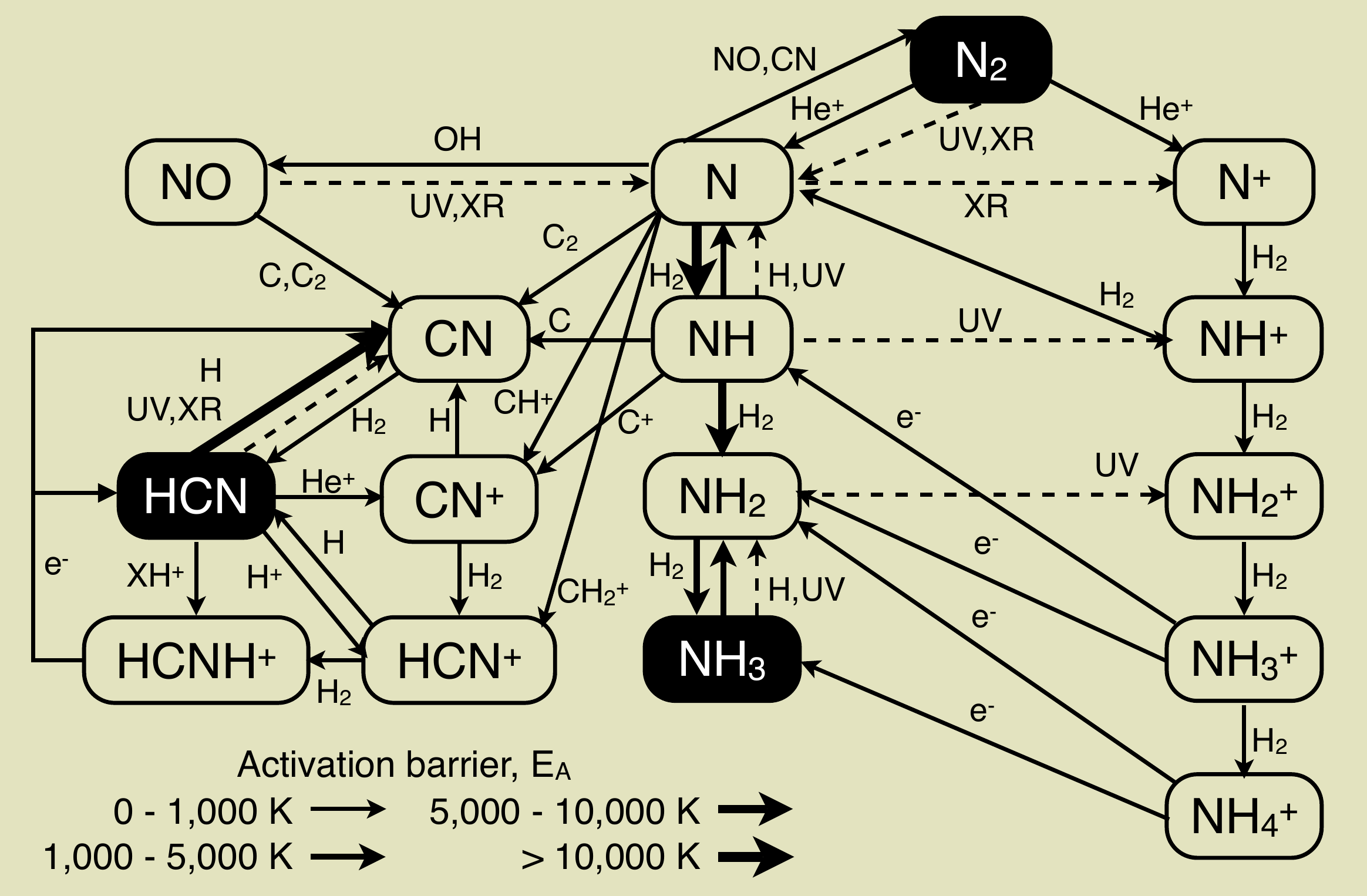}
\caption{Dominant reactions controlling the abundance of  
\ce{HCN} and related species at the position of peak fractional 
abundance in the disk atmospheres at a radius of 1~AU. 
The solid lines indicate gas-phase two-body reactions and the dotted lines 
indicate UV or X-ray-induced reactions.  
The width of the arrows for the gas-phase reactions 
shows the magnitude of the activation barrier.}
\label{figure3b}
\end{figure}

A snapshot of the dominant chemical reactions controlling the abundance of 
\ce{C2H2} and \ce{HCN} (and related species) in the disk atmosphere is given in 
Figures~\ref{figure3a} and \ref{figure3b}, respectively.  
The reactions shown are those which contribute $\gtrsim10$\% to the formation 
and destruction rates at the position of peak fractional abundance 
in the atmosphere at a radius of 1~AU.  
The networks are similar to those presented in \citet{agundez08} and
\citet{bast13} except that we also include the dominant destruction 
mechanisms.

Free carbon and nitrogen (necessary for incorporation into molecules and 
radicals such as CH, NH, CN, and \ce{C2}) 
are released from the main gas-phase reservoirs (CO and \ce{N2}) via 
photodissociation for the Herbig~Ae disk.  
For the two cooler disks, both X-ray-induced dissociation and reactions 
with \ce{He+} (which itself is produced by direct X-ray ionisation) 
are dominant.
Many reactions involving \ce{He+} have measured rate coefficients 
\citep{adams76,anicich77}.  

Whether or not a molecule survives in the disk atmosphere with an 
appreciable abundance requires a delicate balance between 
formation (via neutral-neutral chemistry or ion-molecule chemistry) 
and destruction (via photodissociation or X-ray-induced dissociation).  
The abundance and distribution of \ce{HCN} in the atmosphere of 
all three disks is primarily controlled by formation 
via \ce{H2}~+~\ce{CN} \citep{baulch94}, and destruction 
via photodissociation \citep{vandishoeck06}. 
In the Herbig~Ae disk, destruction via reaction with atomic hydrogen 
also plays a minor role.  
This reaction has a large reaction barrier \citep[12,500~K,][]{tsang91} 
and is only significant in very hot gas ($\gtrsim$~1000~K).  
CN has numerous formation routes via neutral neutral reactions: 
N~+~\ce{C2} \citep{smith04}, NH~+~C \citep{brownsword96}, and 
NO~+~C \citep{chastaing00} where NO is formed via the reaction,  
N~+~OH \citep{wakelam12}.  
Only the formation of NH (via N~+~\ce{H2}) possesses a substantial reaction 
barrier \citep[18,095~K,][]{mcelroy13}. 

For the M~dwarf and T~Tauri disks, photodissociation 
at Lyman-$\alpha$ is more significant than that by the 
FUV continuum background because around 70~--~80\% of the FUV flux 
is contained in the Lyman-$\alpha$ line \citep[see also][]{fogel11}. 
This percentage is in line with that determined towards observations of
M~dwarf and classical T~Tauri stars \citep[see, e.g.,][]{france13,france14}. 
HCN is treated as though it photodissociates via line transitions 
\citep[][see, e.g.,]{lee84} and it 
also has a non-negligible photodissociation cross section at 
1216~$\AA$~\citep[$\sigma$~$=$3~$\times$~10$^{-17}$~cm$^2$, ][]{vandishoeck06}.    

HCN is relatively more abundant in the M Dwarf disk due to the 
weaker FUV flux leading to decreased destruction via photodissociation.  
Ion-molecule reactions also contribute to 
HCN formation in the M~dwarf disk (via \ce{HCNH+}~+~\ce{e-}) 
and destruction (via \ce{HCN}~+~\ce{X+}, where 
\ce{X+} is \ce{H+}, \ce{C+}, \ce{H3+}, \ce{H3O+}, and \ce{He+}) as shown in 
Figure~\ref{figure3b}.  
These reactions all have measured rate coefficients 
\citep{huntress77,clary85,anicich93,semaniak01}.
X-ray-induced photodissociation also contributes to the destruction of 
HCN at the level of 15--~20\%.  
The ion-molecule formation route is triggered by the formation of 
\ce{CN+} and \ce{HCN+} via the reactions, N~+~\ce{CH+}, 
NH~+~\ce{C+}, and N~+~\ce{CH2+} \citep{viggiano80,prasad80}.  
\ce{HCN+} and \ce{HCNH+} are then formed via \ce{CN+}~+~\ce{H2} and 
\ce{HCN+}~+~\ce{H2} \citep{raksit84,huntress77}.  

In all cases, midplane gas-phase HCN is synthesised via 
neutral-neutral chemistry: it is not related to the desorption of 
HCN ice as is the case for traditional snow lines. 
The abundance is mediated by formation via \ce{H2} and \ce{CN} and 
destruction via collisional dissociation \citep{baulch94,tsang91}.  
This formation route has a reaction barrier of 820~K and thus 
requires warm temperatures for activation ($>$~200~K).  
These temperatures are surpassed in the midplane of each disk 
due to the inclusion of heating via viscous dissipation 
\citep[for details see][and references therein]{nomura05}.  
Viscous heating dominates over stellar heating in the midplane
within radii of $\approx$~ 0.40, 2.5, and 1.1~AU for the 
M~dwarf, T~Tauri, and Herbig~Ae disks, respectively. 
The Herbig~Ae transition radius is less than that for the T~Tauri 
disk because the former has significantly stronger stellar heating.  

In Figure~\ref{figure3a}, a snapshot of the chemistry of 
\ce{C2H2} is presented.  
In all three disks, the formation and destruction of \ce{C2H2} in the atmosphere 
is dominated by the neutral-neutral reaction 
\ce{H2}~+~\ce{C2H} \citep{laufer04} and photodissociation \citep{vandishoeck06}, respectively. 
\ce{C2H2} is also preferentially photodissociated at Lyman-$\alpha$ wavelengths 
in both cases \citep[$\sigma$~$\ge$~4~$\times$~10$^{-17}$~cm$^{2}$, ][]{vandishoeck06}, 
for similar reasons as discussed above for HCN.   
In the M~dwarf disk, \ce{C2H2} is also destroyed via reactions with \ce{C}, \ce{H3+}, 
and \ce{He+} \citep{kim75,chastaing99,laufer04}, 
as well as via X-ray-induced photodissociation (at the level of $\approx$~15~--~20\%).   
There are also formation routes via ion-molecule chemistry 
(\ce{C2H3+}~+~\ce{e-}) which contribute at the level of 
a few percent and are barrierless.  

\ce{C2H2} is significantly less abundant in the atmosphere of the 
Herbig~Ae disk than in the other two disks.  
In the absence of efficient ion-molecule pathways (triggered by the formation of 
\ce{C+} via CO~+~\ce{He+}), an important first 
step in the formation of carbon-chain molecules is the 
formation of CH via C~+~\ce{H2} which has a large reaction barrier (11,700~K).  
CH can then react barrierlessly with atomic C to give \ce{C2}.  
In the atmospheres of protoplanetary disks, the gas temperature is controlled 
by the strength of the UV field.  
Carbon-chain growth is impeded in the atmosphere of the Herbig~Ae disk because 
the increased photodissociation counteracts the temperature-activated 
gas-phase chemistry.  
In Figure~\ref{figure3a}, we split the carbon chemistry into that 
dominated by UV radiation (and a higher gas temperature) highlighted in red 
on the left-hand side, and that dominated by X-ray radiation 
(and a lower gas temperature) highlighted in blue on the right-hand side.  
The left-hand side represents the chemistry more dominant in 
Herbig~Ae disks and the right-hand side represents the chemistry more 
important in M~dwarf disks.  
The chemistry tends towards the middle of this reaction scheme in 
environments where the UV radiation is too strong for 
the survival of species other than CO, C, and \ce{C+}.   

In summary, the gas-phase chemistry depends, not only on 
the strength of the FUV radiation (which controls the gas temperature), 
but also on the adopted ionisation sources and spectra and corresponding 
rates propagated throughout the disk.  
This is especially true for the M~dwarf disk in which ion-molecule chemistry 
also plays a role in the formation and destruction of HCN and \ce{C2H2} as
shown in Figures~\ref{figure3a} and \ref{figure3b}. 
The liberation of free carbon and nitrogen is the principle underlying 
reason for the importance of X-ray chemistry. 

\subsubsection{Column densities}

\begin{figure*}
\includegraphics[width=1.0\textwidth]{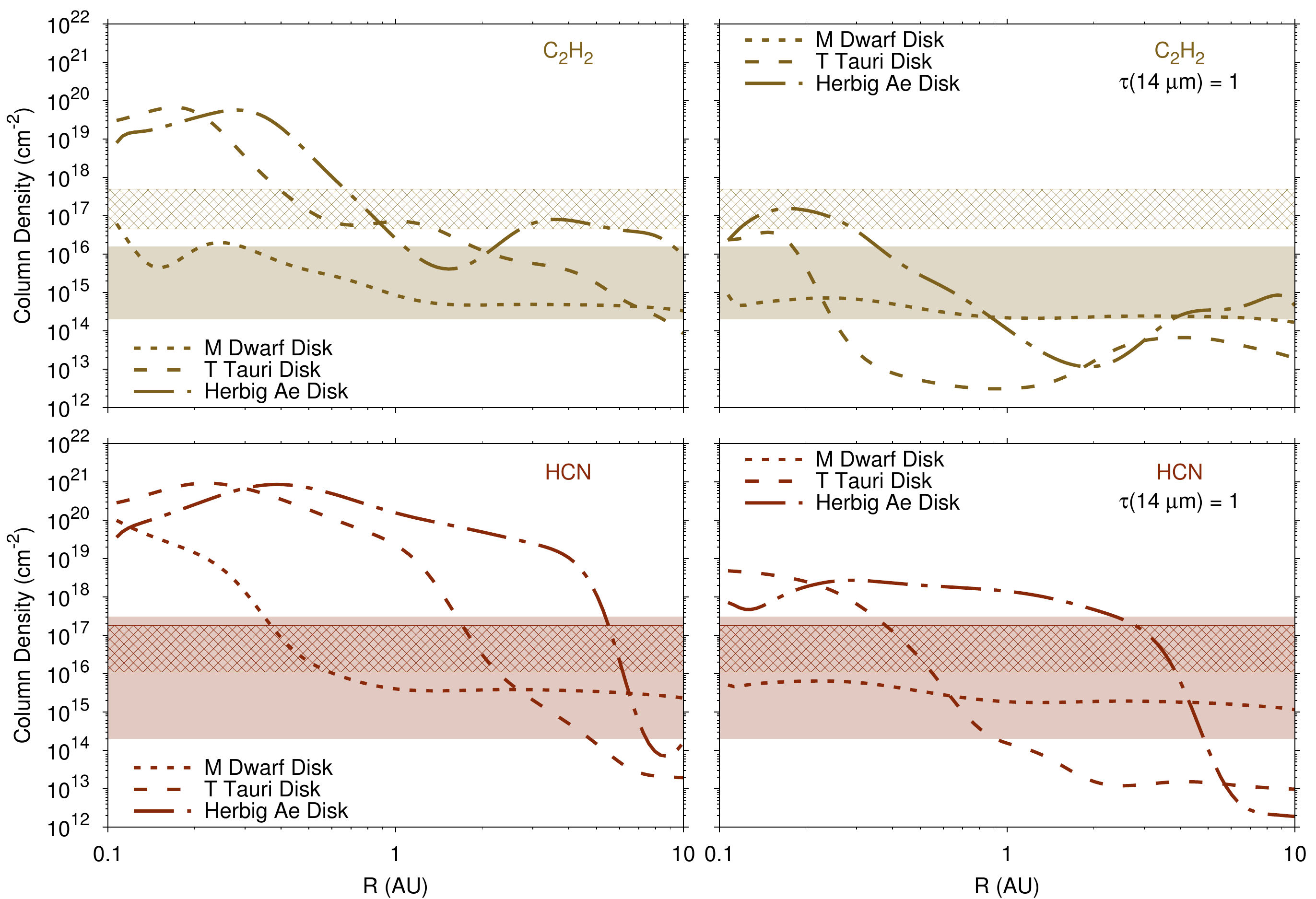}
\caption{Column density of \ce{C2H2} (top row) and \ce{HCN} (bottom row) as a function of radius 
for each disk model for the entire vertical extent of the disk (left-hand column) and 
down to the $\tau$~=~1 surface at 14~$\mu$m (right-hand column).  
The solid coloured and hatched regions indicate 
the range of observed values for the T~Tauri \citep{carr11,salyk11} 
and M~dwarf \citep{pascucci13} disks, respectively.}
\label{figure4}
\end{figure*}

A higher \ce{C2H2} and \ce{HCN} fractional abundance is seen 
in the M~dwarf disk atmosphere compared with the others; however, this 
does not necessarily translate into an observable column density, especially given 
the more tenuous nature of the M~dwarf disk and the fact that the molecules 
peak in fractional abundance higher in the disk atmosphere (where the density is also lower).  
Figure~\ref{figure4} displays the vertically integrated column densities of 
both species as a function of disk radius over the entire vertical extent
(left-hand column) and down to the $\tau$~=~1 surface at 14~$\mu$m (right-hand column).  
There is a general trend that the column density peaks at inner radii, 
then falls off sharply at a particular radius which moves outwards with increasing 
stellar spectral type. 
This fall off occurs at a smaller radius for \ce{C2H2} than for \ce{HCN}. 
This behaviour is also seen in the T Tauri model presented in \citet{agundez08}. 

For \ce{C2H2}, the T~Tauri and Herbig~Ae disks achieve a similar peak column 
density $\sim$~10$^{20}$~cm$^{-2}$ 
at radii of $\approx$~0.2 and 0.3~AU respectively.  
The column density then falls with increasing radius to 
$\lesssim$~10$^{17}$~cm$^{-2}$ beyond 0.5~AU for the T~Tauri star and 
beyond 1~AU for the Herbig~Ae star.  
The total column density for the M~dwarf disk lies orders of magnitude lower, 
reaching a peak value of $\sim$~10$^{17}$~cm$^{-2}$ at 0.1~AU and falling 
to $\lesssim$~10$^{15}$~cm$^{-2}$ beyond 1~AU.  
Comparing with the column densities calculated down to the $\tau$(14~$\mu$m)~=~1 surface, 
for the two warmer disks, the values are lower by between three and four orders of magnitude 
because the total column density is dominated by midplane \ce{C2H2}.  
In contrast, the values for the M~dwarf disk remain comparable beyond 
$\approx$~1~AU because in this case, the total column is dominated 
by atmospheric \ce{C2H2}.  
The M~dwarf disk has a larger column density in the atmosphere than the 
T~Tauri disk beyond 0.2~AU.
Although \ce{C2H2} does not appear to be abundant in the atmosphere 
of the Herbig~Ae disk (according to Figure~\ref{figure3}), 
there is still a significant column density: this is because there is 
a thin layer (only one to two grid cells wide) of relatively abundant \ce{C2H2} 
which overlaps with the $\tau$~=~1 surfaces.  
This is likely a feature of the grid resolution of our model 
and thus is a numerical artefact.  

For HCN, the T~Tauri and Herbig~Ae disks reach a peak column 
density of $\sim$~10$^{21}$~cm$^{-2}$ at similar radii to those for \ce{C2H2}.
The radial behaviour of the column density then follows the spatial extent 
of HCN in the disk midplane (see~Figure~\ref{figure3}).  
Similar to \ce{C2H2}, the peak value for the M~dwarf disk is lower 
($\sim$~10$^{20}$~cm$^{-2}$ at 0.1~AU), 
and the column density remains constant beyond 
$\approx$~0.8~AU at a few $\times$~10$^{15}$~cm$^{-2}$. 
For the corresponding values down to the $\tau$(14~$\mu$m)~=~1 surface, 
the M~dwarf disk has a higher column density than the T~Tauri and 
Herbig~Ae disks beyond 0.6~AU and 4~AU, respectively.  
Although the fractional abundance of HCN is lower in the 
Herbig~Ae disk, the molecular layer is located deeper in the disk 
atmosphere where the density is higher leading to the 
apparently large calculated column density of HCN (for $R<$~4~AU).  

\subsubsection{Comparison with observed trends}

The preferred method for comparing model results with observations 
is the simulation of the line emission; however, this is a non-trivial matter 
involving careful consideration of collisional and radiative excitation 
in the IR \citep[see, e.g.,][]{pontoppidan09,meijerink09,thi13,bruderer15}.  
Moreover, the dust properties and size distribution (including, e.g., grain growth) 
become crucial for the calculation of the emitted spectrum \citep[see, e.g.,][]{meijerink09}.   
This is beyond the scope of the work presented here which is focussed on the 
chemistry; however, this is planned future work.  
Here, we compare the calculated column densities (and ratios) with those 
derived from observations to investigate if the chemical models 
are at least able to reproduce the observed abundances and related trends. 

\begin{figure}
\includegraphics[width=0.5\textwidth]{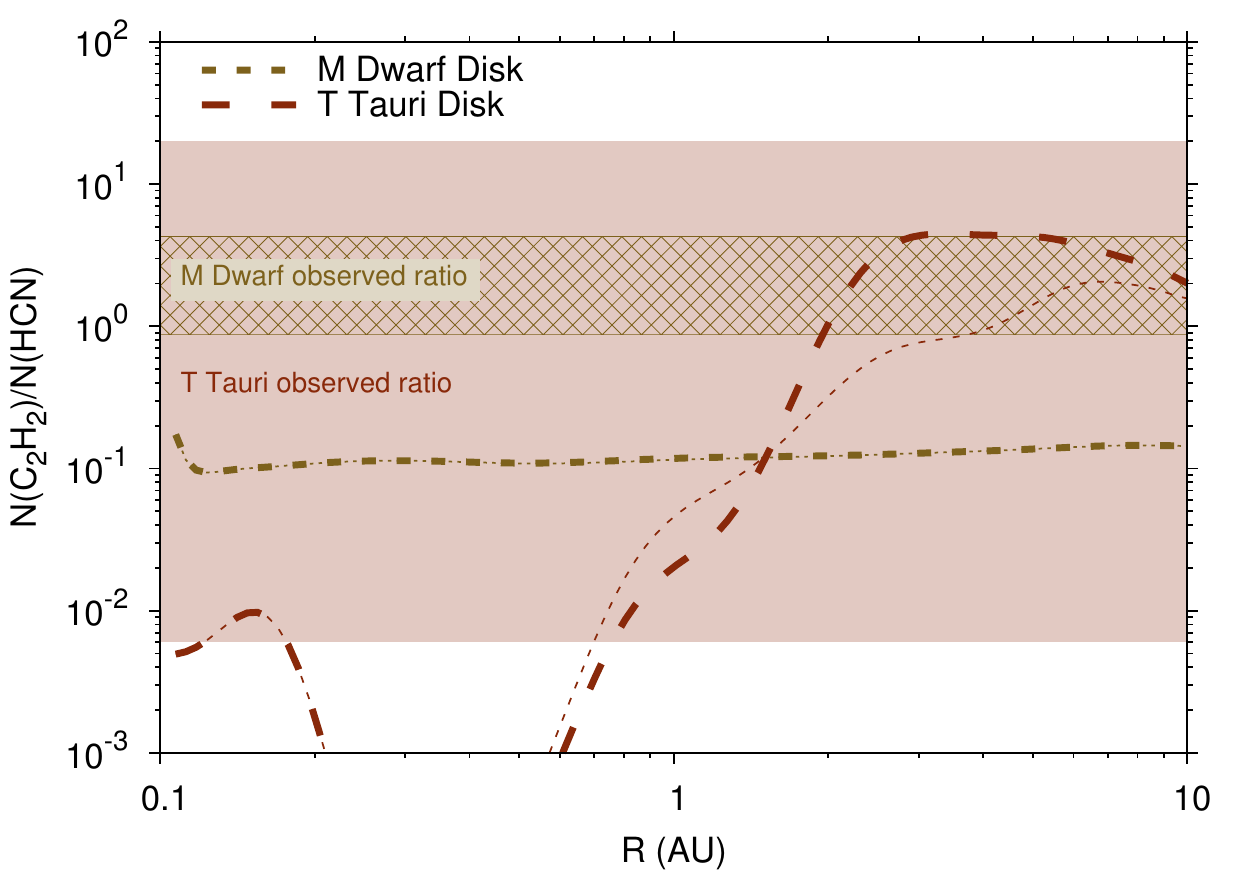}
\caption{Ratio of \ce{C2H2} and \ce{HCN} column densities in the disk atmosphere 
as a function of radius for the M~dwarf disk (gold dotted lines) and T~Tauri disk 
(red dashed lines).  
The solid coloured and hatched regions correspond 
to the range of observed values for T~Tauri \citep{carr11,salyk11} and 
M~dwarf \citep{pascucci13} disks, respectively. 
The thick and fine lines represent the ratios down to the $\tau$~=~1 surface 
at 14 and 3 $\mu$m.}
\label{figure5}
\end{figure}

The \ce{C2H2}/\ce{HCN} column density ratio derived for 
disks around cool stars ranges from 0.87 to 4.3 \citep[see Table 6 in][]{pascucci13}.
In contrast, the \ce{C2H2}/\ce{HCN} column density ratios derived for 
T~Tauri disks lies between 0.006 and 0.43 \citep[see Table 4 in][]{carr11}.  
\citet{salyk11} derive a range from 0.13 to 20 for their sample of 
T~Tauri stars. 
In Figure~\ref{figure5}, the model ratios in the disk atmosphere
are plotted as a function of radius for the T~Tauri disk (red dashed lines) and 
M~dwarf disk (gold dotted lines), overlaid with the observed range.  
The fine lines are the equivalent ratios for the column 
densities integrated down to the $\tau$~=~1 surface at 3~$\mu$m.  
The ratio is flat for the M~dwarf disk ($\sim$~0.1) 
and lies roughly one order of magnitude lower than the observed range
(gold hatched zone).  
On the other hand, the ratio for the T~Tauri disk increases with 
radius reaching a peak of $\approx$~4.  
Also, the T~Tauri ratios beyond 0.7~AU lie well within the wide range of 
observed values (red shaded zone).  
 
For \ce{HCN}, \citet{carr11} derive best-fit column densities for their sample 
of T~Tauri disks between 1.8 and 6.5~$\times$~10$^{16}$~cm$^{-2}$; however, 
assuming extremes in the optical depth of the HCN emission expands this to 
between 0.2 and 31~$\times$~10$^{16}$~cm$^{-2}$.  
For \ce{C2H2}, the derived range is 0.02 to $1.6 \times 10^{16}$~cm$^{-2}$.
\citet{salyk11} derive column densities between 0.05 and 0.63~$\times$~10$^{16}$~cm$^{-2}$ 
and between 0.05 and 1.0~$\times$~10$^{16}$~cm$^{-2}$ for HCN and \ce{C2H2} respectively.  
The observed column density ranges are indicated by the solid colour 
regions in Figure~\ref{figure4}.  
The T~Tauri model results give good agreement with both the absolute column
densities and the ratio of \ce{C2H2}/\ce{HCN} in the vicinity of the 
radius within which the observed emission originates.  

In the inner region of the disk ($\lesssim$~2~AU), the models qualitatively 
reproduce the trend that the \ce{C2H2}/\ce{HCN} ratio is higher in 
M~dwarf disks than in T~Tauri disks; however, the ratio predicted in the 
model lies lower than that observed.  
Hence, the M~dwarf model is either overpredicting the HCN column density
or underpredicting the \ce{C2H2} column density.   
\citet{pascucci13} derive column densities ranging from 
4.5~to~50.1~$\times$~10$^{16}$~cm$^{-2}$ for \ce{C2H2} and from 
1.1~to~18.1~$\times$~10$^{16}$~cm$^{-2}$ for \ce{HCN}.  
The range of observed column densities are indicted by the gold hatched 
regions in Figure~\ref{figure4}. 
The model values for \ce{C2H2} are more than one order of magnitude lower whereas
the values for \ce{HCN} lie within a factor of a few of the observed column densities.    
However, the absolute values will depend to a degree on the adopted model 
parameters, such as disk surface density.  
It is also possible that the excitation mechanisms for \ce{C2H2} and \ce{HCN} 
differ in cool stars relative to T~Tauri stars which means that the relative line emission 
no longer traces the relative abundances nor column densities.
The excitation of HCN in protoplanetary disks was investigated in detail 
by \citet{bruderer15} who conclude that HCN abundances derived assuming LTE 
should differ by no more than a factor of three from those derived assuming non-LTE.   
They also conclude that \ce{C2H2} will behave similarly to HCN because of the 
presence of similar infrared bands through which the excitation can be pumped. 
We plan to test this via simulations of the molecular emission in future work. 

\citet{pascucci13} speculate that the self-shielding of \ce{N2} may play a role 
in determining the \ce{C2H2}/\ce{HCN} ratio: this can potentially lock up a 
greater fraction of atomic nitrogen thereby impeding the production of 
other nitrogen-bearing species, such as HCN.
Here, we have also shown that X-ray-induced chemistry may play 
an important role in both the production and destruction of molecules 
in the atmosphere of disks around cool stars.  
We further investigate the role of \ce{N2} self-shielding and X-ray-induced chemistry 
in Sects.~\ref{N2shielding} and \ref{Xraychemistry}, respectively.  

\subsection{\ce{OH} and \ce{H2O}}
\label{OHandH2O}

\subsubsection{Chemical structure}

\begin{figure*}
\includegraphics[width=1.0\textwidth]{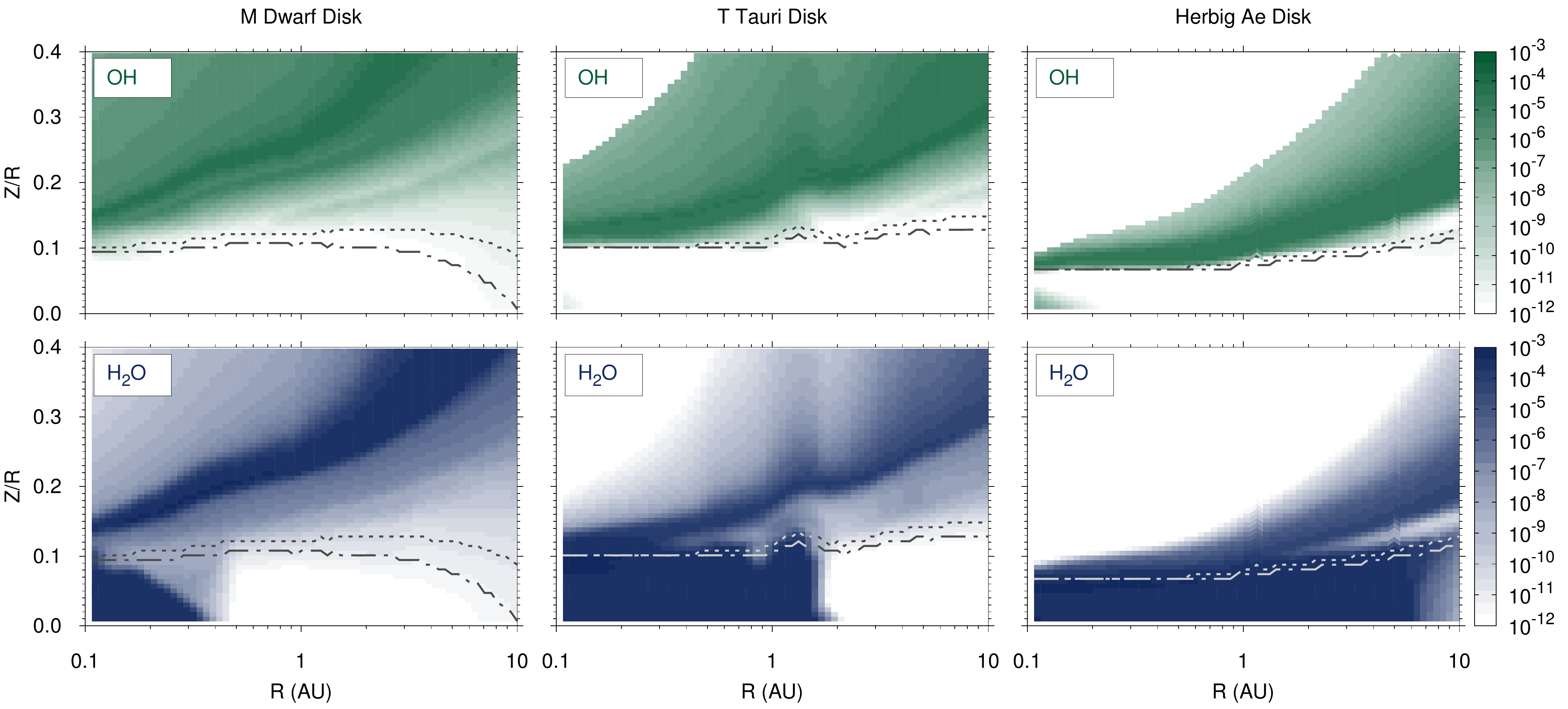}
\caption{Fractional abundance relative to total gas number density of 
\ce{OH} (top row) and \ce{H2O} (bottom row) for the M~dwarf disk (left-hand column), 
T~Tauri disk (middle column), and Herbig~Ae disk (right-hand column). 
The dotted and dot-dashed lines indicate the dust column density (integrated from the surface downwards) 
at which $\tau$~$\approx$~1 at 3~$\mu$m and 14~$\mu$m, respectively.}
\label{figure6}
\end{figure*}

Another interesting trend seen in the IR data is the lack of 
water detections in disks around hotter stars.  
In Figure~\ref{figure6} we display the fractional abundance of OH 
(top row) and \ce{H2O} (bottom row) as a function of radius ($R$) and 
height divided by the radius ($Z$/$R$).  
OH is more extended and resides in a layer slightly higher in the disk than \ce{HCN}, \ce{C2H2}, 
and \ce{H2O}, in line with the hypothesis that the OH chemistry is driven by photodissociation.  
The peak fractional abundance of OH is 8, 5, and 4~$\times$~10$^{-5}$
for the M~dwarf, T~Tauri, and Herbig~Ae disks, respectively. 
The corresponding values for \ce{H2O} are 5, 5, and 4~$\times$~10$^{-4}$, 
again showing a general (albeit very shallow) decline in molecular complexity 
with increasing spectral type.  
The fractional abundance of OH is negligible ($<$~10$^{-11}$) in the midplane 
of all three disks.  
The extent over which \ce{H2O} is abundant 
in the disk midplane is related to the thermal desorption of 
\ce{H2O} ice \citep[we assume a binding energy of 5570~K,][]{fraser01}.  
The \ce{H2O} `snow line' shifts outwards in radius with increasing 
stellar spectral type with positions at 0.35, 1.5, and 6.1~AU for the M~dwarf, T~Tauri, and 
Herbig~Ae disk, respectively.  
Beyond these radii, \ce{H2O} is frozen out on dust grains in the midplane.

\subsubsection{Chemistry of OH and \ce{H2O}}

\begin{figure}
\includegraphics[width=0.5\textwidth]{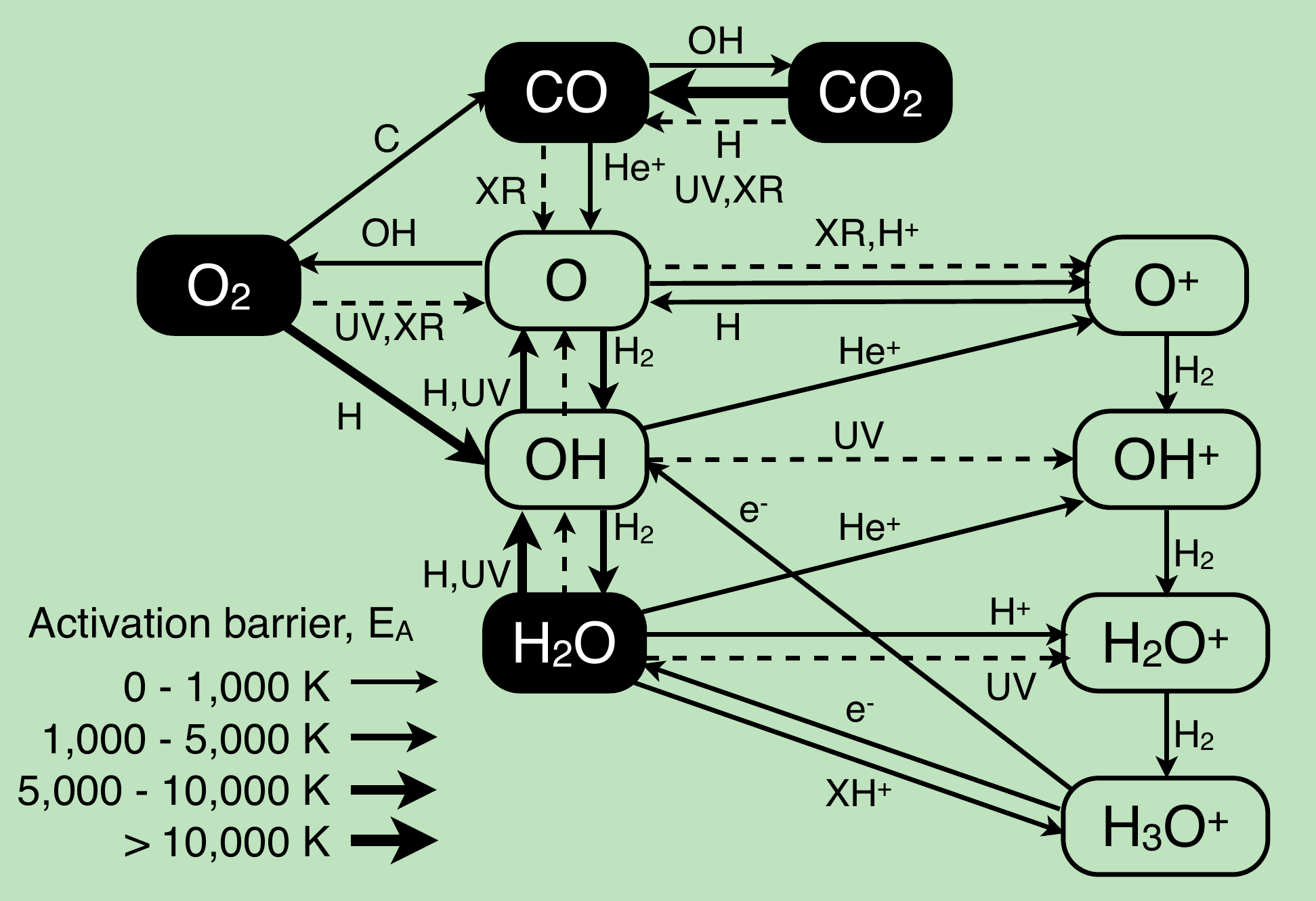}
\caption{Dominant reactions controlling the abundance of  
\ce{OH} and \ce{H2O} and related species at the position of peak fractional 
abundance in the disk atmospheres at a radius of 1~AU. 
The solid lines indicate gas-phase two-body reactions and the dotted lines 
indicate UV or X-ray-induced reactions.  
The width of the arrows for the gas-phase reactions 
shows the magnitude of the activation barrier.}
\label{figure6a}
\end{figure}

Figure~\ref{figure6a} shows a snapshot of the dominant reactions 
contributing to the formation and destruction of OH and 
\ce{H2O} (and related species) at the position of peak abundance 
in the disk atmosphere at a radius of 1~AU. 
Gas-phase \ce{H2O} is predominantly produced in the disk atmosphere 
via the neutral-neutral reaction, \ce{H2}~+~\ce{OH} 
\citep[using the rate coefficient from][]{oldenborg92}.  
In the cooler M~dwarf disk, thermal desorption of water ice dominates at 
small radii ($\approx$~1~AU).  
Destruction is typically via photodissociation: for the M~dwarf and 
T~Tauri disks, this is again dominated by Lyman-$\alpha$ photons \citep{vandishoeck06}.  
In the M~dwarf disk, there are additional destruction routes via 
ion-molecule reactions with \ce{H+} and \ce{H3+} \citep{kim74,smith92}
and X-ray-induced photodissociation contributes at the level of 12~--~15\%  
(similar to that found for \ce{C2H2} and \ce{HCN}). 
Similarly, the gas-phase abundance of OH is mediated 
by production via the photodissociation of \ce{H2O} and 
the reaction between \ce{H2} and \ce{O} \citep{baulch92}, with destruction via 
photodissociation \citep{vandishoeck06} 
and reactions with \ce{H2} and \ce{H} \citep{tsang86,oldenborg92}.  
As in the case for \ce{H2O}, dissociation by Lyman-$\alpha$ 
dominates over that by the background FUV continuum.  
This can be summarised in the following, rather succinct, reaction scheme,  
\begin{equation}
\ce{O} \xrightleftharpoons[\ce{H},\,h\nu]{\ce{H2}} \ce{OH} 
\xrightleftharpoons[\ce{H},\,h\nu]{\ce{H2}} \ce{H2O} 
\xrightleftharpoons[\mathrm{desorption}]{\mathrm{freezeout}} \ce{H2O}_{\mbox{ice}}  
\label{waterformation}
\end{equation}
\citep[see also][and Figure~\ref{figure6a}]{bethell09,woitke09}.
Thus, the water chemistry in the disk atmosphere is rather simple with the 
relative abundances of \ce{O}, \ce{OH}, and \ce{H2O} controlled 
primarily by the relative abundances of H and \ce{H2}, the gas temperature 
(which needs to be sufficiently high to activate the neutral-neutral chemistry)
and the strength of the FUV radiation field (necessary for photodissociation).  
We find that reactions with vibrationally excited or `hot' \ce{H2} 
(see Sect.~\ref{gasphasenetwork}) increase in importance towards the upper atmosphere.  
However, these reactions do not contribute significantly at the position of 
peak abundance and deeper and hence, do not influence the total column density of 
neither OH nor \ce{H2O}.
For a detailed review of water chemistry see \cite{vandishoeck13}. 

\subsubsection{Column densities}

\begin{figure*}
\includegraphics[width=1.0\textwidth]{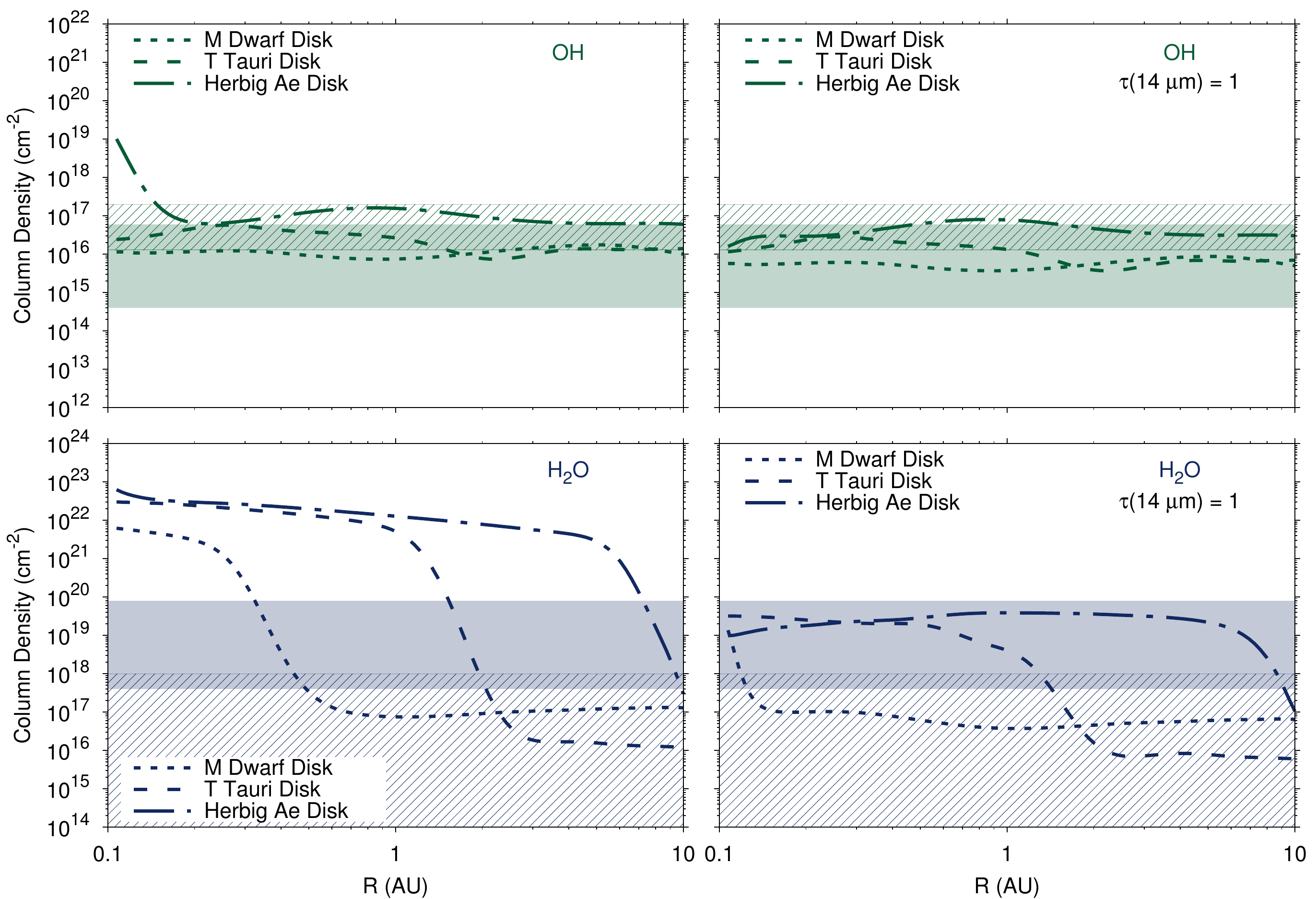}
\caption{Column density of \ce{OH} (top row) and \ce{H2O} (bottom row) as a function of radius 
for each disk model for the entire vertical extent of the disk (left-hand column) and 
down to the $\tau$~=~1 surface at 14~$\mu$m (right-hand column). 
The solid coloured and striped regions indicate 
the range of observed values for the T~Tauri \citep{carr11,salyk11} and Herbig~Ae \citep{fedele11,salyk11} 
disks, respectively.}
\label{figure7}
\end{figure*}

Figure~\ref{figure7} shows the vertically integrated column densities of OH  
(top row) and \ce{H2O} (bottom row) as a function of radius 
over the entire vertical extent of the disk (left-hand columns) 
and down to the $\tau$~=~1 surface at 14~$\mu$m (right-hand column).
The column densities show less structure compared 
with those for \ce{C2H2} and \ce{HCN}.  
Because the majority of \ce{OH} is in the disk atmosphere, the 
column densities for both cases are similar and range between 
a few times 10$^{15}$ to 10$^{17}$~cm$^{-2}$.  
The OH column density also does not vary greatly with radius. 
The Herbig~Ae disk generally has the largest column 
density and the M~dwarf has the lowest, although the values 
for the T~Tauri and M~dwarf disks are similar beyond $\approx$~1.6~AU. 
All three disks show very high total column densities of \ce{H2O} 
($\gtrsim$~10$^{21}$~cm$^{-2}$) within each respective snow line. 
The \ce{H2O} column densities also show a similar trend to 
that for \ce{C2H2} and \ce{HCN}.  
The column densities start high in the inner disk then fall off sharply 
at a distinct radius for each disk (which moves outwards with increasing spectral type). 
This radius generally lies beyond that for \ce{C2H2} and \ce{HCN}. 
Similar to the case for \ce{HCN}, gas-phase \ce{H2O} in the disk 
midplane is likely obscured by dust at near- to mid-IR wavelengths.  
The bottom right-hand panel shows that the column density of `visible' 
\ce{H2O} is on the order of a few times 10$^{19}$~cm$^{-2}$ within 
the snow line for the T~Tauri and Herbig~Ae disks. 
The column density drops to $\sim$~10$^{16}$~cm$^{-2}$ beyond 
1.5~AU for the T~Tauri disk.  
The value for the M~dwarf model remains constant 
over most of the radial extent of the disk at $\sim$~10$^{17}$~cm$^{-2}$.  

\subsubsection{Comparison with observed trends}

Here, we compare the column densities calculated by the models 
with those derived from the observations to test whether the chemical calculations 
are able to reproduce the observed abundances and trends.
For their sample of classical T~Tauri stars in which both OH and \ce{H2O} were detected, 
\citet{salyk11} derive OH column densities from 
0.04 to 6~$\times$~10$^{16}$~cm$^{-2}$.  
The \ce{H2O} column densities
range from 4~$\times$~10$^{17}$ to 1.8~$\times$~10$^{18}$~cm$^{-2}$.  
\citet{carr11} derive a wider range of values 
(4~$\times$~10$^{17}$ to 7.9~$\times$~10$^{20}$~cm$^{-2}$).  
The observed ranges are indicated by the solid colour shaded 
regions in Figure~\ref{figure7}. 
Despite the model T~Tauri disk not being representative of 
any particular source, there is significant overlap between the 
calculated column densities in the disk atmosphere and observed 
values for both species.  

For the Herbig~Ae disks, only OH has been robustly detected at 
IR wavelengths \citep[][with one exception discussed below]{fedele12}.  
The range of observed column densities (derived from VLT/CRIRES data) 
lie between 1.3 and 20~$\times$~10$^{16}$~cm$^{-2}$ and the model values lie within a 
factor of a few of this range 
\citep[indicated by the striped region in Figure~\ref{figure7},][]{fedele11}. 
The maximum model \ce{H2O} column density in the disk atmosphere is more 
than an order of magnitude larger than the maximum upper limits derived 
by both \citet{salyk11} and \citet{fedele11}.  
OH and \ce{H2O} have been detected in the disk around HD~163296 
in {\em Herschel}/PACS data at far-IR wavelengths 
with a relative column density of OH/\ce{H2O}~$\approx$~1 and an 
excitation temperature ranging from 200~--~500~K \citep{fedele12}. 
However, this emission originates further out in the disk 
atmosphere (15~-~20~AU) than that expected at shorter wavelengths 
\citep[see also the analysis in][]{fedele13}.  
  
\begin{figure}[!]
\includegraphics[width=0.5\textwidth]{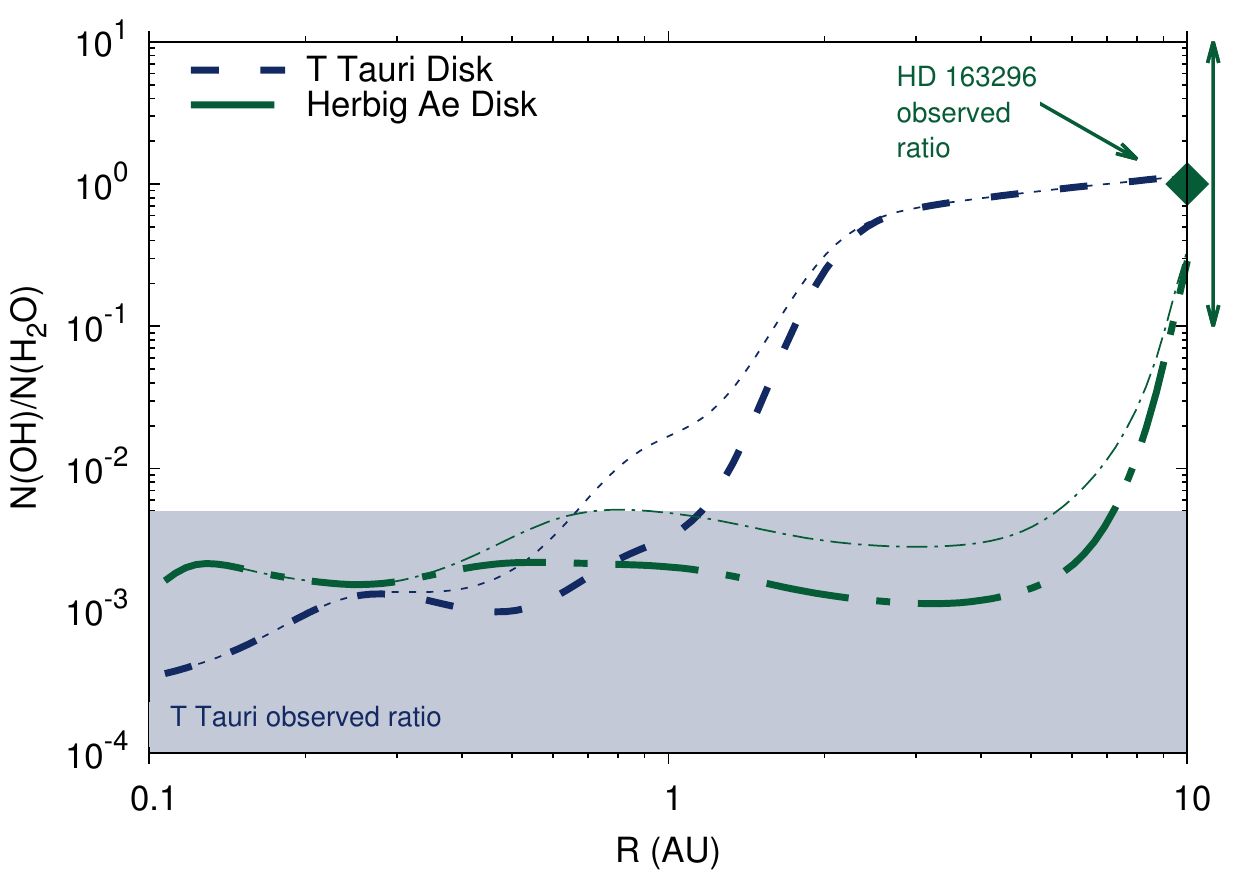}
\caption{Ratio of \ce{OH} and \ce{H2O} column densities in the disk atmosphere 
as a function of radius for the T~Tauri disk (blue dashed lines) and Herbig~Ae 
disk (green dot-dashed lines).  
The solid coloured region corresponds to the range of observed values for 
T~Tauri disks \citep{carr11,salyk11} 
and the green diamond and arrow indicates the ratio range observed 
for HD~163296 \citep{fedele12}. 
The thick and fine lines represent the ratios down to the $\tau$~=~1 surface 
at 14 and 3 $\mu$m.}
\label{figure8}
\end{figure}

The Herbig~Ae model appears to overproduce gas-phase water in the inner 
disk atmosphere relative to observations; however, the derived upper 
limits are dependent on the chosen emitting radius and gas temperature.      
Figure~\ref{figure8} shows the ratio of OH to \ce{H2O} column density as 
a function of disk radius for the T~Tauri model (blue dashed lines) and 
the Herbig~Ae model (green dot-dashed lines).  
The fine lines represent the equivalent ratios for the columns 
integrated down to the $\tau$(3~$\mu$m)~=~1 surface. 
The model T~Tauri ratios lie within the observed range 
for radii less than $\approx$~1~AU.  
The ratio for the Herbig~Ae model has a value $\sim10^{-3}-10^{-2}$ 
within $\approx8$~AU, beyond which it increases and quickly tends to 
$\approx$~1 at larger radii. 
The analysis of the non-detections of OH and \ce{H2O} in \citet{salyk11} 
suggest a lower limit to the \ce{OH}/\ce{H2O} ratio of a few $\times10^{-3}$ 
assuming an emitting radius which ranges from a few AU to several 10's of AU.  
\citet{fedele11} suggest a ratio of $>1-25$ within an emitting 
radius as far out as 30~AU based on near-IR data.
The model values are on the cusp of the \citet{salyk11} lower limit in the 
very inner region 
and tend towards the \citet{fedele11} lower limit at 10~AU.  

The lack of hot water emission from the innermost regions of disks around 
Herbig~Ae stars remains a puzzle, 
especially given the hypothesis that water can shield itself from 
photodissociation by the stellar radiation field 
at column densities, $N(\ce{H2O})\gtrsim 2 \times 10^{17}$~cm$^{-2}$  
\citep[assuming a photodissociation cross section, $\sigma_\ce{H2O}\approx 5 \times 10^{-18}$~cm$^{2}$;][]{bethell09,du14,adamkovics14}.  
The column densities predicted here for the Herbig~Ae disk are in line with those by other 
work \citep[see, e.g.,][]{woitke09}.
One possible explanation is that turbulent mixing within the planet-forming region can 
help sequester water in the midplane where, if the temperature is sufficiently low, 
it can become trapped as ice on dust grains.  
This would require that as the dust grain grow, either via ice mantle growth or 
via coagulation, they become decoupled from the gas and remain in the 
shielded (and cold) midplane \citep[see, e.g., ][]{stevenson88}.  
There are two caveats to this theory: the first is that the temperature of the midplane 
would need to be below $\approx$~150~K within $\approx$~10~AU.  
The midplane temperature in our Herbig~Ae model is generally too high for 
water to reside as ice on dust grains except beyond $\approx$~6~AU (see Figure~\ref{figure2}); 
however, the exact temperature profile of the disk is sensitive to numerous factors including, 
for example, the adopted disk surface density, dust-grain size distribution,  
and degree of flaring (see later).  
It also remains to be demonstrated whether large-scale radial mixing between the warm inner midplane 
and cold outer midplane is a viable mechanism in disks around hotter stars.  
The second caveat is that one would expect the abundance of OH to also be affected 
by vertical mixing in the atmosphere since OH and \ce{H2O} are chemically coupled.  
For a given radiation field, the rate coefficients for the primary routes 
to formation and destruction are known; hence, there may be additional destruction 
routes for gas-phase water, not yet included in the networks. 
One potential explanation which remains to be explored in protoplanetary disks 
is the effect of rotationally excited OH which can be produced via photodissociation 
of \ce{H2O} at Lyman-$\alpha$ wavelengths \citep[see, e.g.,][]{fillion01}.  
The rate coefficient for the OH$(v,j)$ + H reaction can be enhanced 
by several orders of magnitude relative to that for ground state OH 
\citep{li13b}.  
These reactions may be important for shifting the ratio of O/OH/\ce{H2O} 
in the disk atmosphere (see Equation~\ref{waterformation}).  

\citet{pontoppidan10} also extensively discuss several hypotheses for the lack of hot \ce{H2O} 
in Herbig~Ae disks, including (i) an intrinsically lower abundance by an, as yet, 
unknown physical or chemical mechanism; (ii) veiling of the molecular features by 
the strong mid-IR background (if the water line luminosity is a weaker function of 
stellar spectral type than the continuum); and (iii) a well-mixed disk atmosphere 
with the canonical gas-to-dust mass ratio of 100.  
The results presented here show that for a well-mixed atmosphere, the observable column 
density of water vapour remains high.  
However, a second outcome of grain growth in protoplanetary disks 
(in addition to that discussed above) is the 
greater penetration of FUV radiation which can push the molecular layer 
deeper into the disk atmosphere \citep[see, e.g., ][]{aikawa06}; 
hence, a higher fraction of the gas-phase water may be `hidden' from view.  
This remains to be confirmed specifically for the inner regions of disks.  
Another factor to consider, and mentioned previously, is the disk gas and dust structure.  
Disks around Herbig Ae/Be stars have been classified into 
groups based on the shapes of their SED at mid-IR wavelengths \citep{meeus01}.  
Group I disks are postulated to have a flared structure which allows the disk to capture 
more FUV photons which increases the gas and dust temperature throughout the disk, 
whereas group II disks are `flatter', capture less FUV and are thus much colder.  
Dust grain growth and settling has been postulated as the reason behind the 
apparent dichotomy of Herbig Ae/Be disks \citep[see, e.g.,][]{dullemond04} and 
it has also been suggested that group I disks may be transitional in nature, i.e., 
they have evidence (usually confirmed by spatially resolved imaging) of a 
significant gap in the inner disk \citep[see, e.g.,][]{grady15}.
The disk model we have used assumes a flared disk in hydrostatic equilibrium 
without a gap: whether significant grain growth and the presence of an inner 
gap affects the abundance and distribution of gas-phase water remains to be explored. 

\section{Discussion}
\label{discussion}

\subsection{On the importance of \ce{N2} self shielding}
\label{N2shielding}

To quantify the importance of \ce{N2} shielding in the disk atmosphere, 
in Figure~\ref{figure9},  
we show the fractional 
abundance of \ce{N}, \ce{N2}, and \ce{HCN} as a function of $Z/R$ at 
$R$~=~1~AU.  
An equivalent figure for $R$~=~10~AU is shown in the Appendix (Figure~\ref{figure10}). 
Results from three different models are presented: (i) the fiducial model 
(black dashed lines, with \ce{N2} shielding and X-rays included); 
(ii) a model without \ce{N2} shielding (purple lines); and 
(iii) a model without X-ray-induced chemistry (green lines).  
The inclusion of shielding has an affect on the relative abundances 
of \ce{N} and \ce{N2} in a narrow region of the disk only at both 1 and 
10~AU for all disks.  The ratio of N/\ce{N2} is generally 
more affected as the central stellar effective temperature increases.    
Although the fractional abundance of atomic nitrogen can 
vary by more than one order of magnitude, this translates to a 
difference on the order of a factor of a few only for the HCN abundance.  
The results demonstrate that \ce{N2} shielding alone is not able to 
account for the change in \ce{C2H2}/\ce{HCN} column densities and 
line flux ratios seen from M~dwarf to T~Tauri stars.    

The above conclusion holds for models in which the dust and gas 
are assumed to be well mixed.  
If a significant fraction of the dust has 
grown and settled to the midplane \citep[see, e.g.,][]{dominik05}, 
this can lead to a relatively dust-poor disk atmosphere and allow greater 
penetration of FUV radiation.  
In that case, the importance of molecular (or self) shielding increases relative 
to dust shielding \citep[][]{visser09,li13a}. 
Whether the same conclusion holds for M~dwarf disks with advanced 
grain growth and settling remains to be confirmed.

\subsection{On the importance of X-ray-induced chemistry}
\label{Xraychemistry}

\begin{figure*}
\includegraphics[width=1.0\textwidth]{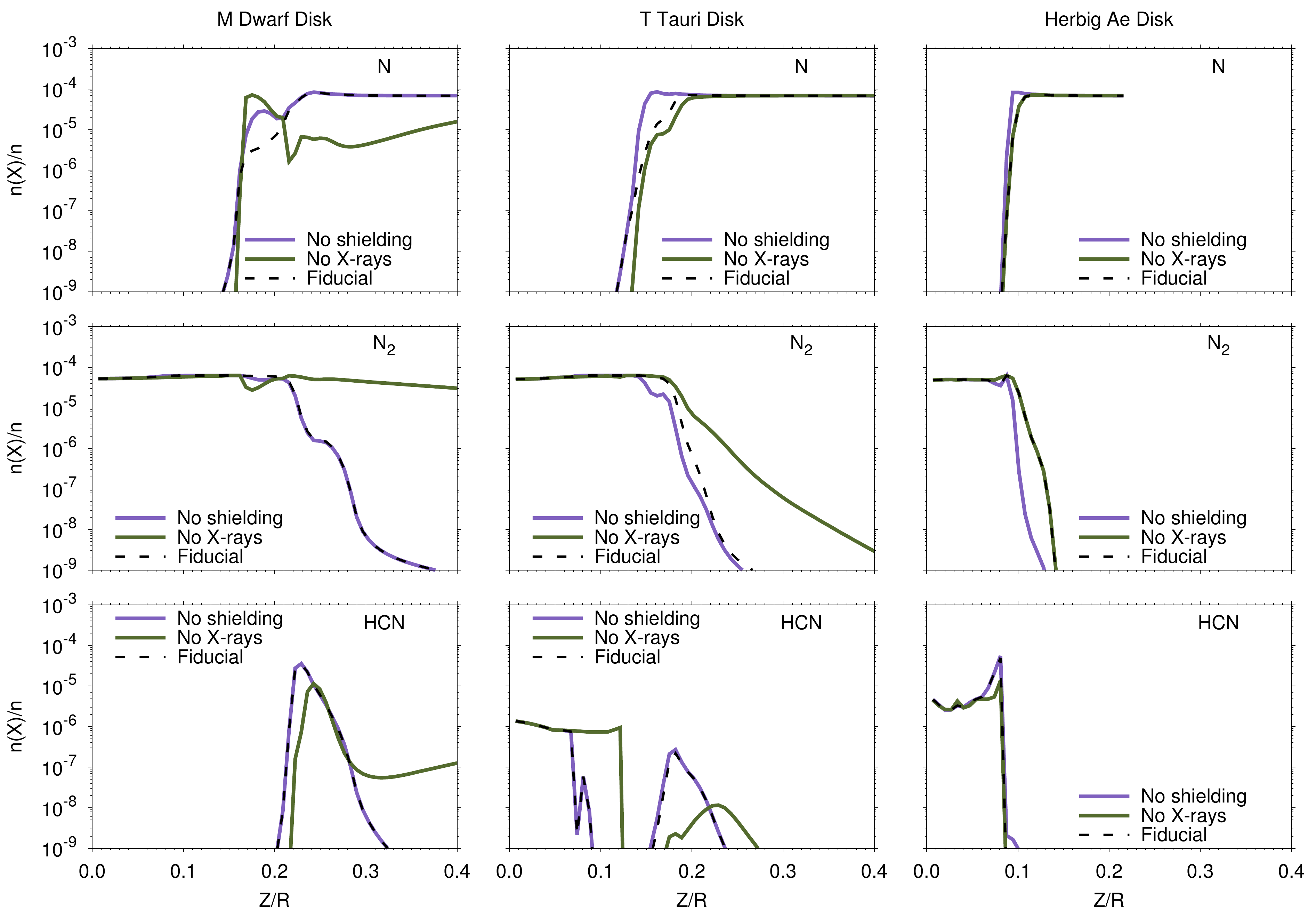}
\caption{Fractional abundance with respect to gas number density of \ce{N} (top row), 
\ce{N2} (middle row), and \ce{HCN} (bottom row) as a function 
of $Z/R$ at $R$~=~1~AU for each disk model.  
The black dashed lines, purple solid lines, and green solid lines represent results from the fiducial 
model (including \ce{N2} shielding and X-rays), the model with \ce{N2} switched off, and 
the model with X-rays switched off, respectively.}
\label{figure9}
\end{figure*}

In Figure~\ref{figure9} (and Figure~\ref{figure10})
models with and without X-ray-induced chemistry are also shown 
(black dashed lines versus green solid lines).  
X-ray-induced chemistry is significantly more important for the 
M~dwarf and T~Tauri disks than for the Herbig~Ae disk.  
In both cases and at both radii, the abundance of \ce{N2} is significantly increased 
in the disk atmosphere when X-ray chemistry is neglected 
which leads to a decrease in the ratio of N/\ce{N2}.  
The abundance of HCN is also significantly perturbed 
by the exclusion of X-ray chemistry.  
The abundance of HCN is increased in the disk atmosphere despite the reduction 
in available free nitrogen indicating that X-rays are
important for HCN destruction even at heights where 
the FUV field is strong.  
The higher penetration depth of X-rays versus 
FUV photons allows an increase in N/\ce{N2} deeper into each disk 
and hence leads to an increase in HCN in the molecular layer 
\citep[see also][]{aikawa99}. 
For the T~Tauri disk at 1~AU, X-rays also help to destroy HCN 
deeper down towards the disk midplane.    

\begin{figure*}
\centering
\includegraphics[width=0.8\textwidth]{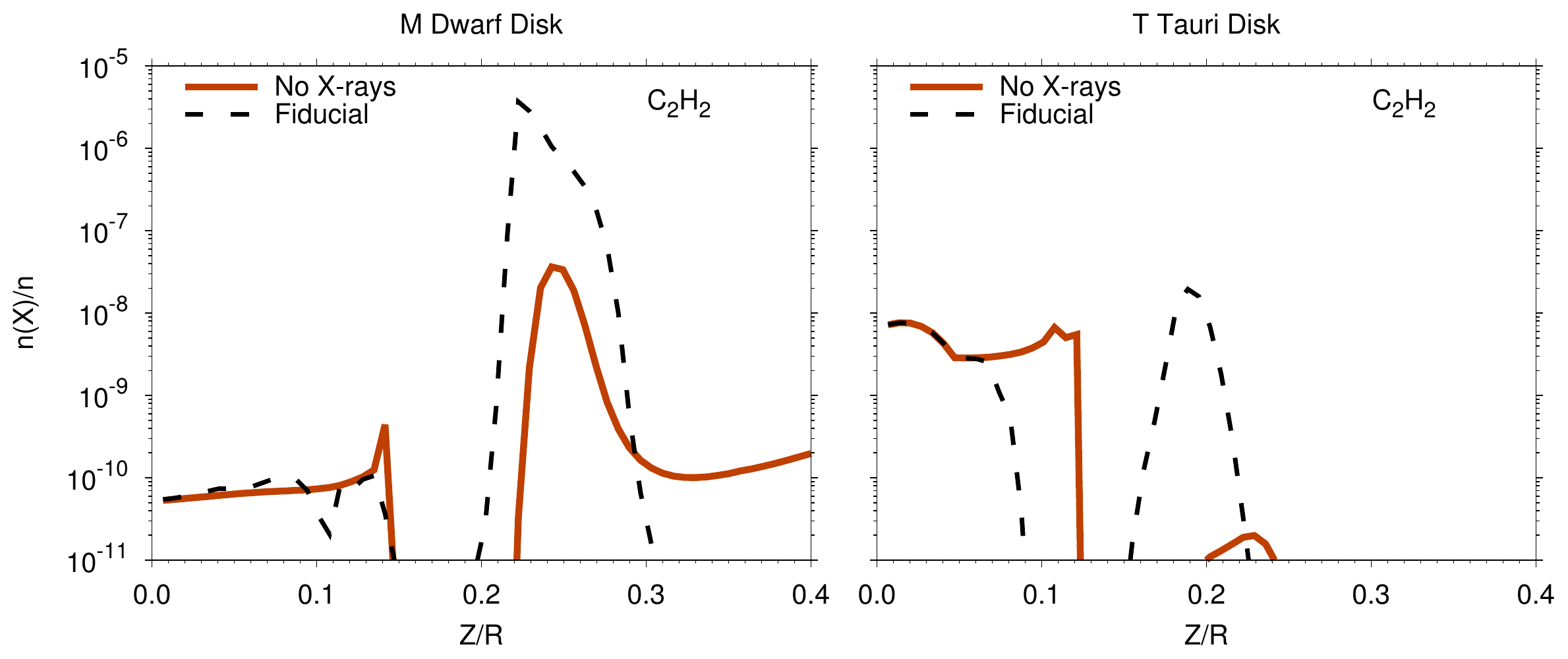}
\caption{Fractional abundance with respect to gas number density of \ce{C2H2} as a function 
of $Z/R$ at $R$~=~1~AU for the M~dwarf and T~Tauri disk models.  
The black dashed lines and orange solid lines represent results from the fiducial 
model (including \ce{N2} shielding and X-rays) and 
the model with X-rays switched off, respectively.}
\label{figure11}
\end{figure*}

The results show that the inclusion or exclusion of X-ray-induced 
chemistry can have a profound affect on the position and the value of peak 
fractional abundance of molecules such as HCN.  
To investigate whether carbon-bearing species are 
similarly affected, in Figure~\ref{figure11}, the fractional abundance of \ce{C2H2} 
is shown as a function of $Z/R$ at a radius of 1~AU for both 
the M~dwarf and T~Tauri disks with and without X-ray-induced chemistry 
(black dashed lines and orange solid lines, respectively).  
The results show that X-rays are important in the disk molecular layer 
for releasing free carbon into the gas phase for incorporation into 
species such as \ce{C2H2}.  
The inclusion of X-rays in the M~dwarf disk increases the peak abundance of 
\ce{C2H2} by more than two orders of magnitude 
and increases the extent over which \ce{C2H2} is relatively abundant 
$\gtrsim~10^{-9}$ with respect to gas number density. 
The results for the T~Tauri disk are even more extreme with 
a three orders of magnitude increase in the peak fractional abundance 
from $\approx~10^{-11}$ to $\gtrsim~10^{-8}$ when X-ray-induced chemistry 
is included.   

As discussed in Sect.~\ref{C2H2andHCN}, the efficacy of the X-ray-induced 
chemistry is because of the generation of \ce{He+} 
which in turn reacts with those molecules robust to photodissociation, CO and \ce{N2}.  
This creates a steady supply of free and reactive atomic atoms and ions  
for incorporation into other molecules via more traditional 
ion-molecule chemistry \citep[see, e.g.,][]{herbst95}.  
\ce{C2H2} is more reliant on efficient ion-molecule chemistry 
for its formation than HCN because neutral-neutral pathways to carbon-chain 
growth have significant activation barriers (see Figure~\ref{figure3a}).  
The barriers en route to HCN, in comparison, 
are lower (as discussed in Sect.~\ref{C2H2andHCN}).  
Thus, switching off X-ray chemistry has a larger effect on the magnitude of the 
peak abundance reached by \ce{C2H2} in the atmosphere as both ion-molecule and 
neutral-neutral pathways are inhibited in the cooler disks.

\subsection{On the influence of initial nitrogen reservoirs}
\label{initialnitrogen}

The results have so far suggested that X-ray-induced chemistry is crucial
for efficient molecular synthesis in the disk atmosphere in the 
planet-forming regions of protoplanetary disks around cool stars.  
An additional scenario to consider is the effect of the assumed initial abundances 
at the beginning of the calculation.  
In the results presented thus far, a set of initial abundances from 
the output of a dark cloud model were used (see Table~\ref{table2}).  
The calculations begin with a ratio of 
N:\ce{N2}:\ce{N2}$_\mathrm{ice}$:\ce{NH3}$_\mathrm{ice}$ equal to 
1.0:0.21:0.39:0.26.  
Taking inspiration from the recent work by \citet{schwarz14}, 
we run an additional set of calculations in which we assume all 
species are in atomic form and 
(i) nitrogen is also in atomic form; 
(ii) nitrogen begins as \ce{N2} gas; 
(iii) nitrogen begins as \ce{N2} ice; and 
(iv)  nitrogen begins as \ce{NH3} ice.  
In this way, we investigate the degree of chemical processing 
in the disk atmosphere for each initial nitrogen reservoir.  

\begin{figure*}
\includegraphics[width=1.0\textwidth]{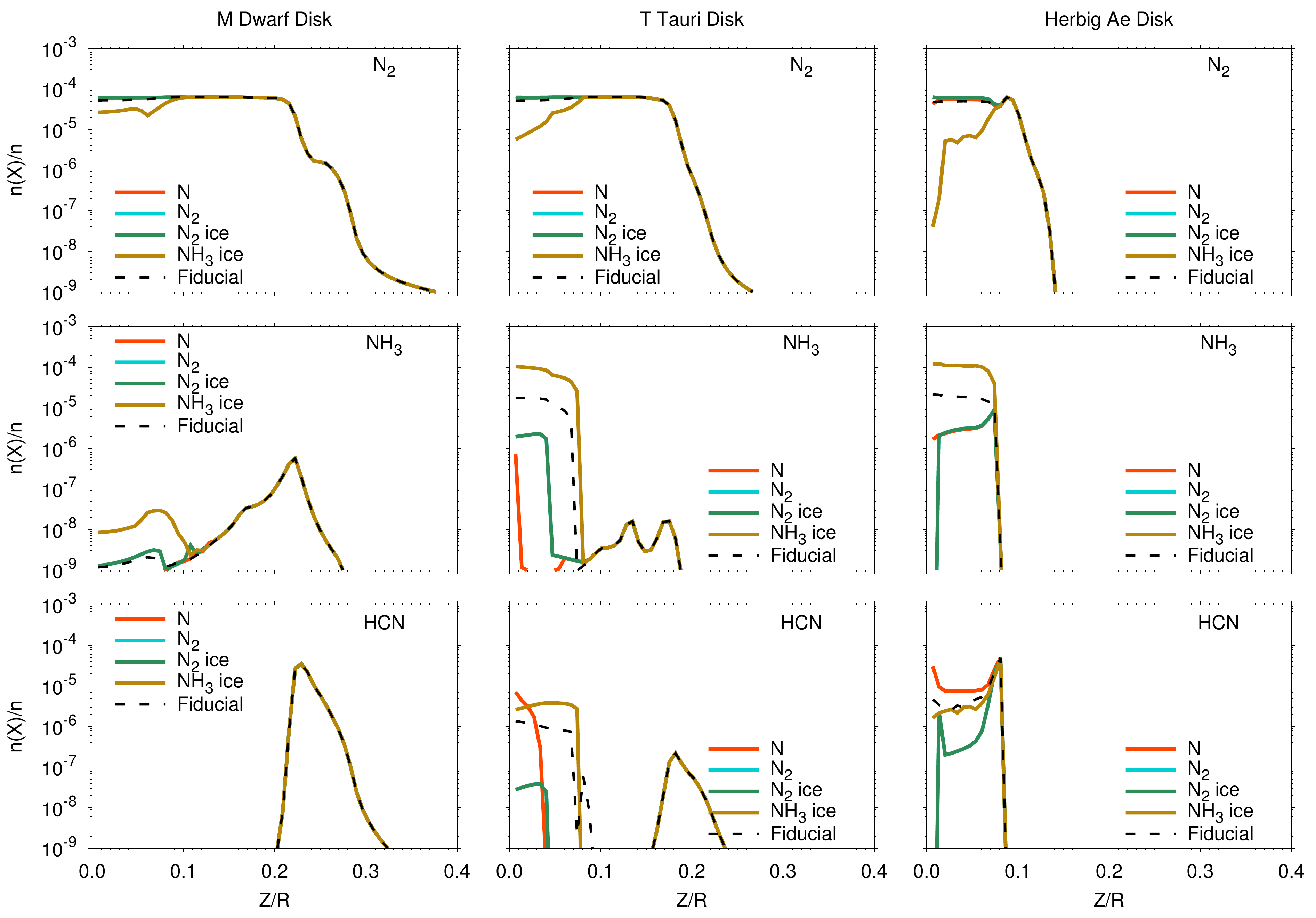}
\caption{Fractional abundance with respect to gas number density of \ce{N2} (top row), 
\ce{NH3} (middle row), and \ce{HCN} (bottom row) as a function 
of $Z/R$ at $R$~=~1~AU for each disk model.  
The black dashed lines, orange lines, cyan lines, green lines, and gold lines represent results 
from the fiducial model and a model in which (i) all species are initially atomic; (ii) all nitrogen is 
in \ce{N2} gas; (iii) all nitrogen is in \ce{N2} ice; 
and (iv) all nitrogen is in \ce{NH3} ice, respectively.}
\label{figure12}
\end{figure*}

In Figure~\ref{figure12}, the fractional abundances
of gas-phase \ce{N2} (top row), \ce{NH3} (middle row), 
and \ce{HCN} (bottom row) are shown as a function of $Z/R$ at 1~AU.  
An equivalent plot for $R$~=~10~AU is presented in the Appendix (Figure~\ref{figure13}).  
The results show that the calculated abundances in the disk atmosphere 
($Z/R$~$\gtrsim$~0.1) are independent of the form of the initial 
nitrogen reservoir at a radius of both 1 and 10~AU for all models. 
The chemistry in the disk atmosphere has achieved steady state 
by $10^{6}$~years and has `forgotten' its origins.  
On the other hand, the abundances in the disk midplane are very sensitive to 
the initial nitrogen reservoir.  
The results for cases (ii) and (iii) show that similar abundances 
are achieved regardless of whether \ce{N2} begins in gas or ice form.  
At 1~AU the dominant factor is whether nitrogen begins as \ce{NH3} ice (case (iv)).  
Here, the abundance of \ce{N2} in the midplane is lower relative to the 
fiducial model with the difference increasing with increasing spectral type.  
Correspondingly, the abundance of gas-phase \ce{NH3} is significantly higher 
increasing by around one order of magnitude compared with the fiducial model.  
At the other extreme, beginning with all species in atomic form 
generates the lowest abundance of \ce{NH3} in the disk midplane, especially 
in the case of the M~dwarf and T Tauri disks with differences 
between 1 and 6 orders of magnitude when compared with the model in which 
nitrogen begins as ammonia ice.  
\ce{NH3} ice is thought to be produced in situ on the surfaces of grains 
or on or within the ice mantle; hence, 
the efficiency of the conversion from atomic N to \ce{NH3} is very sensitive 
to temperature and a sufficient flux of 
both atomic N and H must reside on the surface for the reaction to proceed.  

The initial nitrogen reservoir also has an effect on the midplane 
HCN abundance.  
For the T~Tauri and Herbig~Ae disks, at 1~AU,  
HCN reaches the highest abundance for the atomic model, followed by the model in 
which N begins as ammonia ice.  
Both models which have nitrogen initially in the form of \ce{N2} 
produce the lowest abundance of HCN in the disk midplane.  
The results at 10~AU show less spread except in the case of the 
Herbig~Ae disk.  
\ce{NH3} and \ce{HCN} are both significantly enhanced in the 
midplane for the model in which N begins as \ce{NH3} ice.  

At elevated temperatures, HCN is formed via the reaction between 
\ce{H2} and \ce{CN} \citep{baulch94}, 
the latter of which has various routes to formation via atomic nitrogen and 
nitrogen hydrides, e.g., \ce{N}~+~\ce{CH} or \ce{C}~+~\ce{NH}.  
Both examples are barrierless reactions \citep{brownsword96,smith04,daranlot13}.  
Conversely, reactions which directly produce \ce{CN} or \ce{HCN} from \ce{N2} 
(e.g., \ce{C}~+~\ce{N2} or \ce{CH}~+~\ce{N2}) 
have large reaction barriers \citep[$\gg$~10,000~K,][]{baulch94,rodgers96}.  
In the dense, warm midplane, \ce{NH3} is more easily broken apart 
by cosmic-ray-induced photodissociation than \ce{N2} thus releasing 
nitrogen hydrides into the gas phase \citep{gredel89,heays14}.  
There is also a direct (and barrierless) route to HCN from \ce{NH3},
\begin{equation}
\ce{NH3} + \ce{CN} \longrightarrow \ce{HCN} + \ce{NH2}
\end{equation}
\citep{sims94}. 
Hence, the nitrogen chemistry in the disk midplane 
(and resultant abundance and distribution of N-bearing molecules) 
is sensitive to whether nitrogen begins in the form of \ce{N2} or 
\ce{NH3} with the latter resulting in an increase 
in N-bearing species in the midplane.

\subsection{Comparison with previous models}

Astrochemical models of the inner regions of protoplanetary disks have often 
been neglected in favour of the outer disk ($>>$~10~AU), 
motivated by the larger molecular inventory observed via emission at 
(sub)mm wavelengths \citep[see, e.g., the recent review by][]{dutrey14}.  
However, focus will return to the planet-forming regions of 
protoplanetary disks driven by spatially-resolved observations of molecules 
at (sub)mm wavelengths with the ALMA Large Millimeter/Submillimeter Array 
\citep[ALMA, see, e.g.,][]{qi13} and the launch of the {\em James Webb Space Telescope} 
(JWST) in 2018 \citep[see, e.g.,][]{gardner06}.  

Early models of the planet-forming region spanned a wide range of complexity in 
both physics and chemistry and have primarily focussed 
on disks around T~Tauri stars. 
\citet{willacy98} used a one-dimensional dynamical model to 
determine the chemical composition of the midplane from 0.1 to 100~AU.  
They used a chemical network which included gas-phase chemistry and 
gas-grain interactions (freezeout and desorption).  
They conclude that neutral-neutral chemistry is more important than 
ion-molecule chemistry for controlling the abundances in the midplane, similar to 
that found here.  
They also concluded that the chemical composition of the inner midplane 
was not dependent on the initial molecular abundances.  
This is a different conclusion to that presented here where we find that  
the composition of the initial ice reservoir plays a crucial role 
in the subsequent gas-phase chemistry; however, the model presented here 
uses a chemical network which is significantly more expansive than that 
adopted in \citet{willacy98} and includes grain-surface chemistry and 
ice mantle processing.  

\citet{markwick02} calculated the two-dimensional chemical composition 
of the inner region assuming a disk 
heated internally by viscous heating only, and
using a similar chemical network to that from \citet{willacy98}.  
They also included a simple prescription for the 
X-ray ionisation rate throughout the vertical extent.  
In general, \citet{markwick02} compute significantly 
higher column densities for \ce{C2H2} than found in this work 
and they also do not produce OH in the disk surface layer.  
This is because they neglected photodissociation and photoionisation 
by photons originating from the central star and the external interstellar 
radiation field, processes that we find are important for governing 
the chemistry in the `observable' molecular layer in the inner regions.  

More recently, \citet{agundez08} explored the chemistry in the inner 
regions of a disk around a T~Tauri star using a model similar to that 
adopted for photon-dominated regions (PDRs).  
They find similar conclusions to here: temperature-activated 
neutral-neutral chemistry helps to build chemical complexity in the 
disk atmosphere.  
However, we also find that X-ray driven chemistry is potentially very 
important for building additional complexity, by releasing 
free atomic (or ionic) carbon and nitrogen into the gas deeper into the 
disk atmosphere.    
Qualitatively, we see the same behaviour in the column densities in the disk 
atmosphere with radius: the column density starts higher then decreases 
at a radius specific to each molecule.  
\ce{C2H2} decreases first followed by HCN, then \ce{H2O}.  
OH, on the other hand remains flat.
The peak column densities calculated for \ce{C2H2} and \ce{HCN} 
by \citet{agundez08} ($\sim10^{16}$~cm$^{-2}$) are somewhat lower 
than those computed here for the T~Tauri disk. 
We also find much larger column densities 
for both \ce{OH} and \ce{H2O} in the atmosphere.  
The reason for the particularly low OH column density is unclear but 
\citet{agundez08} do neglect heating via UV excess emission from the star 
and also assume that the gas and dust temperatures are equal; whereas we 
find that the gas is significantly hotter than the dust in the region of the 
inner disk atmosphere where the molecules reside.  
The column densities that we calculate for \ce{OH} and \ce{H2O} for the T~Tauri disk 
are also in line with those determined in the work by \citet{bethell09} and \citet{glassgold09}; 
however, unlike \citet{bethell09}, we do not consider dust grain settling and so do not 
need to invoke \ce{H2O} self shielding as a mechanism for 
explaining the survival of gas-phase water in the disk atmosphere.  
It is interesting that \citet{bethell09} (and follow up work by \citealt{du14}) 
also predict an increase in column density of \ce{H2O} in the disk atmosphere with 
increasing stellar FUV luminosity (and corresponding hardening of the radiation field).  

\subsection{A connection between the disk atmosphere and the planet-forming midplane?}
\label{midplanelink}

Currently, the only means to probe the inner planet-forming regions 
of protoplanetary disks is via near- to mid-infrared observations.  
Future high-angular-resolution observations at (sub)mm wavelengths with ALMA 
may also elucidate the molecular composition of the inner region; however, due to the 
high column densities in the inner disk, dust opacity begins to affect even (sub)mm 
line emission.  
The higher sensitivity and spectral resolution of JWST will also allow 
the measurement of absorption features by gases and ices other than water in nearby 
edge-on protoplanetary disk systems.  

Given that forming planets sweep up material primarily from the disk midplane, 
it is worth exploring to what degree the composition of the disk atmosphere 
reflects that of the midplane.  
During the planet-formation process, the molecules accreted (whether gas or ice) 
are ultimately reprocessed in the planet atmosphere.  
Recent population synthesis models suggest that the main contribution to the heavy 
element content in the atmospheres of forming planets are ices accreted during 
the formation of the planetary embryo and icy planetesimals which 
are captured during the gas accretion stage and vaporised in the atmosphere 
\citep[see, e.g.,][]{thiabaud14}.

\begin{figure*}
\includegraphics[width=1.0\textwidth]{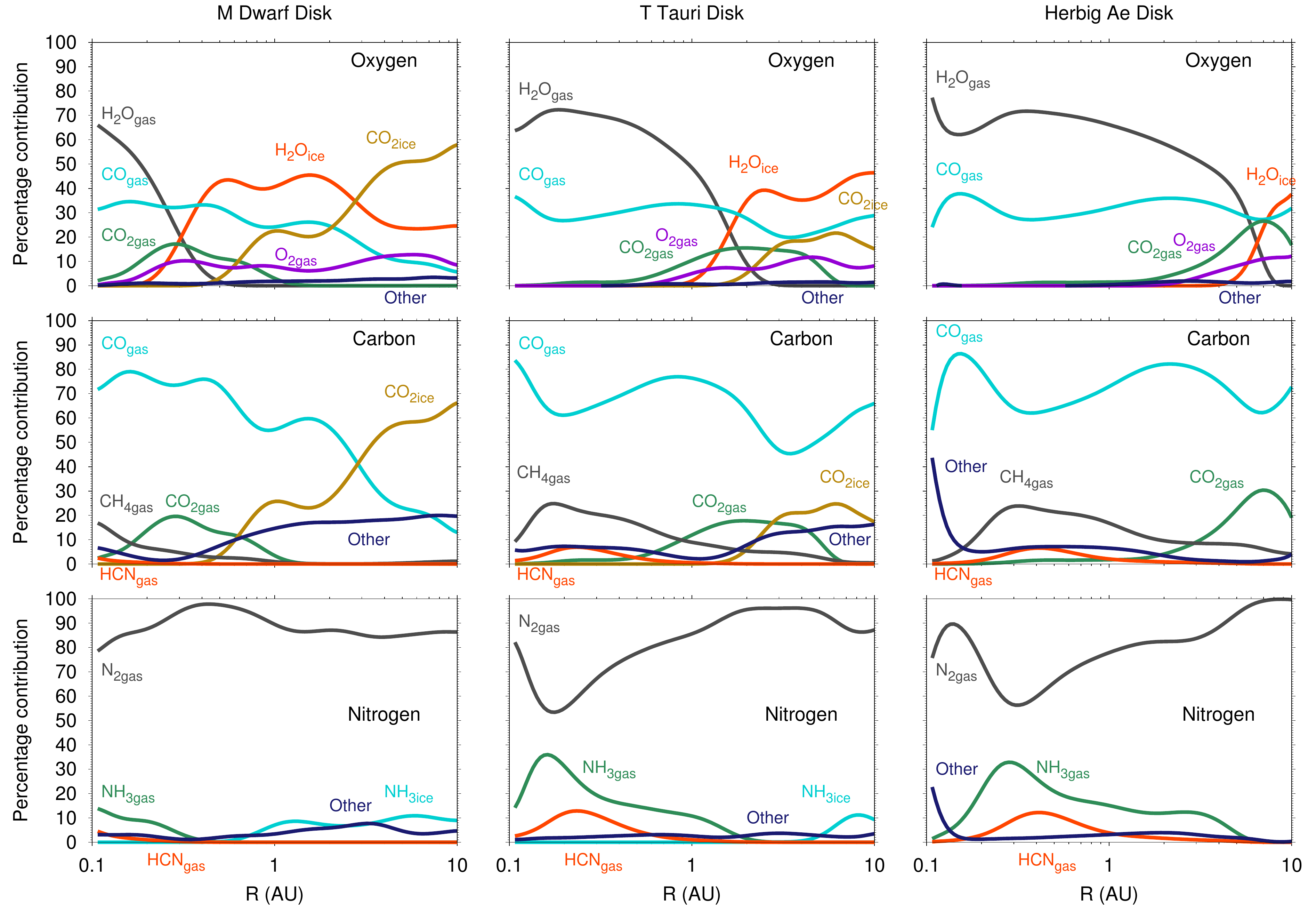}
\caption{Percentage contribution of the dominant molecular carriers to the total elemental oxygen, carbon, and nitrogen 
column densities, as a function of radius.}
\label{figure14}
\end{figure*}

\begin{figure*}
\includegraphics[width=1.0\textwidth]{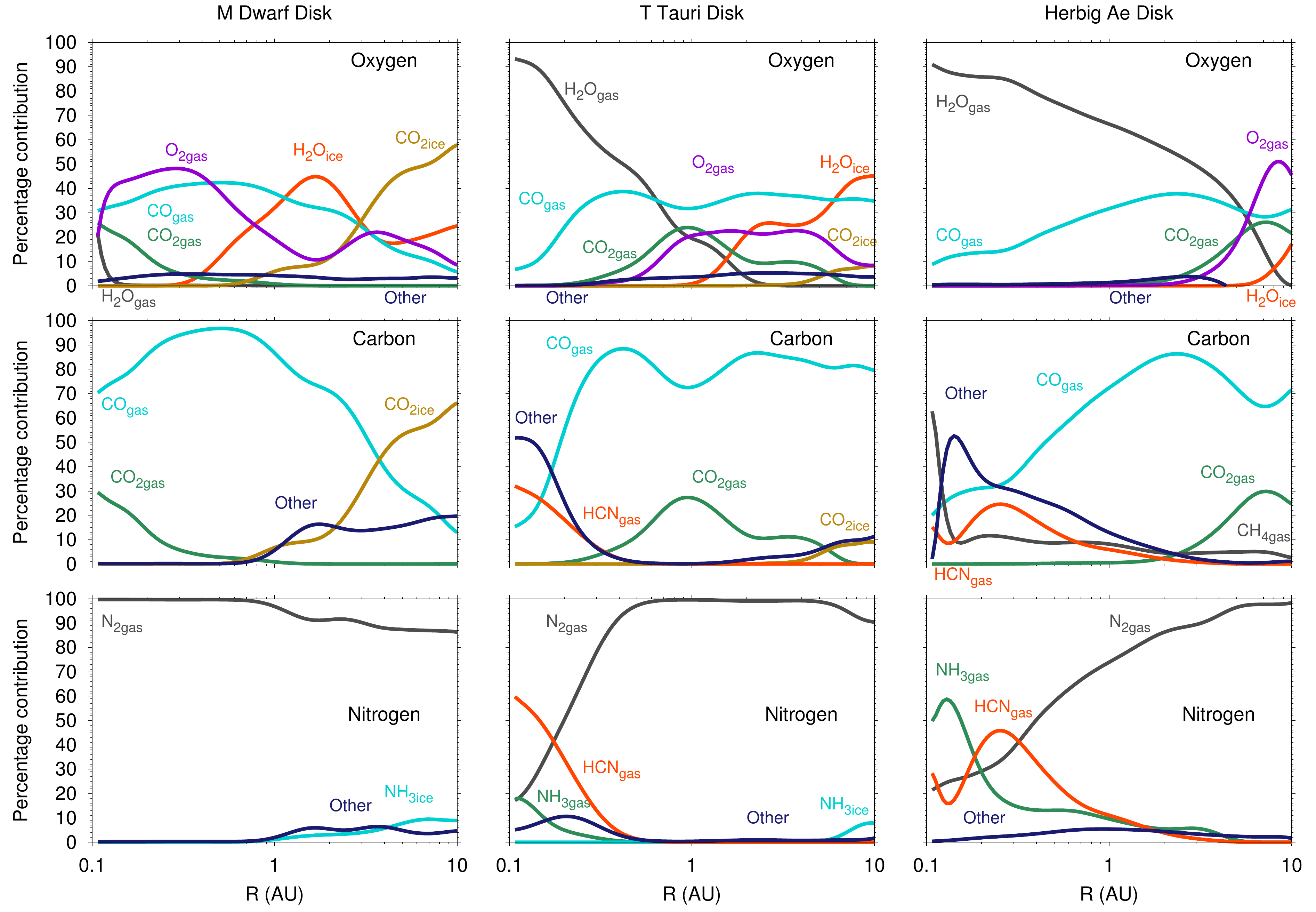}
\caption{Same as Figure~\ref{figure14} for the disk atmosphere only.}
\label{figure15}
\end{figure*}

In Figure~\ref{figure14}, the percentage contribution of each of the dominant 
molecular carriers of oxygen (top row), carbon (middle row), and nitrogen (bottom row) 
are plotted as a function of radius.  
In Figure~\ref{figure15}, we present the equivalent data for the disk 
atmosphere only.  
The dominant carriers are identified as those species which contribute most 
to the total column density of each element (for the data plotted in 
Figure~\ref{figure14}) and to the column density in the disk atmosphere, 
down to the $\tau$(14~$\mu$m)~=~1 surface (for the case of the data plotted in Figure~\ref{figure15}).  
The data identified as `Other' refers to the summation over all 
other species which contribute to the elemental abundance, but which 
individually do not contribute significantly.  
The identity of `Other' depends on the location in the disk 
and the particular disk model but are typically complex organic ices 
(e.g., \ce{H2CO}, \ce{CH3OH}, etc.) 
in the outer regions of the two cooler disks, and gas-phase hydrocarbons 
(e.g., C$_\mathrm{n}$H$_\mathrm{m}$) in the inner regions 
of the two warmer disks.   
For the most part, the plotted species contribute $\approx$~90~--~100\% to
the total elemental abundance.

The dominant carriers throughout the vertical extent 
is set by the locations of snow lines: gas-phase 
\ce{H2O} in the inner regions is superseded by \ce{H2O} ice 
once the snow line is surpassed.  
An interesting result for the M~dwarf disk is that \ce{CO2} 
ice carries most of the oxygen beyond $\approx$~3~AU 
indicating an efficient conversion from \ce{H2O} ice to \ce{CO2} 
ice on and within ice mantles.  
This is not seen in the two warmer disks because the higher temperatures 
make it difficult for CO to reside on the grains sufficiently long 
for reaction with OH.  
Beyond the \ce{H2O} snowline, the dominant gas-phase carrier of 
elemental oxygen is \ce{CO}: thus the gas-phase C/O ratio seemingly 
tends towards 1, whereas the ice mantle remains oxygen-rich.  
However, in each disk, there are regions where molecular oxygen 
has a non-negligible contribution to the total oxygen column, on the 
order of 10\%.   
The two-dimensional fractional abundances of important 
gas-phase oxygen-bearing molecules not discussed in detail in the text, 
such as \ce{CO}, \ce{O2}, and \ce{CO2}, are presented in Figure~\ref{figureA.1} 
in the Appendix. 

The dominant carbon carrier is CO; in the outer disk of the cooler 
M~dwarf, \ce{CO2} ice and other complex organic ices such 
as \ce{CH3OH} take over.  
In the inner region of the T~Tauri and Herbig~Ae disks, 
gas-phase \ce{CH4} and other hydrocarbons begin to 
contribute at the level of 30~--~40\%: the physical conditions in the 
inner regions of the hotter disks are such that the chemistry tends towards 
thermochemical equilibrium in which hydrocarbons dominate over CO.  
In the outer regions of the two hotter disks, again either \ce{CO2} ice or 
gas also contributes at the level of 30\%.  
For the disk around TW~Hya, \citet{bergin14} suggest that 
gas-phase CO is an order of 
magnitude lower in abundance than that expected if 100\% of the 
freely available (i.e., not contained in refractory material) elemental 
carbon were locked up in gas-phase CO.  
Our results for the M~dwarf disk show a depletion in CO in the 
outermost regions consistent with this level of depletion; 
however, we find that significantly more carbon is contained in CO gas in the 
warmer disks ($\approx40-50$\%, at least within a radius of 10~AU).  

The picture for nitrogen is more simple: in all cases, 
gas-phase \ce{N2} is the primary carrier.  
In the M~dwarf disk, \ce{NH3} contributes at the level of 10\% 
whereas in the two hotter disks, there is a region where the contribution 
from gas-phase \ce{NH3} approaches 50\% which is temperature dependent. 
This is again because the conditions in the inner disk midplane 
approach thermochemical equilibrium.  
The low abundance of \ce{NH3} ice in this region suggests that planetary atmospheres 
which gain the bulk of their heavy elements from planetesimal 
accretion will be depleted in nitrogen relative to carbon and oxygen.
The two-dimensional fractional abundances of important 
gas-phase nitrogen-bearing molecules not discussed in detail in the text, 
such as \ce{N2} and \ce{NH3}, are presented in Figure~\ref{figureA.4} in the Appendix. 

In Figure~\ref{figure15}, the equivalent values for the disk 
atmosphere are plotted.  
For the oxygen carriers, the main difference is the 
increased contribution of gas-phase \ce{O2} to the 
oxygen budget in the atmosphere (up to $\approx$~50\%).  
Where \ce{O2} dominates the oxygen budget depends on the 
disk model, moving from the innermost regions of the M~dwarf 
disk to the outermost region of the Herbig~Ae disk.   
\ce{O2} is formed in the atmosphere via reaction between 
atomic oxygen and \ce{OH} and destroyed via reaction with 
atomic hydrogen and carbon, with photodissociation increasing in 
importance as the spectral type of the star increases.   
Figure~\ref{figure6a} shows the dominant formation and 
destruction mechanisms for \ce{O2} in the disk atmosphere, and 
the two-dimensional abundance distributions and column densities of atomic and 
molecular oxygen are given in the Appendix. 
Where \ce{O2} dominates, \ce{CO2} also makes a non-negligible 
contribution in the atmosphere (up to $\approx$~30\%).  

The story for carbon is more simple: CO dominates in the disk atmosphere 
over much of the radial extent of all disks, with some contribution 
from \ce{CO2}, as found for the full column values.  
In the two warmer disks, the gas-phase carbon carriers in the disk atmosphere 
switch to `Other', in this case, various gas-phase hydrocarbons, 
C$_\mathrm{n}$H$_\mathrm{m}$, which is not representative of the 
total column.  
Gas-phase HCN also has a non-negligible contribution to both the carbon 
and nitrogen budget in the atmosphere (up to 30\% and 60\% respectively).  

Similarly, gas-phase \ce{N2} dominates the nitrogen budget throughout most 
of the disk atmosphere; however, in the innermost regions of the two warmer disks, 
gas-phase \ce{NH3} and \ce{HCN} take over.  
Again, this is not representative of the nitrogen budget 
throughout the disk vertical extent which is dominated by \ce{N2} ($\approx$~90\%). 
In Figure~\ref{figure3b} we also show those reactions responsible 
for the formation and destruction of \ce{N2} and \ce{NH3} in the disk atmosphere. 

\begin{figure*}
\includegraphics[width=1.0\textwidth]{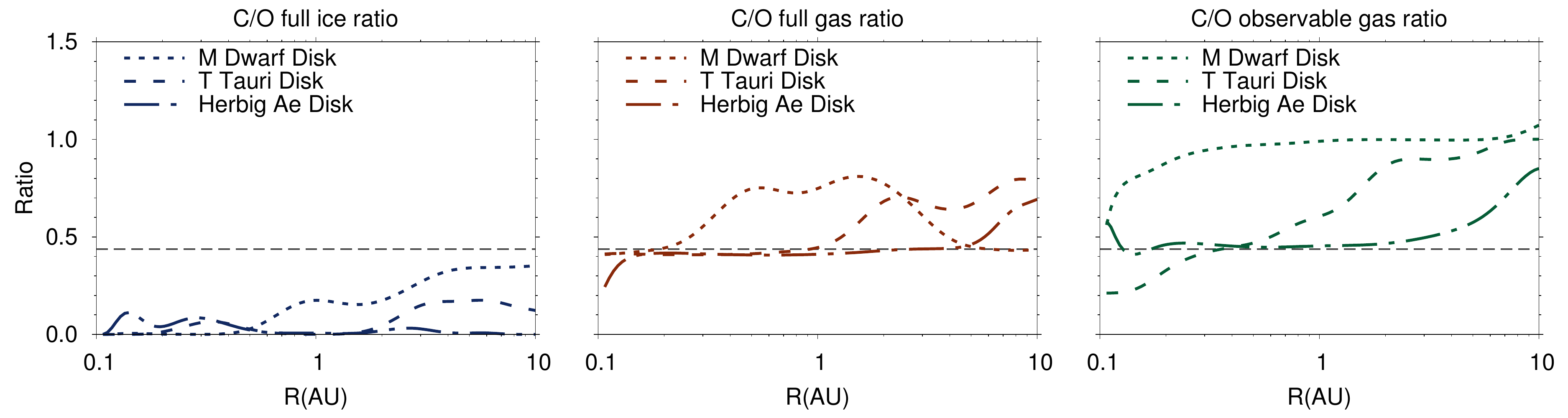}
\caption{C/O ratio as a function of radius for the total ice column density (left panel), the 
total gas column density (middle panel), and for the `observable' disk atmosphere only (right panel).  
The gray dashed line indicates the underlying C/O elemental ratio (0.44).}
\label{figure16}
\end{figure*}

In Figure~\ref{figure16}, the C/O ratio is plotted as a function of radius for three 
cases.  Also plotted is the assumed underlying elemental ratio (C/O~=~0.44, gray dashed lines).  
The left-hand panel shows the ratio when summed over the dominant ice 
reservoirs (\ce{H2O} and \ce{CO2} ice).  
The ice is more oxygen rich than the underlying elemental ratio; 
however, for the two cooler disks, as \ce{CO2} begins to freezeout and/or form, the C/O 
ratio of the ice tends towards the underlying ratio.  
On the other hand, the gas is either representative of the elemental ratio or more 
carbon rich (as shown in the middle panel).  
The behaviour of the C/O ratio depends on the locations of the \ce{H2O} and 
\ce{CO2} snow lines which move outwards with increasing stellar spectral type.    
When \ce{H2O} ice freezes out and/or forms, the C/O ratio in the gas increases as oxygen is removed.  
Once \ce{CO2} ice begins to freeze out and/or form, the gas C/O ratio begins to decrease again 
and the ice correspondingly becomes relatively more carbon rich.  
This is most clearly seen for the M~dwarf disk.  
In the final panel, the C/O ratio calculated using `observable' tracers only
is shown (CO, \ce{CO2}, \ce{H2O}, \ce{C2H2}, \ce{CH4}, and \ce{HCN} gas).  
Comparing the final panel with the middle panel, for the two cooler disks, the 
gas in the disk atmosphere appears significantly more carbon rich when calculated 
using just the listed tracers. 
Notably, the C/O ratio for the M~dwarf disk appears $\gtrsim1$ throughout most 
of the disk atmosphere ($\gtrsim 1$~AU).  
This is primarily due to the presence of gas-phase \ce{O2} in the 
atmosphere which is another `hidden' reservoir of atomic oxygen 
(in addition to \ce{H2O} ice and \ce{CO2} ice).  

The chemical model results suggest that disks around cooler stars might appear 
more carbon rich without the need for additional sinks (neither chemical nor physical) 
to account for the depletion of oxygen. 
However, we stress that detailed radiative transfer calculations are necessary to 
confirm definitively whether the chemical models replicate the trends seen in the observations.  
Because this is a non-trivial calculation for a two-dimensional 
disk structure requiring careful consideration of the 
dust structure and size distribution, this is beyond the scope of the work presented 
here but is planned future work.  
Assuming that the trends in the chemical models do translate into trends in 
the simulated emission, the
derivation of the C/O ratio using CO, \ce{CO2}, \ce{H2O}, \ce{CH4}, \ce{C2H2}, 
and \ce{HCN} alone, 
may overestimate the underlying C/O ratio by up to a factor of two.  
Similarly, the C/N ratio may be overestimated by between one and several orders 
of magnitude if observations of HCN and \ce{NH3} in the disk atmosphere alone are 
used to determine the underlying C/N ratio. 
In the case that the dominant source of heavy elements in a planetary atmosphere is 
icy planetesimals rather than the gas, then this overestimation increases to a factor of 10.  
Although less extreme than the estimation of the C/N ratio, a factor of two is sufficient 
to incorrectly assume the formation of carbon-rich versus 
oxygen-rich planetary atmospheres, and the dominant carbon and oxygen carriers subsequently 
found therein \citep[see, e.g.,][]{madhusudhan13}. 

\subsection{On the abundance of \ce{O2} in protoplanetary disks}
\label{O2chemistry}

The models predict that gas-phase \ce{O2} is a 
significant oxygen reservoir in the inner regions of protoplanetary disks.  
\ce{O2} may contain up to 10\% of the total available oxygen over the full vertical 
extent of the disk, and this increases to up to 50\% when the disk atmosphere only 
is considered (see Figures~\ref{figure14} and \ref{figure15}).  

\ce{O2} has proven to be a somewhat elusive molecule in interstellar and circumstellar 
environments.  
Upper limits determined by SWAS (Submillimeter Wave Astronomy Satellite) and 
{\em Odin} towards nearby molecular clouds constrain 
\ce{O2}/\ce{H2}~$\lesssim 10^{-7}$ \citep{goldsmith00,pagani03}. 
On the other hand, ISO upper limits for the abundance of \ce{O2} ice in dark clouds 
are much more conservative \citep[\ce{O2}/\ce{H2O}$<0.6$,][]{vandenbussche99}.  
However, gas-phase \ce{O2} has been successfully detected towards two 
warmer sources, the dense core,  $\rho$~Oph A \citep[][]{larsson07,liseau12}, 
and Orion \citep{goldsmith11}.  
Furthermore, \citet{yildiz13} report a deep {\em Herschel}-determined upper limit towards the 
low-mass Class~0 protostar, NGC~1333-IRAS~4A (\ce{O2}/\ce{H2}~$<6\times10^{-9}$). 
The authors find that \ce{O2} is absent in both the outer cold envelope and 
inner hot core and conclude that the material entering 
protoplanetary disks is likely poor in molecular oxygen (gas and ice).  

Early gas-phase only chemical models routinely overpredicted the 
abundance of \ce{O2} in dark clouds, which was postulated to form 
primarily via ion-molecule chemistry \citep[see, e.g.,][]{bergin00}.  
More modern and sophisticated gas-grain models are able to reproduce the low 
gas-phase abundance of \ce{O2} provided the conversion of \ce{O2} ice into 
\ce{H2O} ice is included, and the chemistry is allowed to 
evolve for sufficiently long time scales \citep[see, e.g.,][]{bergin00,roberts02,yildiz13}.  
Recent laboratory experiments have shown that the hydrogenation of \ce{O2} 
ice is rapid at low temperatures \citep{ioppolo08,miyauchi08}.

The origin of gas-phase \ce{O2} in the inner regions of protoplanetary disks is 
different to that expected in dark clouds.   
In both cases, the vital reaction is formation via O~+~OH. 
This reaction has been well studied\footnotemark~across the temperature 
range of interest for circumstellar environments.   
However, the origin of OH in dark clouds is 
via the dissociative recombination of \ce{H3O+} which is 
generated via successive proton-donation reactions originating from 
O~+~\ce{H3+}.   
In the inner regions of disks, OH is generated via the reaction between 
O~+~\ce{H2}, and atomic oxygen, in turn is released from 
CO via photodissociation, X-ray-induced dissociation, or 
via reaction with \ce{He+}.  
\ce{O2} is destroyed via photodissociation and reactions with 
C and H to yield CO and OH, respectively (see Figure~\ref{figure6a}). 
These latter two reactions have well-constrained rate coefficients 
\citep[][]{geppert00,miller05}. 
Gas-phase \ce{O2} is able to persist in the disk atmosphere 
for the same reasons as gas-phase \ce{H2O}: the gas temperature is 
sufficiently high to activate the required neutral-neutral chemical 
reactions.   
The model T Tauri disk generates column densities of OH and \ce{H2O} in good 
agreement with those observed towards T Tauri stars (see Figure~\ref{figure7}); 
hence, given that the rate coefficients for the important reactions are 
well studied, the prediction that gas-phase \ce{O2} may also be relatively abundant 
in T Tauri disks is substantive. 

\footnotetext{http://kida.obs.u-bordeaux1.fr/}

\section{Summary}
\label{summary}

In this work, the chemistry and resulting molecular composition of the planet-forming 
regions ($<$~10~AU) of protoplanetary disks has been explored, with 
the aim to investigate potential reasons for the trends seen in near- to 
mid-infrared observations.  
The results demonstrate that, as the effective temperature of the central star increases, 
the molecular complexity of the disk atmosphere decreases, showing 
that the FUV luminosity of the host star plays an important role in determining 
the molecular composition of the disk atmosphere.  
The weaker FUV flux impinging upon disks hosted by M~dwarf stars allows molecules 
to reach relatively high abundances in the atmosphere: X-ray-induced chemistry can further 
increase molecular complexity by driving a rich ion-molecule chemistry, helping to 
qualitatively explain the large column densities of small organic molecules, 
such as, \ce{C2H2} and \ce{HCN} seen in M~dwarf disk atmospheres.  
The key process is the liberation of free carbon and nitrogen from their 
main molecular reservoirs (CO and \ce{N2}, respectively) via X-rays. 
The results shown here, in conjunction with the mid-IR observations, suggest that 
these objects are good targets for ALMA which can further help 
elucidate chemistry in disks towards the low-mass low-luminosity regime.  

The chemical models suggest that the gas in the inner regions of M~dwarf 
disks is generally more carbon rich than that in disks around T~Tauri stars.  
When the C/O gas-phase ratio is calculated 
using only observable tracers in the disk atmosphere, then the ratio 
appears larger than it actually is (and C/O~$\to1$).  
This is because the models predict that gas-phase \ce{O2} is a significant reservoir of 
oxygen in the disk atmosphere beyond the water snowline.  

The results also demonstrate a degree of chemical decoupling between the 
disk atmosphere and the midplane.
The gas is generally more carbon-rich than the midplane ices (see Figure~\ref{figure16}).    
This is further corroborated by our studies on the importance of 
the initial nitrogen reservoir (whether atomic nitrogen, molecular nitrogen, or ammonia).  
We find that this does not play a role in determining the resulting composition of the 
observable molecular layer: the chemistry is at steady state. 
However, the initial reservoir is important for determining the composition 
in the disk midplane where, generally, the chemical timescales are longer. 
For example, icy planetesimals forming in disks where much of the initial nitrogen 
is locked up in \ce{NH3} ice may be more nitrogen rich than those for which nitrogen 
was contained primarily in the more volatile \ce{N2} \citep[see also][]{schwarz14}. 

Whether the trends seen in the chemical models can reproduce those derived from 
observations remains to be confirmed via calculations of the molecular emission 
and will be conducted in future work.     
Assuming that the chemical trends do translate into observable trends, 
near- to mid-IR observations of the dominant tracers in the 
atmosphere may overestimate the underlying C/O and C/N ratios of the gas and ice in the 
region in which forming planets sweep up most of their material by up to a factor of 10 
and more than an order of magnitude, respectively.  

Despite the results qualitatively demonstrating some of the observed trends, 
several issues remain.  
The large column densities of water vapour and correspondingly strong
water emission lines at near- to mid-IR wavelengths predicted by models of 
Herbig~Ae disks have not been confirmed by observations, indicating that there 
may be a heretofore unconsidered destruction mechanism for gas-phase water at 
high temperatures. 
For example, reactions of rovibrationally excited OH with atomic hydrogen 
may shift the ratio of O/OH/\ce{H2O} in the atmosphere. 
Alternative hypotheses are that the molecular line emission is 
veiled by the strong stellar continuum emission \citep[as discussed in][]{pontoppidan10} 
or that the disk dust structure plays an important role. 

A further issue is that the M~dwarf disk model predicts a \ce{C2H2}/\ce{HCN} 
ratio which is an order of magnitude lower relative to the observations.  
An increase in the underlying C/O or C/N elemental ratio in the disk atmosphere may help 
explain the high \ce{C2H2}/\ce{HCN} ratio in M~dwarf disks with the enrichment in carbon 
relative to oxygen and nitrogen caused by vertical or radial mixing.  
In this scenario, less volatile species (e.g., \ce{H2O} and \ce{NH3}) are transported 
to regions where it is sufficiently cold for freezeout onto dust grains 
\citep[see, e.g.,][]{stevenson88,meijerink09}. 
If the ice-coated dust grains are sufficiently large, they become decoupled 
from the gas and settle to the disk midplane.  
In this way, the gas can become enriched in more volatile species, e.g., \ce{CO}, which 
alters the underlying elemental balance of the atmosphere.  
However, as shown in the results here, molecular nitrogen (which is volatile)
is significantly more abundant in the disk atmosphere than \ce{NH3}.  
Hence, this mechanism is unlikely to enrich the disk atmosphere 
in carbon relative to nitrogen but perhaps the chemistry of nitrogen-bearing 
species is perturbed by the depletion of oxygen via this mechanism. 

The gas-phase chemical network used for \ce{HCN} and \ce{C2H2} is (to our knowledge) 
relatively complete, with experimentally measured and/or critically reviewed 
reaction rate coefficients used where available. 
We have shown that self-shielding of \ce{N2} alone is not sufficient to explain 
\ce{C2H2}/\ce{HCN} in the inner regions and has a more significant effect in disks 
around stars with a higher FUV luminosity.  
It remains to be confirmed whether this also holds for disks in which 
grain growth and settling have generated a relatively dust-poor atmosphere for which
molecular shielding dominates over dust shielding.  
Instead, X-ray-induced chemistry is more important for releasing 
atomic nitrogen from \ce{N2} in the M~dwarf disk.  
Furthermore, we find that the exclusion of X-ray-induced chemistry only increases 
the discrepancy with observation, because the formation of \ce{C2H2} is primarily driven 
by ion-molecule chemistry in the M~dwarf disk atmosphere, 
as shown in Figures~\ref{figure3a} and \ref{figure11}. 
However, all chemical networks suffer from a degree of uncertainty; hence, a systematic 
sensitivity study of the relative 
abundance of these two species over the parameter space of physical conditions 
in protoplanetary disks around cool stars, is worthy of further exploration.  
This would also confirm the hypothesis presented here, that is, 
that X-ray chemistry is responsible for the higher \ce{C2H2}/\ce{HCN} 
ratio observed in cool stars. 

We finish by stating that the future is bright for near- to mid-IR astronomy 
with MIRI \citep[Mid-InfraRed Instrument,][]{wright04} 
on JWST and METIS \citep{brandl14}, currently being developed for 
installation on the European Extremely Large Telescope (E-ELT).  
Both facilities will have significantly higher spectral resolution 
($\sim$~3000 and 100,000 respectively) across the 
wavelength range of interest for probing the molecular composition of the 
planet-forming regions of nearby protoplanetary disks.  

\begin{acknowledgements}
The authors thank Drs. Alan Heays, Kenji Furuya, and Mihkel Kama for useful discussions 
and an anonymous referee for their insightful comments.    
C.W. acknowledges support from the European Union A-ERC grant 291141
CHEMPLAN and from the Netherlands Organisation for 
Scientific Research (NWO, program number 639.041.335). 
H.~N.~acknowledges the Grant-in-Aid for Scientific Research 23103005 and 25400229.  
She also acknowledges support from the Astrobiology Project of the CNSI, 
NINS (grant numbers AB261004, AB261008).  
A portion of the numerical calculations were carried out on SR16000
at YITP in Kyoto University.
\end{acknowledgements}

\begin{appendix}

\section{Supplementary figures}

\begin{figure*}
\includegraphics[width=1.0\textwidth]{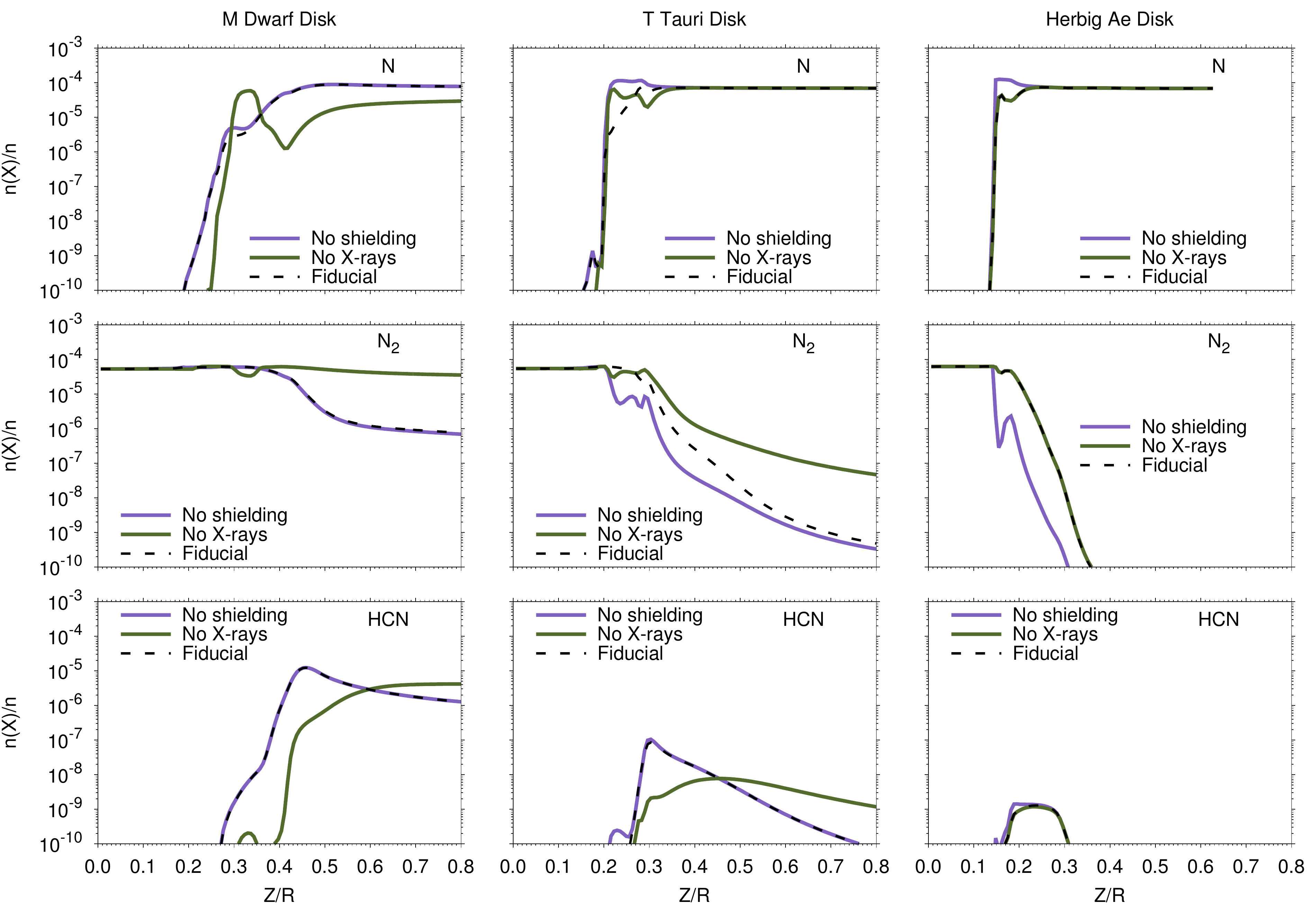}
\caption{Fractional abundance with respect to gas number density of \ce{N} (top row), 
\ce{N2} (middle row), and \ce{HCN} (bottom row) as a function 
of $Z/R$ at $R$~=~10~AU for each disk model.  
The black dashed lines, purple solid lines, and green solid lines represent results from the fiducial 
model (including \ce{N2} shielding and X-rays), the model with \ce{N2} switched off, and 
the model with X-rays switched off, respectively.}
\label{figure10}
\end{figure*}

\begin{figure*}
\includegraphics[width=1.0\textwidth]{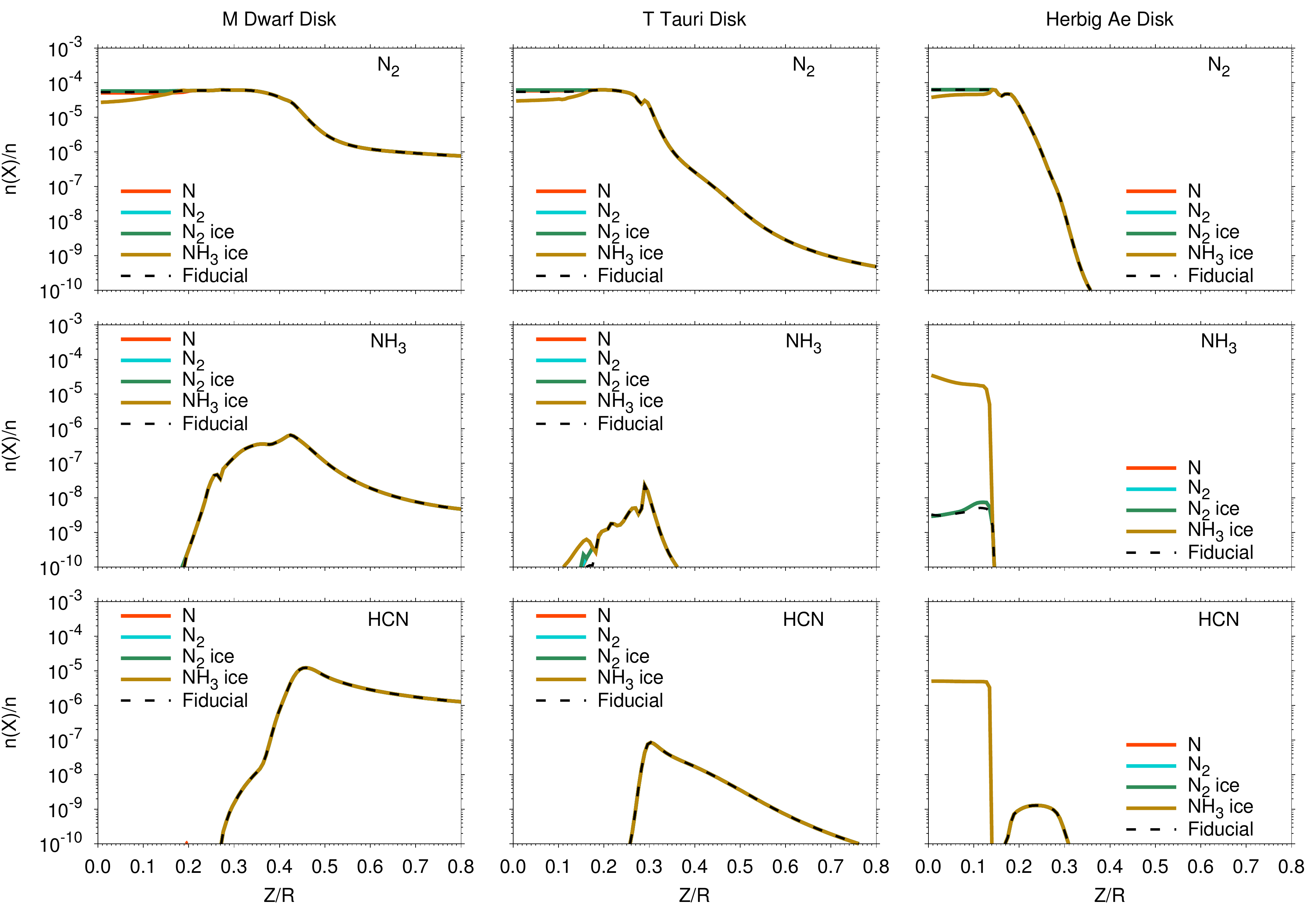}
\caption{Fractional abundance with respect to gas number density of \ce{N2} (top row), 
\ce{NH3} (middle row), and \ce{HCN} (bottom row) as a function 
of $Z/R$ at $R$~=~10~AU for each disk model.  
The black dashed lines, orange lines, cyan lines, green lines, and gold lines represent results 
from the fiducial model and a model in which (i) all species are initially atomic; (ii) all nitrogen is 
in \ce{N2} gas; (iii) all nitrogen is in \ce{N2} ice; 
and (iv) all nitrogen is in \ce{NH3} ice, respectively.}
\label{figure13}
\end{figure*}

In Figures~\ref{figure10} and \ref{figure13}, we show results at a radius of 10~AU 
from the same models described in Figures~\ref{figure9} and \ref{figure12}.

\begin{figure*}
\centering
\subfigure{\includegraphics[width=1.0\textwidth]{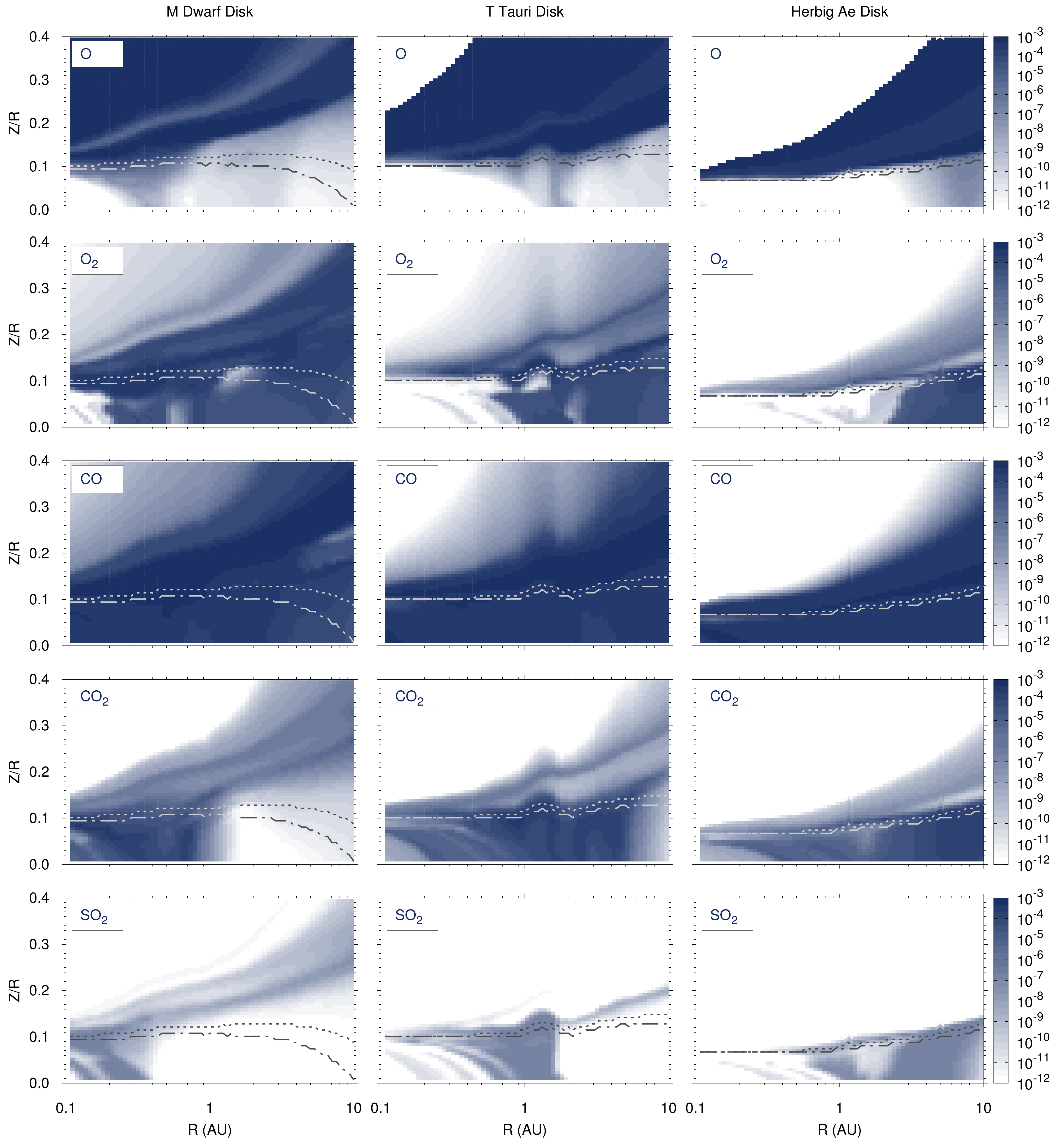}}
\caption{Fractional abundance relative to total gas number density of 
oxygen-bearing species for the M~dwarf disk (left-hand column), 
T~Tauri disk (middle column), and Herbig~Ae disk (right-hand column). 
The dotted and dot-dashed lines indicate the dust column density (integrated from the surface downwards) 
at which $\tau$~$\approx$~1 at 3~$\mu$m and 14~$\mu$m, respectively.
\label{figureA.1}}
\end{figure*}

\begin{figure*}
\centering
\subfigure{\includegraphics[width=1.0\textwidth]{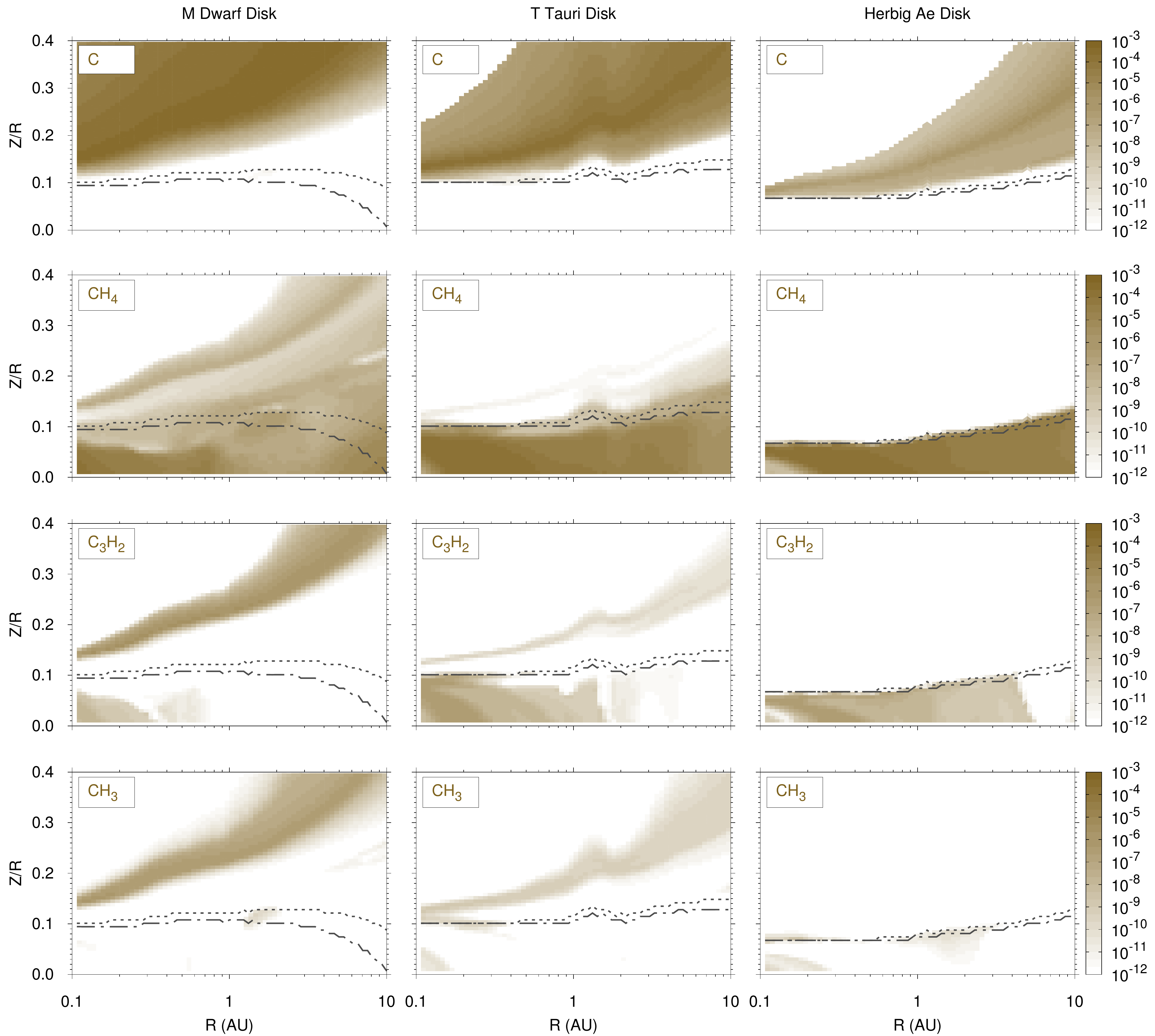}}
\caption{Fractional abundance relative to total gas number density of 
carbon-bearing species for the M~dwarf disk (left-hand column), 
T~Tauri disk (middle column), and Herbig~Ae disk (right-hand column). 
The dotted and dot-dashed lines indicate the dust column density (integrated from the surface downwards) 
at which $\tau$~$\approx$~1 at 3~$\mu$m and 14~$\mu$m, respectively.
\label{figureA.2}}
\end{figure*}

\begin{figure*}
\centering
\subfigure{\includegraphics[width=1.0\textwidth]{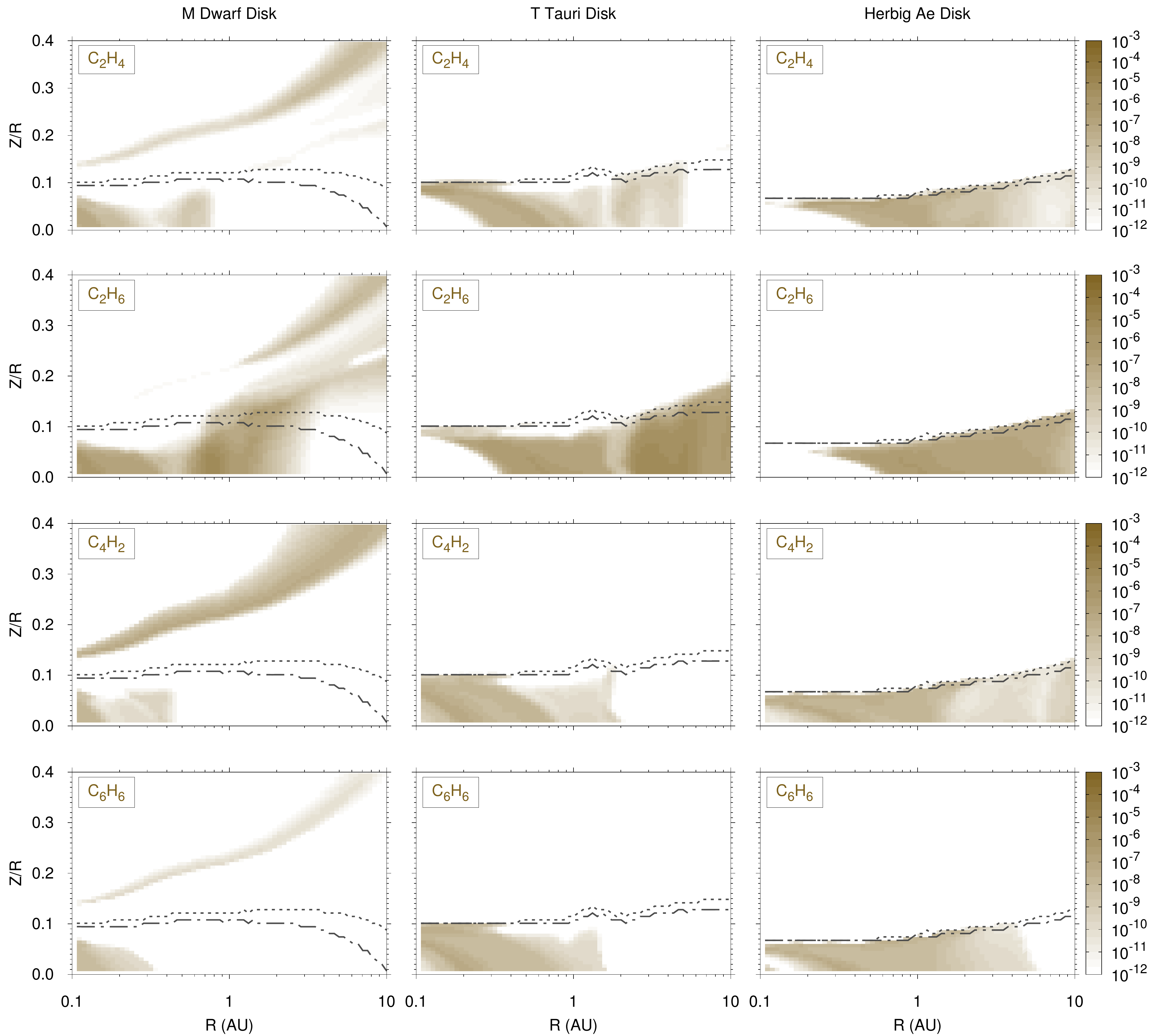}}
\caption{Fractional abundance relative to total gas number density of 
carbon-bearing species for the M~dwarf disk (left-hand column), 
T~Tauri disk (middle column), and Herbig~Ae disk (right-hand column). 
The dotted and dot-dashed lines indicate the dust column density (integrated from the surface downwards) 
at which $\tau$~$\approx$~1 at 3~$\mu$m and 14~$\mu$m, respectively.
\label{figureA.3}}
\end{figure*}

\begin{figure*}
\centering
\subfigure{\includegraphics[width=1.0\textwidth]{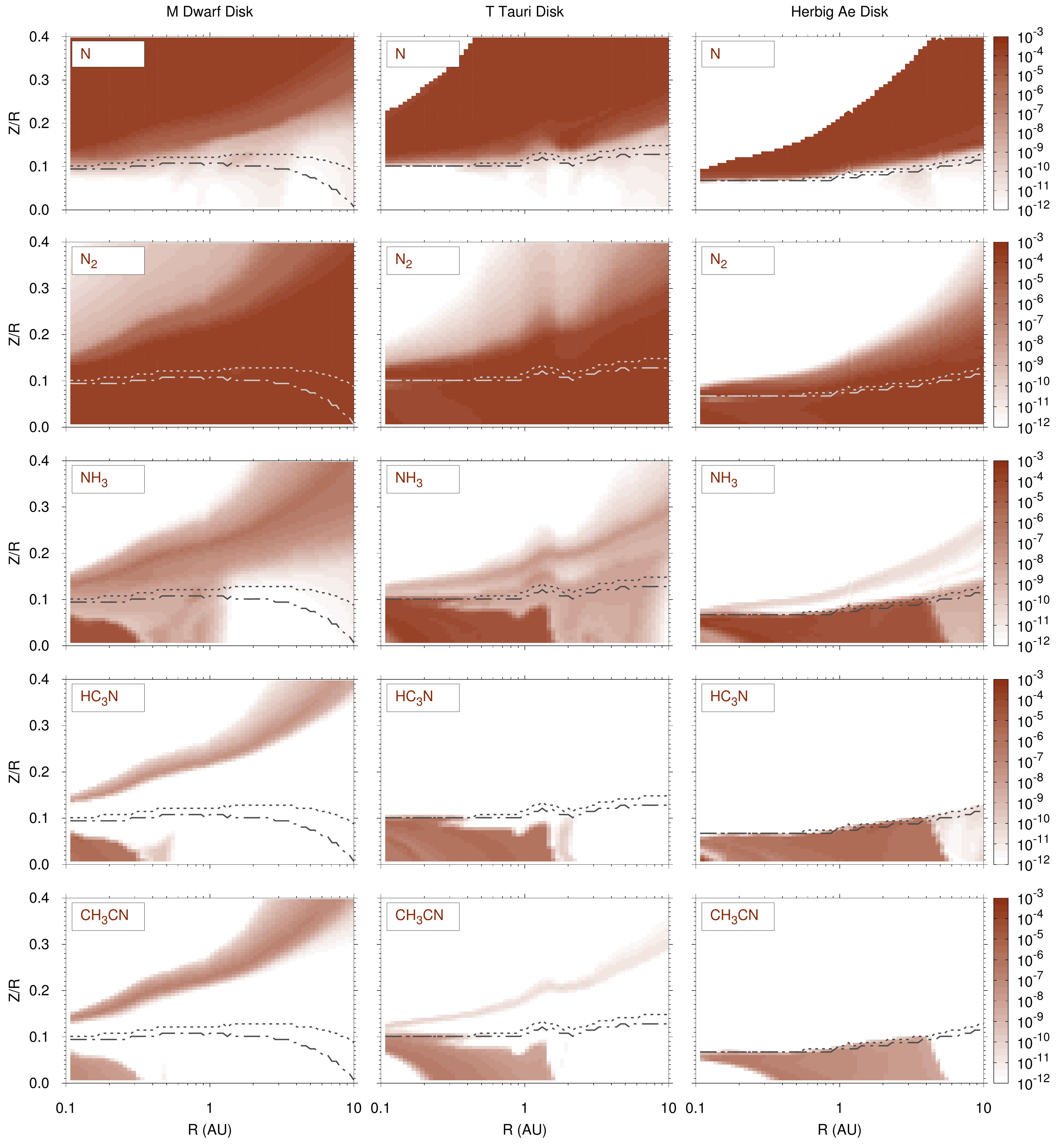}}
\caption{Fractional abundance relative to total gas number density of 
nitrogen-bearing species for the M~dwarf disk (left-hand column), 
T~Tauri disk (middle column), and Herbig~Ae disk (right-hand column). 
The dotted and dot-dashed lines indicate the dust column density (integrated from the surface downwards) 
at which $\tau$~$\approx$~1 at 3~$\mu$m and 14~$\mu$m, respectively. 
\label{figureA.4}}
\end{figure*}

In Figures~\ref{figureA.1} to \ref{figureA.4} we show the fractional abundances
as a function of radius, $R$, and height scaled by the radius, $Z/R$, for 
those species not discussed in detail in the text.  
These include important oxygen- (e.g., \ce{O2}, \ce{CO}, and \ce{CO2}), 
carbon- (e.g., C$_\mathrm{n}$H$_\mathrm{m}$) 
and nitrogen-bearing species (e.g., \ce{N2} and \ce{NH3}). 

\end{appendix}


\begin{thebibliography}{}
\bibitem[Acke et al.(2005)]{acke05} Acke, B., van~den~Ancker, M.~E., \& Dullemond, C.~P. 2005, \aap, 436, 209 
\bibitem[\'{A}d\'{a}mkovics et al.(2014)]{adamkovics14} \'{A}d\'{a}mkovics, M., Glassgold, A.~E., Najita, J.~R. 2014, \apj, 786, 135
\bibitem[Adams \& Smith(1979)]{adams76} Adams, N.~G. \& Smith, D. 1976, Int. J. Mass Spectrom. Ion Phys., 21, 349
\bibitem[Ag\'{u}ndez et al.(2008)]{agundez08} Ag\'{u}ndez, M., Cernicharo J., \& Goicoechea, J.~R. \aap, 2008, 483, 831 
\bibitem[Aikawa \& Herbst(1999)]{aikawa99} Aikawa, Y. \& Herbst, Y. 1999, \aap, 351, 233
\bibitem[Aikawa \& Nomura(2006)]{aikawa06} Aikawa, Y. \& Nomura, H. 2006, \apj, 642, 1152
\bibitem[Anicich et al.(1977)]{anicich77} Anicich, V.~G., Laudenslager, J.~B., Huntress, W.~T., \& Futrell, J.~H. 1977, J.~Chem.~Phys., 67, 4340
\bibitem[Anicich et al.(1993)]{anicich93} Anicich, V.~G. 1993, J.~Phys.~Chem.~Ref.~Data, 22, 1469
\bibitem[Bast et al.(2013)]{bast13} Bast, J.~E., Lahuis, F., van~Dishoeck, E.~F., \& Tielens, A.~G.~G.~M. 2013, \aap, 551, A118
\bibitem[Baulch et al.(1992)]{baulch92} Baulch, D.~L., Cobos, C.~J., Cox, R.~A., et al. 1992, J.~Phys.~Chem.~Ref.~Data, 21, 411
\bibitem[Baulch et al.(1994)]{baulch94} Baulch, D.~L., Cobos, C.~J., Cox, R.~A., et al. 1994, J.~Phys.~Chem.~Ref.~Data, 23, 847
\bibitem[Baulch et al.(2005)]{baulch05} Baulch, D.~L., Bowman, C.~T., Cobos, C.~J., et al. 2005, J.~Phys.~Chem.~Ref.~Data, 34, 757
\bibitem[Bethell \& Bergin(2009)]{bethell09} Bethell, T. \& Bergin, E.~A. 2009, Science, 326, 1675
\bibitem[Bohlin et al.(1978)]{bohlin78} Bohlin, R.~C., Savage, B.~D., \& Drake, J.~F. 1978, \apj, 224, 132
\bibitem[Bergin et al.(2000)]{bergin00} Bergin, E.~A., Melnick, G.~J., Stauffer, J.~R., et al. 2000, \apj, 538, L129 
\bibitem[Bergin et al.(2014)]{bergin14} Bergin, E., Cleeves, L.~I., Crockett, N., \& Blake, G. 2014, Faraday~Discus., 168, 61 
\bibitem[Brandl et al.(2014)]{brandl14} Brandl, B.~R., Feldt, M., Glasses, A., et al. 2014, in SPIE Conf. Ser. 9147, ed. S.~Ramsey, I.~S.~McLean, H.~Takami, 914721
\bibitem[Brownsword et al.(1996)]{brownsword96} Brownsword, R.~A., Gatenby, S.~D., Herbert, L.~B., et al. 1996, J. Chem. Soc. Faraday Trans., 92, 723
\bibitem[Bruderer et al.(2012)]{bruderer12} Bruderer, S., van~Dishoeck, E.~F., Doty, S.~D., \& Herczeg, G.~J. 2012, \aap, 541, 91 
\bibitem[Bruderer et al.(2015)]{bruderer15} Bruderer, S., Harsono, D., \& van~Dishoeck, E.~F. 2015, \aap, 575, A94
\bibitem[Cardelli et al.(1991)]{cardelli91} Cardelli, J.~A., Savage, B.~D., \& Ebbets, D.~C. 1991, \apj, 383, L23
\bibitem[Cardelli et al.(1996)]{cardelli96} Cardelli, J.~A., Meyer, D.~.M., Jura, M., Savage, B.~D. 1996, \apj, 467, 334
\bibitem[Carr \& Najita(2008)]{carr08} Carr, J.~S. \& Najita, J.~R. 2008, Science, 319, 1504
\bibitem[Carr \& Najita(2011)]{carr11} Carr, J.~S. \& Najita, J.~R. 2011, \apj, 733, 102
\bibitem[Chang et al.(2007)]{chang07} Chang. Q, Cuppen, H.~M., \& Herbst, E. 2007, \aap, 469, 973 
\bibitem[Chastaing et al.(1999)]{chastaing99} Chastaing, D., James, P~.L., Sims, I.~R., \& Smith, I.~W.~M. 1999, Phys.~Chem.~Chem.~Phys., 1, 2247
\bibitem[Chastaing et al.(2000)]{chastaing00} Chastaing, D., Le Picard, S.~D., \& Sims, I.~R. 2000, J.~Chem.~Phys., 112, 8466
\bibitem[Clary et al.(1985)]{clary85} Clary, D.~C., Smith, D., \& Adams, N.~G. 1985, Chem.~Phys.~Lett., 119, 320 
\bibitem[Cuppen et al.(2010a)]{cuppen10a} Cuppen, H. M., Ioppolo, S., Romanzin, C., \& Linnartz, H. 2010a, PCCP, 12, 12077
\bibitem[Cuppen et al.(2010b)]{cuppen10b} Cuppen, H.~M., Kristensen, L.~E., \& Gavardi, E. 2010b, \mnras, 406, L11 
\bibitem[D'Alessio et al.(2006)]{dalessio06} D'Alessio, P., Calvet, N., Hartmann, L., Franco-Hern\'{a}ndez, R., \& S\'{e}rvin, H. 2006, \apj, 638, 314
\bibitem[Daranlot et al.(2013)]{daranlot13} Daranlot, J., Hu, X., Xie, C., et al. 2013, Phys.~Chem.~Chem.~Phys., 15, 13888
\bibitem[Dominik et al.(2005)]{dominik05} Dominik, C., Blum, J., Cuzzi, J.~N., \& Wurm, G. 2007, Protostars \& Planets V, 783
\bibitem[Dullemond \& Dominik(2004)]{dullemond04} Dullemond, C.~P. \& Dominik, C. 2004, \aap, 417, 159
\bibitem[Du \& Bergin(2014)]{du14} Du, F. \& Bergin, E.~A. 2014, \apj, 792, 2
\bibitem[Dutrey et al.(2014)]{dutrey14} Dutrey, A., Semenov, D., Chapillon, E., et al. 2014, Protostars \& Planets VI, 317
\bibitem[Fedele et al.(2011)]{fedele11} Fedele, D., Pascucci, I., Brittain, S., et al. 2011, \apj, 732, 106
\bibitem[Fedele et al.(2012)]{fedele12} Fedele, D., Bruderer, S., van~Dishoeck, E.~F., et al. 2012, \aap, 544, L9
\bibitem[Fedele et al.(2013)]{fedele13} Fedele, D., Bruderer, S., van~Dishoeck, E.~F., et al. 2013, \aap, 559, A77
\bibitem[Federman et al.(1979)]{federman79} Federman, S.~R., Glassgold, A.~E., \& Kwan, J. 1985, \apj, 227, 466
\bibitem[Fillion et al.(2001)]{fillion01} Fillion, J.~H., van~Harrevelt, R., Ruiz, J., et al. 2001, J.~Phys.~Chem.~A, 105, 11414
\bibitem[Fogel et al.(2011)]{fogel11} Fogel, J.~K.~J., Bethell, T.~J., Bergin, E.~A., Calvet, N., \& Semenov, D. 2011, \apj, 726, 29  
\bibitem[France et al.(2013)]{france13} France, K., Froning, C.~S., Linsky, J.~L., et al. 2013, \apj, 763, 149 
\bibitem[France et al.(2014)]{france14} France, K., Schindhelm, E., Bergin, E.~A., Roueff, E., \& Abgrall, H. 2014, \apj, 784, 127
\bibitem[Fraser et al.(2001)]{fraser01} Fraser, H.~J., Collings, M.~P., McCoustra, M.~R.~S., \& Williams, D.~A. 2001, \mnras, 327, 1165
\bibitem[Gardner et al.(2006)]{gardner06} Gardner, J.~P., Mather, J.~C., Clampin, M. et al. 2006, Space Sci.~Rev., 123, 485
\bibitem[Garrod et al.(2008)]{garrod08} Garrod, R.~T., Widicus Weaver, S.~L., \& Herbst, E. 2008, \apj, 682, 283
\bibitem[Garrod \& Pauly(2011)]{garrod11} Garrod, R.~T. \& Pauly, T. 2011, \apj, 735, 15
\bibitem[Geppert et al.(2000)]{geppert00} Geppert, W.~D., Reignier, D., Stoecklin, T., et al. 2000, Phys.~Chem.~Chem.~Phys., 2, 2873
\bibitem[Gibb et al.(2007)]{gibb07} Gibb, E.~L., van~Brunt, K.~A., Brittain, S.~D., \& Rettig, T.~W. 2007, \apj, 660, 1572
\bibitem[Gibb \& Horne (2013)]{gibb13} Gibb, E.~L. \& Horne, D. 2013, \apjl, 776, L28
\bibitem[Glassgold et al.(1985)]{glassgold85} Glassgold, A.~E., Huggins, P.~J., \& Langer, W.~D. 1985, \apj, 290, 615 
\bibitem[Glassgold et al.(1997)]{glassgold97} Glassgold, A.~E., Najita, J., Igea, J. 1997, \apj, 480, 344 
\bibitem[Glassgold et al.(2009)]{glassgold09} Glassgold, A.~E., Meijerink, R., \& Najita, J.~R. 2009, \apj, 701, 142
\bibitem[Goldsmith et al.(2000)]{goldsmith00} Goldsmith, P.~F., Melnick, G.~J., Bergin, E.~A., et al. 2000, \apjl, 539, L123
\bibitem[Goldsmith et al.(2011)]{goldsmith11} Goldsmith, P.~F., Liseau, R., Bell, T.~A., et al. 2011, \apj, 737, 96
\bibitem[Graedel et al.(1982)]{graedel82} Graedel, T.~E., Langer, W.~D., Frerking, M.~A. 1982 \apjs, 48, 321
\bibitem[Grady et al.(2015)]{grady15} Grady, C., Fukugawa, M., Maruta, Y., et al. 2015, \apss, 355, 253
\bibitem[Gredel et al.(1989)]{gredel89} Gredel, R., Lepp, S., Dalgarno, A., \& Herbst, E. 1989, \apj, 347, 289
\bibitem[Hamaguchi et al.(2005)]{hamaguchi05} Hamaguchi, K., Yamauchi, S., \& Koyama, K. 2005, \apj, 618, 360
\bibitem[Hasegawa et al.(1992)]{hasegawa92} Hasegawa, T.~I., Herbst, E., \& Leung, C.~M. 1992, \apj, 82, 167
\bibitem[Herbst(1995)]{herbst95} Herbst, E. 1995, ARPC, 46, 27
\bibitem[Heays et al.(2014)]{heays14} Heays, A.~N., Visser, R., Gredel, R., et al. 2014, \aap, 562, A61
\bibitem[Herczeg \& Hillenbrand(2009)]{herczeg09} Herczeg, G.~J. \& Hillenbrand L.~A. 2009, \apj, 681, 594 
\bibitem[Huntress(1977)]{huntress77} Huntress, W.~T. 1997, \apjs, 33, 495
\bibitem[Ioppolo et al.(2008)]{ioppolo08} Ioppolo, S., Cuppen, H.~M., Romanzin, C. van Dishoeck, E.~F., \& Linnartz, H. 2008, \apj, 686, 1474 
\bibitem[Lahuis et al.(2006)]{lahuis06} Lahuis, F., van~Dishoeck, E.~F., Boogert, A.~C.~A., et al. 2006, \apj, 636, L145
\bibitem[Lamberts et al.(2013)]{lamberts13} Lamberts, T., Cuppen, H.~M., Ioppolo, S., \& Linnartz, H. 2013, PCCP, 15, 8287
\bibitem[Larsson et al.(2007)]{larsson07} Larsson, B., Liseau, L., Pagani, L., et al. 2007, \aap, 466, 999
\bibitem[Laufer \& Fahr(2004)]{laufer04} Laufer, A.~H. \& Fahr, A. 2004, Chem. Rev., 104, 2813 
\bibitem[Lee(1994)]{lee84} Lee, L.~C. 1984, \apj, 282, 172
\bibitem[Lee et al.(1996)]{lee96} Lee, H.~H., Herbst, E., Pineau de For\^{e}ts, G., Roueff, E., \& Le~Bourlot, J. 1996, \aap, 311, 690
\bibitem[Li et al.(2013b)]{li13b} Li, X., Arasa, C., van~Dishoeck, E.~F., van Hemert, M. 2013, J.~Phys.~Chem.~A, 117, 12889 
\bibitem[Li et al.(2013a)]{li13a} Li, X., Heays, A.~N., Visser, R., et al. 2013, \aap, 555, A14 
\bibitem[Liseau et al.(2012)]{liseau12} Liseau, R., Goldsmith, P.~F., Larsson, B., et al. 2012, \aap, 541, A73
\bibitem[Karssemeijer \& Cuppen(2014)]{karssemeijer14} Karssemeijer, L.~J. \& Cuppen, H.~M. 2014, \aap, 569, A107
\bibitem[Kim et al.(1974)]{kim74} Kim, J.~K., Theard, L.~P., \& Huntressm W.~T. 1974, Int.~ J.~Mass~Spectrom.~Ion~Phys., 15, 223
\bibitem[Kim \& Huntress(1975)]{kim75} Kim, J.~K. \& Huntress, W.~T. 1975, Int.~ J.~Mass~Spectrom.~Ion~Phys., 16, 451
\bibitem[Madhusudhan et al.(2013)]{madhusudhan13} Madhusudhan, N., Amin, M.~A., \& Kennedy, G.~M. 2013, \apjl, 794, L12
\bibitem[Maloney et al.(1996)]{maloney96} Maloney, P.~R., Hollenbach, D.~J., \& Tielens, A.~G.~G.~M. 1996, \apj, 466, 561
\bibitem[Mandell et al.(2012)]{mandell12} Mandell, A.~M., Bast, J., van~Dishoeck, E.~F., et al. 2012, \apj, 747, 92
\bibitem[Markwick et al.(2002)]{markwick02} Markwick, A., Ilgner, M., Millar, T.~J., \& Henning, Th. 2002, \aap, 385, 632
\bibitem[Meeus et al.(2001)]{meeus01} Meeus, G., Waters, L.~B.~F.~M., Bouwman, J. et al. 2001, \aap, 365, 476
\bibitem[Meeus et al.(2012)]{meeus12} Meeus, G., Montesinos, B., Mendigut\'{i}a, I., et al. 2012, \aap, 544, A78
\bibitem[Meijerink et al.(2009)]{meijerink09} Meijerink, R., Pontoppidan, K.~M., Blake, G.~A., Poelman, D.~R., \& Dullemond, C.~P. 2009, \apj, 704, 1471
\bibitem[Meyer et al.(1998)]{meyer98} Meyer, D.~M., Jura, M., \& Cardelli, J.~A., 1998, \apj, 493, 222
\bibitem[McElroy et al.(2013)]{mcelroy13} McElroy, D., Walsh, C., Markwick, A.~J., et al. 2013, \aap, 550, A36
\bibitem[Miyauchi et al.(2008)]{miyauchi08} Miyauchi, N., Hidaka, H., Chigai, T., Nagaoka, A., Watanabe, N., \& Kouchi, A. 2008, Chem.~Phys.~Lett., 456, 27 
\bibitem[Miller et al.(2005)]{miller05} Miller, J.~A., Pilling, M.~J., \& Troe, J. 2005, in Proceedings of the Combustion Institute, 30, 43
\bibitem[Najita et al.(2011)]{najita11} Najita, J.~R., \'{A}da\'{a}mkovics, M., \& Glassgold, A.~E. 2011, \apj, 743, 147
\bibitem[Najita et al.(2013)]{najita13} Najita, J.~R., Carr, J.~S., Pontoppidan, K.~M., et al. 2013, \apj, 766, 134
\bibitem[Noble et al.(2012)]{noble12} Noble, J., Theul\'{e}, P., Borget, F., et al. 2011, \mnras, 428, 3232
\bibitem[Nomura \& Millar(2005)]{nomura05} Nomura, H. \& Millar, T. J. 2005, \aap, 438, 923
\bibitem[Nomura et al.(2007)]{nomura07} Nomura, H., Aikawa, Y., Tsujimoto, M., Nakagawa, Y., \& Millar, T.~J. 2007, \apj, 661, 334
\bibitem[\"{O}berg et al.(2009a)]{oberg09a} \"{O}berg, K.~I., van~Dishoeck, E.~F., \& Linnartz, H. 2009, \aap, 496, 281 
\bibitem[\"{O}berg et al.(2009b)]{oberg09b} \"{O}berg, K.~I., Linnartz, H., Visser, R., \& van~Dishoeck, E.~F. 2009, \apj, 693, 1209
\bibitem[\"{O}berg et al.(2009c)]{oberg09c} \"{O}berg, K.~I., Garrod, R.~T., van~Dishoeck, E.~F., \& Linnartz, H. 2009, \aap, 504, 891
\bibitem[\"{O}berg et al.(2011)]{oberg11} \"{O}berg, K.~I., Boogert, A.~C.~A., Pontoppidan, K.~M., et al. 2011, \apj, 740, 109
\bibitem[Oldenborg et al.(1992)]{oldenborg92} Oldenborg, R.~C., Loge, G.~W., Harradine, D.~M., \& Winn, K.~R. 1992, J.~Phys.~Chem., 96, 8426
\bibitem[Pagani et al.(2003)]{pagani03} Pagani, L. Olofsson, A.~O.~H., Bergman, P. et al. 2003, \aap, 402, L77 
\bibitem[Pascucci et al.(2008)]{pascucci08} Pascucci, I., Apai, D., Hardegree-Ullman, E.~E., et al. 2008, \apj, 673, 477
\bibitem[Pascucci et al.(2009)]{pascucci09} Pascucci, I., Luhman, K., Henning, Th., et al. 2009, \apj, 696, 143
\bibitem[Pascucci et al.(2013)]{pascucci13} Pascucci, I., Herczeg, G., Carr, J.~S., \& Bruderer, S. 2013, \apj, 779, 178
\bibitem[Pontoppidan et al.(2009)]{pontoppidan09} Pontoppidan, K.~M., Meijerink, R., Dullemond, C.~P., \& Blake, G.~A 2009, \apj, 704, 1482
\bibitem[Pontoppidan et al.(2010)]{pontoppidan10} Pontoppidan, K.~M., Salyk, C., Blake, G.~A., et al. 2010, \apj, 720, 887
\bibitem[Prasad \& Huntress(1980)]{prasad80} Prasad, S.~S. \& Huntress, W.~T. 1980, \apj, 43, 1 
\bibitem[Preibisch et al.(2005)]{preibisch05} Preibisch, T., Kim, Y.-C., Favata, F., et al. 2005, \apjs, 160, 401
\bibitem[Qi et al.(2013)]{qi13} Qi, C., \"{O}berg, K.~I., Wilner, D.~J., et al. 2013, Science, 341, 630
\bibitem[Raksit et al.(1984)]{raksit84} Raksit, A.~B., Schiff, H.~I., \& Bohme, D.~K. 1984, Int.~ J.~Mass~Spectrom.~Ion~Phys., 56, 321
\bibitem[Roberts \& Herbst(2002)]{roberts02} Roberts, H. \& Herbst, E. 2002, \aap, 395, 233
\bibitem[Rodgers \& Smith(1996)]{rodgers96} Rodgers, A.~S. \& Smith, G.~P. 1996, Chem.~Phys.~Lett., 253, 313
\bibitem[Salyk et al.(2008)]{salyk08} Salyk, C., Pontoppidan, K.~M., Blake, G.~A., et al. 2008, \apj, 676, L49
\bibitem[Salyk et al.(2011)]{salyk11} Salyk, C., Pontoppidan, K.~M., Blake, G~.A., Najita, J.~R., \& Carr, J.~S. 2011, \apj, 731, 130
\bibitem[Schwarz \& Bergin(2014)]{schwarz14} Schwarz, K.~R. \& Bergin, E.~A. 2014, \apj, 797, 113
\bibitem[Semaniak et al.(2001)]{semaniak01} Semaniak, J., Minaev, B.~F., Derkatch, A.~M., et al. 2001, \apjs, 135, 275  
\bibitem[Sha et al.(2005)]{sha05} Sha, X., Jackson, B., Lemoine, D., Lepetit, B. 2010, \jcp, 122, 14709 
\bibitem[Sims et al.(1994)]{sims94} Sims, I.~R., Queffelec, J.,-L., Defrance, A. et al. 1994, \jcp, 100, 4229 
\bibitem[Smith et al.(1992)]{smith92} Smith, D., Spanel, P., \& Mayhew, C.~A. 1992, Int.~ J.~Mass~Spectrom.~Ion~Phys., 117, 457
\bibitem[Smith et al.(2004)]{smith04} Smith, I.~W.~M., Herbst, E., \& Chang, Q. 2004, \mnras, 350, 323 
\bibitem[St\"{a}uber et al.(2005)]{stauber05} St\"{a}uber, P., Doty, S.~D., van~Dishoeck, E.~F., \& Benz, A.~O. 2005, \aap, 440, 949 
\bibitem[Stevenson \& Lunine(1988)]{stevenson88} Stevenson, D. J. \& Lunine, J.~I. 1988, Icarus, 75, 146
\bibitem[Thi et al.(2013)]{thi13} Thi, W.-F., Kamp, I., Woitke, P., et al. 2013, \aap, 551, A49
\bibitem[Thiabaud et al.(2014)]{thiabaud14} Thiabaud, A., Marboeuf, U., Alibert, Y., Leya, I., \& Mezger, K. 2014, \aap, 574, A138  
\bibitem[Tielens \& Hagen(1982)]{tielens82} Tielens, A.~G.~G.~M. \& Hagen, W. 1982, \aap, 114, 245
\bibitem[Tsang \& Herron(1991)]{tsang91} Tsang, W. \& Herron, J.~T. 1991, J.~Phys.~Chem.~Ref.~Data, 20, 609
\bibitem[Tsang et al.(1986)]{tsang86} Tsang, W. \& Hampson, R.~F. 1986, J.~Phys.~Chem.~Ref.~Data, 15, 1087
\bibitem[Vandenbussche et al.(1999)]{vandenbussche99} Vandenbussche, B., Ehrenfreund, P., Boogert, A.~C.~A., et al. 1999, \aap, 346, L57
\bibitem[van Dishoeck \& Black(1988)]{vandishoeck88} van~Dishoeck, E.~F. \& Black, J,~H, 1988, \apj, 334, 771
\bibitem[van Dishoeck et al.(2006)]{vandishoeck06} van~Dishoeck, E.~F., Jonkheid,  B., \& van Hemert, M.~C. 2006, Faraday Discuss., 133, 231
\bibitem[van Dishoeck et al.(2013)]{vandishoeck13} van~Dishoeck, E.~F., Herbst, E., \& Neufeld, D.~A. 2013, Chem. Rev., 113, 9043
\bibitem[Vasyunin \& Herbst(2013)]{vasyunin13} Vasyunin, A.~I. \& Herbst, E. 2013, \apj, 762, 86
\bibitem[Viggiano et al.(1980)]{viggiano80} Viggiano, A.~A., Howorka, F., Albritton, D.~L., et al. 1980, \apj, 236, 492
\bibitem[Visser et al.(2009)]{visser09} Visser, R., van~Dishoeck, E.~F., \& Black, J.~H. 2009, \aap, 503, 323
\bibitem[Visser et al.(2011)]{visser11} Visser, R., Doty, S.~D., \& van~Dishoeck, E.~F. 2011, \aap, 534, A132
\bibitem[Wakelam et al.(2012)]{wakelam12} Wakelam, V., Herbst, E., Loison, J.-C., et al. 2012, \apj, 199, 21
\bibitem[Walsh et al.(2010)]{walsh10} Walsh, C., Millar, T.~J., \& Nomura, H. 2010, \apj, 722, 1607
\bibitem[Walsh et al.(2012)]{walsh12} Walsh, C., Nomura, H., Millar, T.~J., \& Aikawa, Y. 2012, \apj, 747, 114 
\bibitem[Walsh et al.(2014)]{walsh14} Walsh, C., Millar, T.~J., Nomura, H. et al. 2014, \aap, 563, A33
\bibitem[Weingartner \& Draine(2001)]{weingartner01} Weingartner, J.~C. \& Draine, B.~T. 2001, \apj, 548, 296 
\bibitem[Willacy et al.(1998)]{willacy98} Willacy, K., Klahr, H.~H., Millar, T.~J., \& Henning, Th. 1998, \aap, 338, 995
\bibitem[Willacy \& Woods(2009)]{willacy09} Willacy, K. \& Woods, P.~M. 2009, \apj, 703, 479
\bibitem[Williams \& Cieza(2011)]{williams11} Williams, J.~P. \& Cieza, L.~A. 2011, \araa, 49, 67
\bibitem[Woitke et al.(2009)]{woitke09} Woitke, P.,  Thi, W.-F., Kamp, I., \& Hogerheijde, M. 2009, \aap, 501, L5
\bibitem[Woods \& Willacy(2009)]{woods09} Woods, P.~M. \& Willacy, K. 2009, \apj, 693, 1360
\bibitem[Wright et al.(2004)]{wright04} Wright, G.~S., Reike, G., van Dishoeck, E.~F., et al. 2004, in SPIE Conf. Ser. 5487, ed.~J.~C.~Mather, 653
\bibitem[Yildiz et al.(2013)]{yildiz13} Yildiz, U.~A., Acharyya, K., Goldsmith, P.~F., et al. 2013, \aap, 558, A58
\bibitem[Zinnecker \& Preibisch(1994)]{zinnecker94} Zinnecker. H \& Preibisch, Th. 1994, \aap, 292, 152
\end{thebibliography}
\end{document}